\documentclass[preprint]{aastex}

\usepackage{amsmath}    
\usepackage{epsfig}
\usepackage{natbib}                
\usepackage{lscape}
\usepackage{verbatim}              
\usepackage{graphicx}              
\usepackage{amsmath,amsthm,amsfonts,amsopn,amssymb} 
\usepackage{afterpage}
\usepackage{rotating}					
\usepackage{ulem}

\shorttitle{Stellar Diameters and Temperatures IV} 

\shortauthors{Boyajian, van Belle, \& von Braun}

\begin{document}
\title{Stellar Diameters and Temperatures \\
IV. Predicting Stellar Angular Diameters}

\author{Tabetha S. Boyajian\altaffilmark{1},
Gerard van Belle\altaffilmark{2},
Kaspar von Braun\altaffilmark{3}
}

\altaffiltext{1}{Department of Astronomy, Yale University, New Haven, CT 06511}
\altaffiltext{2}{Lowell Observatory, Flagstaff, AZ 86001}
\altaffiltext{3}{Max Planck Institute for Astronomy, K\"{o}nigstuhl 17, 69117 Heidelberg, Germany}



\begin{abstract}

The number  of stellar angular diameter measurements has greatly increased over the past few years due to innovations and developments in the field of long baseline optical interferometry (LBOI). We use a collection of high-precision angular diameter measurements for nearby, main-sequence stars to develop empirical relations that allow the prediction of stellar angular sizes as a function of observed photometric color. These relations are presented for a combination of 48 broad-band color indices.  We empirically show for the first time a dependence on metallicity to these relations using Johnson $(B-V)$ and Sloan $(g-r)$ colors. Our relations are capable of predicting diameters with a random error of less than 5\% and represent the most robust and empirical determinations to stellar angular sizes to date. 

\end{abstract}




\section{Introduction}
\label{sec:introduction}

Empirical determination of stellar radius provides essential constraints in a variety of fields within astrophysics. For instance, stellar models or indeed any calibration equations that involve stellar radii have to be able to reproduce stellar radius measurements based on, e.g., interferometry or the study of eclipsing binary stars \citep{tor10}. Exoplanetary parameters are functions of stellar astrophysical parameters \citep{von12}. Any ability to reliably predict stellar sizes would be of great use for the studies of microlensing events \citep{cal10}, searches for KBOs \citep{wan10}, asteroseismology \citep{hub12}, binary stars \citep{sou05a}, and exoplanet transits \citep{ass09}, and many more.

It has long been known that there exists a relationship between stellar surface brightness and stellar broad-band colors \citep{wes69}.  From the Stephan-Boltzmann Law, $L \propto R^2 T_{\rm eff}^4$, it can also be established that surface brightness is dependent upon apparent magnitude and stellar angular size \citep[e.g., see the discussion in the introduction of ][]{bar76}.  These two relations can be combined to show that stellar angular size can be predicted on the basis of multi-color photometry. Stellar color -- based on two bands -- can be calibrated to produce an `absolute' (constant distance) angular size, i.e., the angular size that a star would have if it were moved to a distance at which its apparent magnitude were zero. The apparent stellar magnitude in one of the two bands can then be used to provide the scale between that `zero-magnitude' angular size and the measured angular size at its actual distance. \citet{bar76} demonstrate the gross insensitivity of these techniques to interstellar reddening, greatly extending their general applicability throughout astrophysics, e.g., for calibration of Cepheid period-luminosity relations \citep{dib95,fou97,ker04d,gro07,sto11a,sto11b} or distances to eclipsing binaries \citep{lac77b, sou05a}.

Since stellar apparent magnitudes are direct observables, the methods described above are simple in their application once they are calibrated.  Until recently, however, the key missing element was a sufficient body of data with which to carry out those empirical calibrations. Stellar photometry was available, but empirically measured stellar angular sizes were not.  Optical interferometry provides the most direct measures of stellar angular diameters - where only the transformation from uniform disk to limb-darkened angular diameter rely on model calculations.  Interferometry has provided angular diameters for hundreds of evolved stars over the past 15 years \citep[eg.,][]{van99a,moz03}, but similarly sized homogenous data sets for main sequence stars have only become available in the last few years with our concerted CHARA Array program \citep{boy08,boy09,boy12a,boy12b, boy13a}.

This paper provides an updated account of, as well as an extension to, empirically derived surface brightness calibrations presented in earlier work \citep{hin89,dib98,van99b,moz03,ker04c,dib05}, specifically using the non-linear construction of their functions employed within \citet{bon06} and \citet{ker08a}.  Section~\ref{sec:data} describes the sample and input data, and Section~\ref{sec:discussion} presents our results with a discussion on their applications. We summarize and conclude in \S \ref{sec:conclusion}.


\section{Data and Analysis}
\label{sec:data}


\subsection{Definitions}
\label{sec:definitions}

\begin{figure}										
\centering
\epsfig{file=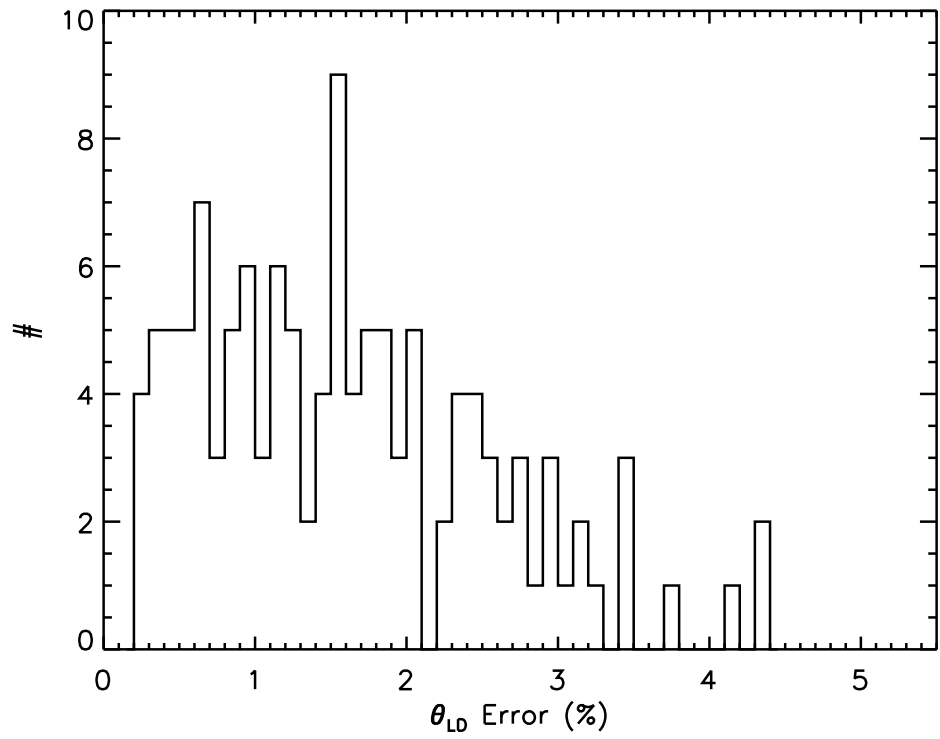, width=0.9\linewidth, clip=} 
  \caption[ ] {A histogram showing the distribution of the angular diameter errors for stars used in this sample.  See Section~\ref{sec:definitions} for details.}
  \label{fig:LDerrors_histogram}
  \end{figure}

Data for our analysis are taken from the compilation of angular diameters, broad-band photometry, and metallicities presented in \citet{boy12b} and \citet{boy13a}.  This anthology of angular diameter measurements come from an assortment of interferometers: the CHARA Array \citep{bai08, boy12a, lig12, dif04, big11, von11a, cre12, big06, baz11,hub12,cre12, bai12}, the Palomar Testbed Interferometer PTI \citep{van09}, the Very Large Telescope Interferometer VLTI \citep{ker03b, dif04, the05, ker03a, chi12, ker04b}, the Sydney University Stellar Interferometer SUSI \citep{dav11}, the Narrabri Stellar Intensity Interferometer NSII \citep{han74}, Mark~III \citep{moz03}, and the Navy Prototype Optical Interferometer NPOI \citep{nor99, nor01}.  The interferometer measures a uniform disk angular diameter, which is transformed into a limb-darkened angular diameter using calculated coefficients from model atmospheres (e.g. \citealt{cla00}) that depend on the wavelength of observation as well as the stellar photospheric properties.  The majority of the observations were made in the infrared where the correction from uniform disk to limb-darkened diameter is small, $\sim 2-3$\%, and the errors in the correction for limb darkening contribute $<<0.1$\% to the total angular diameter error budget.  We impose a limit to only include objects with angular diameters measured to better than 5\%.  The resulting sample has errors spanning a range of $0.2$ to $4.3$\% with a median error of $1.5$\%.  This distribution of angular diameter errors is plotted in Figure~\ref{fig:LDerrors_histogram}.  

\citet{boy13a} use the interferometric sample described above to build an assortment of color-temperature relations.  For our analysis, we use the broad-band photometry measurements compiled in \citet{boy13a} from the Johnson $(BVR_{\rm J}I_{\rm J}JHK)$, Cousins $(R_{\rm C}I_{\rm C})$, Kron $(R_{\rm K}I_{\rm K})$, Sloan $(griz)$, and WISE $(W_{\rm 3}W_{\rm 4})$ photometric systems.  Uncertainties in the photometric magnitudes are not used in the analysis, and reference consistency for each bandpass was minded where ever possible (see \citealt{boy13a} for references and details).  Unlike the analysis in \citet{boy13a}, we refrain from using {\it 2MASS} $(JHK)$ photometry in our analysis because it is saturated for the majority of these bright stars and thus introducing several percent more uncertainty in the results.  

We also adopt the metallicities compiled in \citet{boy12b} and \citet{boy13a} to use for our analysis. It was noted in \citet{boy12b} and \citet{boy13a} that there is no uniform source of metallicity measurements for the complete sample of interferometrically observed stars in the anthology.  This is problematic because although the precision of measuring metallicity within a method is very good ($\pm 0.03$~dex; \citealt{val05}), the accuracy between references is severely lacking (up to 0.2~dex; \citealt{tor12}).   Where systematic offsets between different groups using different analysis methods exist, the sensitivity of using the data for the empirical calibrations is greatly reduced, ultimately compromising the precision of our relations.  This absence of a complete {\it and} uniform data set prompted \citet{boy13a} to reference the metallicities for the anthology stars from the \citet{and11} catalog, where the values are an average over numerous references available in the literature.  This crude approach is currently the best option to unify the data set. However, using the averages makes accurate characterization of the metallicity errors difficult, particularly on a object - to - object basis.  For these reasons, we do not include the errors of metallicity in our analysis.  

In summary, the sample we use consists of 124 main-sequence stars with measured limb-darkened angular diameters of better than 5\% precision.  The sample has spectral types ranging from A to M, with metallicities fairly evenly distributed around solar $\pm 0.5$~dex.  All stars reside in the local neighborhood ($\sim< 50$~pc), thus no corrections for reddening was applied to their published photometric magnitudes.   

The zero-magnitude angular diameter, $\theta_{m_{\lambda}=0}$ in Equation \ref{eq:zmld_log}, represents the angular diameter that are star would have if it were at a distance at which its apparent magnitude equals zero (see \S \ref{sec:introduction}); it is thus a wavelength-dependent quantity. It is defined as:

\begin{eqnarray}
\label{eq:zmld_log}
	\log \theta_{m_{\lambda}=0} = \log \theta_{\rm LD} +  0.2 m_{\lambda},
\end{eqnarray}

\noindent where $m_{\lambda}$ is the apparent magnitude for a star in filter $\lambda$, and $\theta_{\rm LD}$ is the angular diameter of the star, corrected for limb-darkening. Full derivations of this equation can be found in \citet{dib93, fou97, ker04c}, and references therein\footnote{Equation~\ref{eq:zmld_log} is also functionally identical to the definition found in \citet{hin89} (equation 5): $S_V = 5 log \theta + V_0$, although this formulation omits the intermediary surface brightness term, $S_V$, an issue to be addressed in a forthcoming paper.}.  

Historically, the values for $\theta_{m_{\lambda}=0}$ in different filters were linearly modeled with respect to broad-band color-index.  These relations, initially developed in \citet{bar76} and coined the \textquotedblleft Barnes-Evans relations\textquotedblright, would, for example, take the form of $a + b X$, where $X$ is some color index $(\lambda-\lambda^{\prime})$.  More recently, non-linear functions have preferentially been used to map literature photometry data to predict stellar angular size (e.g. from \citealt{bon06,ker08a}).  In order to model the data in this work, we continue the use of a polynomial:


\begin{eqnarray}
\label{eq:poly3}
\log \theta_{m_{\lambda}=0} = \sum_{i=0}^{n} a_i X^i
\end{eqnarray}

\noindent where $X$ is the observed color index.  


\subsection{Application}
\label{sec:application}


\subsubsection{Angular Diameters from 2-Band Photometry}
\label{sec:photometry}

We solve for the results defined by Equation \ref{eq:poly3} using the non-linear least squares fitting function \texttt{MPFIT.pro} \citep{mar09} in IDL, weighting the data by the errors in the limb-darkened angular diameters.  For every color index we model, we use the resulting $\chi^{2}$ and degrees of freedom to calculate a F-value for successive $i$ versus $i+1$ polynomial models in order to determine the statistical significance of the additional parameter. The F-test probability is computed using IDL (using \texttt{MPFTEST.pro}, also available in the \citealt{mar09} suite of IDL routines), where we deem whether the new fit with additional parameters is statistically significant if $<0.05$.  Section~\ref{sec:metallicity} introduces metallicity as an additional variable into the equation, and where applicable, those solutions should trump the results derived from Equation~\ref{eq:poly3}.

In Table~\ref{tab:SB_poly3_coeffs}, we show the Equation~\ref{eq:poly3} coefficients for all variations of the observable photometric magnitudes and color-indices. The total number of stars used in the fit, $N$,  deviates from 124 because photometric observations are not available for some stars in some bandpasses.  The magnitude range listed in Table~\ref{tab:SB_poly3_coeffs} applies only to the value of the color-index, and these relations should not be extrapolated past these limits.  The two last columns of Table~\ref{tab:SB_poly3_coeffs} give the reduced $\chi^2$ and standard deviation of the residuals in percent, calculated by StdDev~$[(\theta_{m_{\lambda}=0; \rm Obs.} - \theta_{m_{\lambda}=0; \rm Calc.})/\theta_{m_{\lambda}=0; \rm Obs.}] \times 100 \%$.

In Figures~\ref{fig:relations1} through \ref{fig:relations7} we show the data and the polynomial fits of the solutions with the format of Equation \ref{eq:poly3} whose coefficients are given in Table~\ref{tab:SB_poly3_coeffs}.  In each plot, the color of the data point reflects the metallicity of the star, as depicted in the legends.  The bottom panel in each panel of Figures \ref{fig:relations1} through \ref{fig:relations7} shows the fractional residuals of the data to the fit, where the horizontal dotted line marks a zero deviation.  


\subsubsection{Metallicity Dependence}
\label{sec:metallicity}

The metal abundance in a star changes the opacity due to line-blanketing, and thus changes the total flux emitted at different wavelengths, particularly in the bluest of colors.  While intermediate-band photometric systems (e.g. Str{\"o}mgren; \citealt{str63}) have shown to be useful to determine stellar metallicities consistent with spectroscopically derived values, broad-band photometric systems are less sensitive to such measurements in comparison (for example, see \citealt{arn10}, and references therein).  

The dependence of surface-brightness relations on stellar metallicity has been tested but remains empirically unproven \citep{ker08a, ker04c}.  This is a challenging task with the data at hand because not only are the errors in metallicity large, but there are only a modest range in metallicities within the sample.  However, visual inspection of the residuals in the $(B-V)$ model (left hand side of Figure~\ref{fig:BmV_V_FeH}) show a remarkable correlation with metallicity.  The top panel of Figure~\ref{fig:BmV_V_FeH_residuals_log_SB} confirms the trend with the residuals with respect to metallicity in the $(B-V)$ model.  

As such, we test a model that includes an additional parameter to account for a metallicity dependence in each of the modeled color indexes.  We choose a multi-variate polynomial function of the form:


\begin{eqnarray}
\label{eq:poly3_cm}
	 \log \theta_{m_{\lambda}=0} = \sum_{i=0}^{n} a_i X^i + b Y 
\end{eqnarray}


\noindent where stellar color and stellar metallicity [Fe/H] are represented by $X$ and $Y$, respectively.  We also impose limits restricting the data to exclude M-dwarfs ($T_{\rm eff} < 4000$~K) on the account that metallicity uncertainties for the latest-type stars are a factor of ten greater than those for solar-type stars.  

To model each color index, we follow the approach described above (Section~\ref{sec:photometry}) by increasing the number of additional parameters in Equation~\ref{eq:poly3_cm} until no improvement is shown in the test statistics.  The results for this model are then compared with a model which excludes the last term in Equation~\ref{eq:poly3_cm} to determine the probability of random improvement by including metallicity.  Since the magnitude of line-blanketing is a known function of stellar temperature \citep{san69}, we also used this method to test whether adding cross-term parameter (i.e. $c X Y$) to Equation~\ref{eq:poly3_cm} would change the results of the analysis and found that it did not.  

We detect a statistically significant metallicity dependence in the $(B-V$) and the $(g-r)$ color index, which is no surprise since these colors are the shortest filter wavelengths included in this analysis, and thus the most influenced by changes in metallicity.  We find that the $(B-V)$ surface brightness relation has an extremely significant solution (p-value of $\sim 10^{-18}$) when including metallicity as an additional parameter.   This solution is explicitly expressed in Equation~\ref{eq:poly3_cm_bmv}.  It is restricted to be valid for stars with $-0.02 < (B-V) < 0.95$, has a reduced $\chi^2 = 23.5$ (total of $N=100$ stars), and has a standard deviation of the residuals of 4.5\%.

\begin{eqnarray}
\label{eq:poly3_cm_bmv}
	\log \theta_{m_{V}=0} 	&	=	&	(0.52005 \pm     0.00121) + \nonumber \\
							&		&	(0.90209 \pm  0.01348) (B-V) + \nonumber \\
							&		&	(-0.67448 \pm      0.03676) (B-V)^2 + \nonumber \\
							&		&	(0.39767 \pm      0.02611) (B-V)^3 + \nonumber \\
							&		&	(-0.08476 \pm 0.00161) [{\rm Fe/H}] 
\end{eqnarray}

The analysis of the Sloan $(g-r)$ colors leads to calculated a p-value of  $\sim 10^{-5}$, a detection much less outstanding than the $(B-V)$ color, but still a quite significant improvement. The metallicity dependent form of the relation is expressed by Equation~\ref{eq:poly3_cm_gmr}.  The fit uses $N=66$ stars, has a reduced $\chi^2 = 25.5$, and the standard deviation of the residuals is 5.8\%.

\begin{eqnarray}
\label{eq:poly3_cm_gmr}
	\log \theta_{m_{V}=0} 	&	=	&	(0.66645  \pm  0.00138) + \nonumber \\
							&		&	(0.74459  \pm  0.00315) (g-r) + \nonumber \\
							&		&	(-0.08276 \pm  0.00346) [{\rm Fe/H}] 
\end{eqnarray}

In Figures~\ref{fig:BmV_V_FeH} and \ref{fig:gmr_V_FeH} we show side by side the solutions with and without metallicity for the $(B-V)$ and $(g-r)$ colors. Figure~\ref{fig:BmV_V_FeH_residuals_log_SB} shows the trends detected in the surface brightness relation residuals with respect to the star's metallicity.   This improvement when including metallicity in the model argues that the inclusion of metallicity cannot be ignored when using the $(B-V)$ or $(g-r)$ color indexes.  This is the first time such a dependence on metallicity has been found from empirical data on surface brightness relations (see discussion in Section~\ref{sec:discussion}).


\section{Discussion}
\label{sec:discussion}

We compare our results to the works from  \citet{van99b}, \citet{bon06} and \citet{ker08a}, all of which use predictions based on empirical values to determine stellar angular sizes.

The \citet{van99b} calibration for main-sequence stars uses a limited data set based on \citet{bro74} -- all that was available at the time -- to derive their calibrations.  This data set consisted of 11 main-sequence B- and A-type stars ($-0.5 \lesssim (V-K) \lesssim +0.5$), and is extrapolated to the Sun's position at $(V-K) \simeq 1.5$.   We plot their ($V-K$) relation (green triple dot-dashed line) in Figure~\ref{fig:relations5}.  This shows only a slight systematic offset of the \citet{van99b} relation to smaller diameters, however, considering their small sample size used in the calibration, this offset is statistically insignificant with the quoted one sigma errors of the relation in \citet{van99b}.

The \citet{bon06} work, also referred to as the The JMMC Stellar Diameters Catalog or JSDC\footnote{http://www.jmmc.fr/catalogue\_jsdc.htm}, is the backbone to the The SearchCal catalog, managed by the JMMC working group\footnote{http://www.jmmc.fr/searchcal\_page.htm}.  It is a tool to allow an observer to search for suitable interferometric calibrators in an interface tailored to their individual needs.  The application allows for the observable photometric colors and estimated angular sizes to be used in conjunction with observatory and instrument configuration in order to derive instrumental visibility estimates for each source.  The method used to derive the predicted angular diameters is outlined and defined in \citet{bon06}.  To calibrate the \citet{bon06} relations, a broad source of data available from eclipsing binaries, interferometric measurements, as well as lunar occultations are used \citep{bar78, and91, seg03, moz03}.  The \citet{bon06} calibration data cover the full extent of luminosity classes (V to I) and spectral types (O to M).  They use $4^{th}$ order polynomials to model the data for the $(B-V)$ and $(V-K)$ relations, and a $5^{th}$ order polynomial to model the data for $(V-R)$, where the solutions to their fits yield uncertainties of 8\%, 10\%, and 7\% for $(B-V)$, $(V-R)$, and $(V-K)$, respectively.  

We show the $(B-V)$, $(V-R)$, and $(V-K)$ model fits from \citet{bon06} along with our data in Figure~\ref{fig:BmV_V_FeH}, \ref{fig:relations4}, and \ref{fig:relations5}, respectively  (blue dash-dot line). The $(B-V)$ and $(V-K)$ data and models show that the \citet{bon06} solution is indistinguishable from the one presented here.  Major differences in our model to the \citet{bon06} model are only relevant within the $(V-R)$ relations.  We suspect that this is possibly due to the \citet{2002yCat.2237....0D} $R$-band photometric systems used in \citet{bon06} not being being identical to the Johnson $R$-band used here. In Figure~\ref{fig:Compare_JMMC_diameters_new} we show a direct comparison of the interferometrically measured angular diameters to the final angular diameters in the JMMC Catalog.  We find that $\theta_{\rm Interferometry}/\theta_{\rm JMMC} = 0.999 \pm 0.095$, undoubtedly excellent agreement.  

The \citet{ker08a} relations are calibrated using interferometrically measured stellar angular diameters.  Similar to our sample selection, they use only stars with diameters measured to better than 5\% precision, which reside on or near the main-sequence. At the time \citet{ker08a} was published, the sample size available for calibrating their relations consisted of 34 stars.  Since then, major advances in the field of interferometry have made it possible for several interferometric diameter surveys of nearby stars to contribute to the field (e.g. \citealt{van09, boy12a, boy13a, hub12}). Aside from the advantage of statistical improvement by virtue of possessing a larger sample size, we are afforded the luxury of being able to treat the data for our analysis in a variety of ways without sacrificing significance.  For instance, \citet{ker08a} consider repeated angular diameter measurements of the same star as unique sources. In order to avoid any bias in our results, we are able to use the weighted averages presented in the \citet{boy12b} and \citet{boy13a} compilations, if multiple measurements exist\footnote{If we considered every radius measurement an independent quantity, a sample size of $n=137$ stars would be available for analysis.  The averaging of repeated sources brings this sample size down to $n=124$ stars.}. 

We show the \citet{ker08a} $(B-V)$, $(V-R_C)$, and $(V-I_C)$ relations in Figures~\ref{fig:BmV_V_FeH} and \ref{fig:relations4} (red dashed line). The \citet{ker08a} $(V-R_C)$, and $(V-I_C)$ relations are consistent with our own for stars earlier than $\sim$~K0 ($(V-R_C) \lesssim 0.7$, and $(V-I_C) \lesssim 1.2$), but over-overestimate diameters by $\sim 10$\% for the latest-type stars, a difference of $\sim 2$-$\sigma$.  The \citet{ker08a} $(B-V)$ relation compared to the one presented here as well as the one in \citet{bon06} differs for stars earlier than $\sim$~F3 (or $(B-V) \lesssim 0.35$), by $\sim 15$\%, also $\sim 2$-$\sigma$.  Stars later than $\sim$~M0 (or $(B-V) \gtrsim 1.5$) also show disagreement between our solution and \citet{ker08a}, however these differences can be attributed to the sparse sampling of data used in \citet{ker08a} with respect to color-index, especially on the endpoints of the relation, as well as the increased amount of scatter in the full data set used in both calibrations.

Nonetheless, compared to the \citet{ker08a} treatment, our solutions do not improve the precision of the (metallicity independent) surface brightness relations.  Specifically, the scatter in the $(B-V)$ relations both yield just below 8\%, although the sample size is different: $n=124$ and $n=42$ for this work and \citet{ker08a}, respectively.  The $(V-R_{\rm C})$ and $(V-I_{\rm C})$ relations show contrasting improvement with the new data added to the sample: $\sigma (V-R_{\rm C}) =4.9, 4.5$\% and $\sigma (V-I_{\rm C}) = 5.1, 5.6$\% for this work and \citealt{ker08a}, respectively.  However, we note that the inclusion of metallicity in our $(B-V)$ model knocks the scatter down to 4.5\%.  

We detect a significant dependence of the stellar metallicity on the $(B-V)$ and $(g-r)$ surface brightness relations.  But  why was this not detected before now?   
The linear surface brightness relations in \citet{ker04c} showed no correlation with metallicity on the visible to infrared calibrations (($B-L$); see their figure~5).  Our detection is a likely consequence of several factors:
\begin{itemize}
\item Metallicities of the 29 stars used in the \citet{ker04c} calibrations come from {\it nine} different references; concerns with systematics between data sets is relevant at this level of inconsistency.
\item Over the past decade, M-dwarf metallicities have been refined using more sophisticated techniques and calibrations.  In fact, {\it all} metallicities for the M-dwarfs cited in \citet{ker04c} have been refined, exceeding differences greater than $1.0$~dex.  The  M-dwarfs comprise 25\% of the 29 stars in the \citet{ker04c} sample and they use these M-dwarfs to draw their conclusions on metallicity.  Our decision to reject M-dwarfs decreases our sample size by two dozen (also $\sim 25$\%), however we have a healthy size sample remaining (N=100) to draw statistically significant results
\item The visual-to-infrared $(B-L)$ color index is not sensitive to metallicity. 
\item The sample size has increased by a factor of 4 since the \citet{ker04c} calibrations, allowing for statistically robust analysis.
\end{itemize} 

The more recent, non-linear calibrations using visible photometry presented in \citet{ker08a} did not explore the possible impact of metallicity on the relations, but it was suggested to the reader that it was expected to be small compared to the intrinsic dispersion of the relations.  Our solution for $(B-V)$ and $(g-r)$ show that this is not the case, and even though the typical metallicity of our sample is close to solar with a dispersion of $\pm 0.5$~dex, including metallicity in these solutions drops the scatter of the fits to be similar of those that do not show an [Fe/H] dependence (Table~\ref{tab:SB_poly3_coeffs}).  We are undoubtedly helped by the larger sample size, as well as a more consistent method of treating the  available for this work for such a clear trend to become apparent in these two color indexes.  

It is still the case that we are not able to break much below 5\% precision on the surface brightness relations, even with the inclusion of metallicity as an additional parameter. We attempted two further sieves of the data where we rejected all stars with angular diameter errors $> 2$~\% and all with errors $> 1$~\%, leaving a total of $N=86$ and $N=40$ calibration stars, respectively.  Unfortunately in no case does this improve the $\chi^2$ nearly enough to be statistically significant below a p-value of $\sim 0.99$ when comparing to the full data set.  This is expected however since the fits we present are weighted to the errors in angular diameter.  At this point, it is unclear whether the large reduced $\chi^2$'s of the fits are astrophysical in nature, or purely measurement error in observed metallicities and/or photometric magnitudes.  The importance of a complete and uniform database of these calibration star's stellar properties can not be understated.


\section{Summary and Conclusion}
\label{sec:conclusion}

\begin{figure}										
\centering
\begin{tabular}{cc}
\epsfig{file=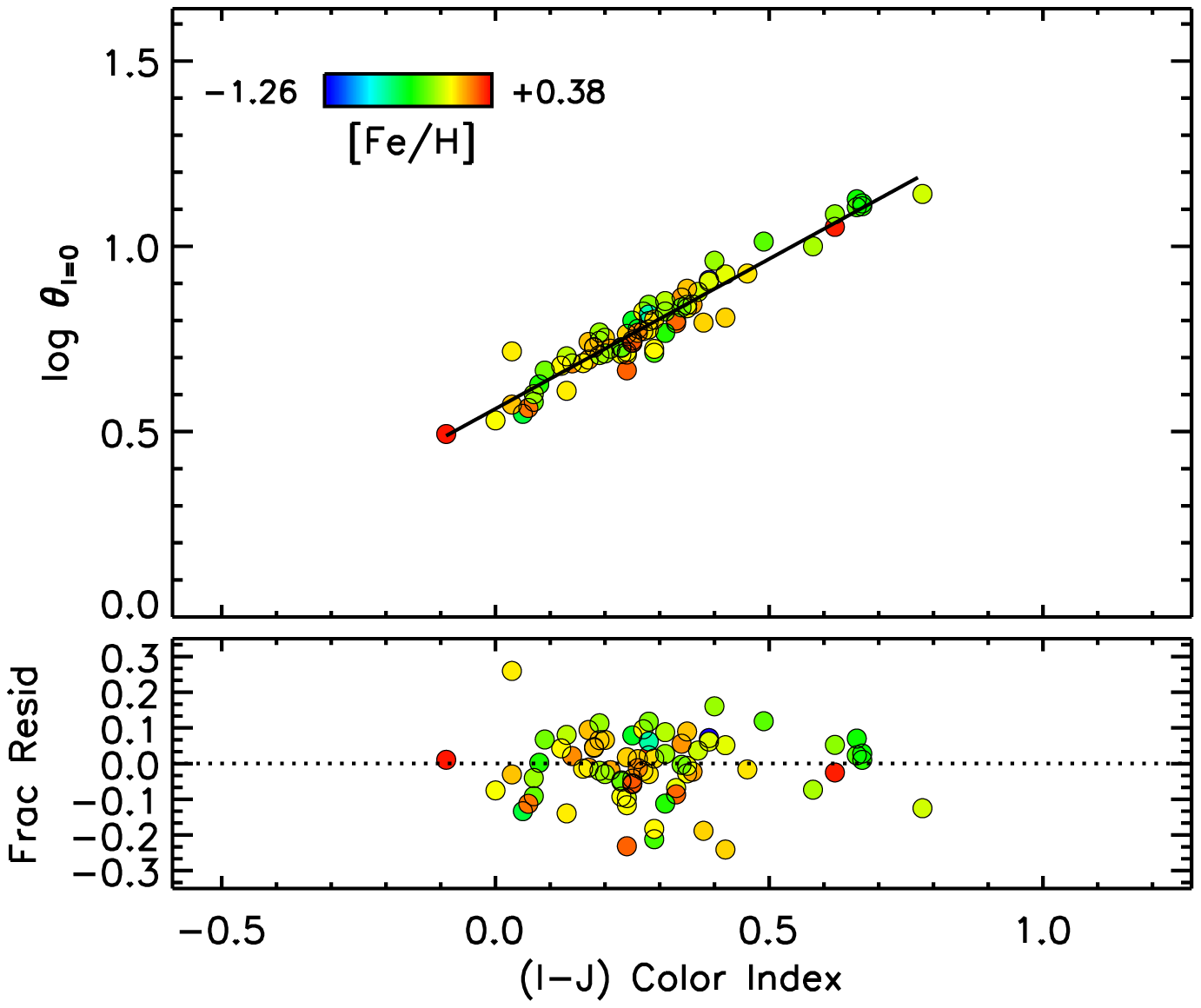, width=0.5\linewidth, clip=}	&
\epsfig{file=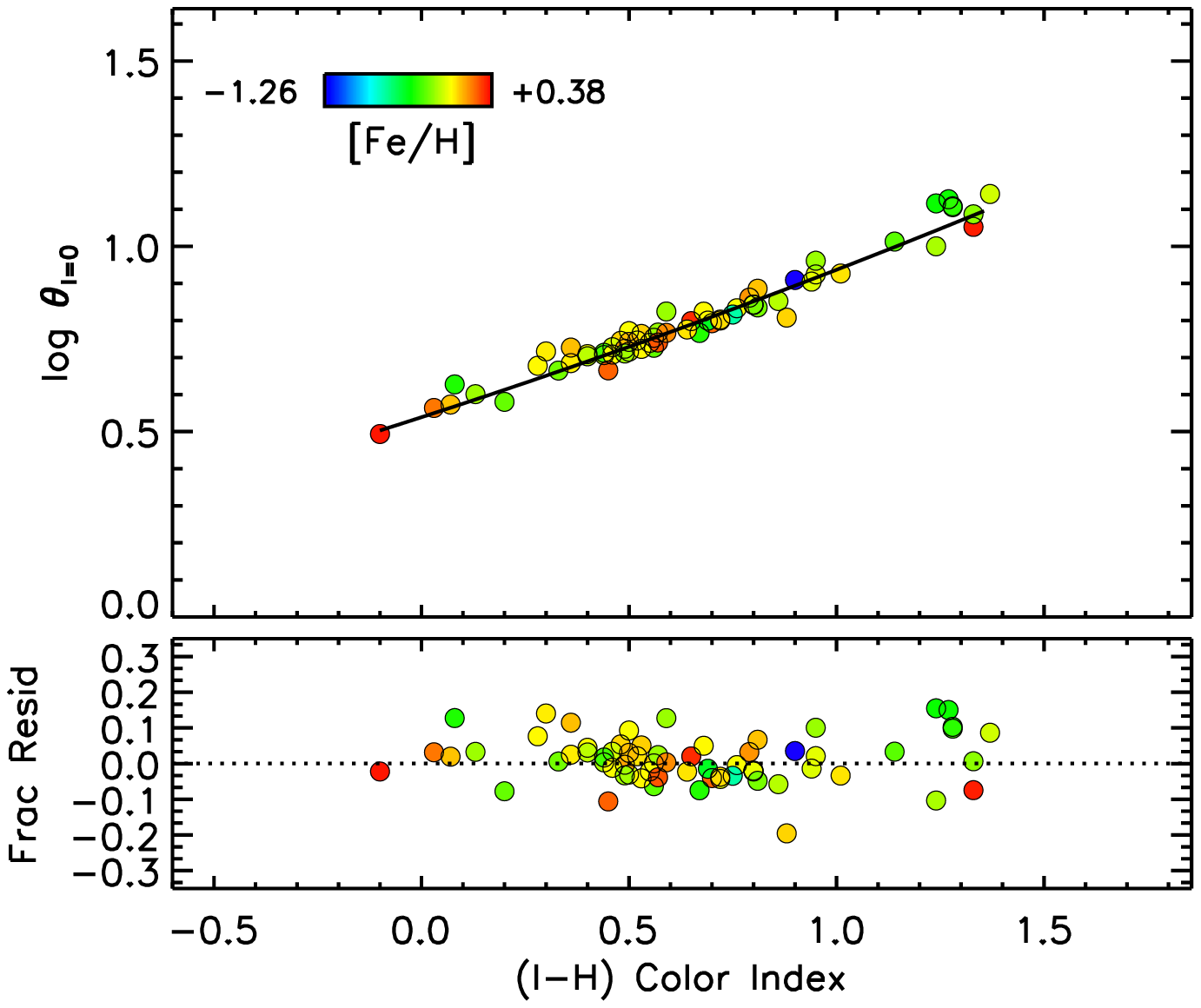, width=0.5\linewidth, clip=}	\\
\epsfig{file=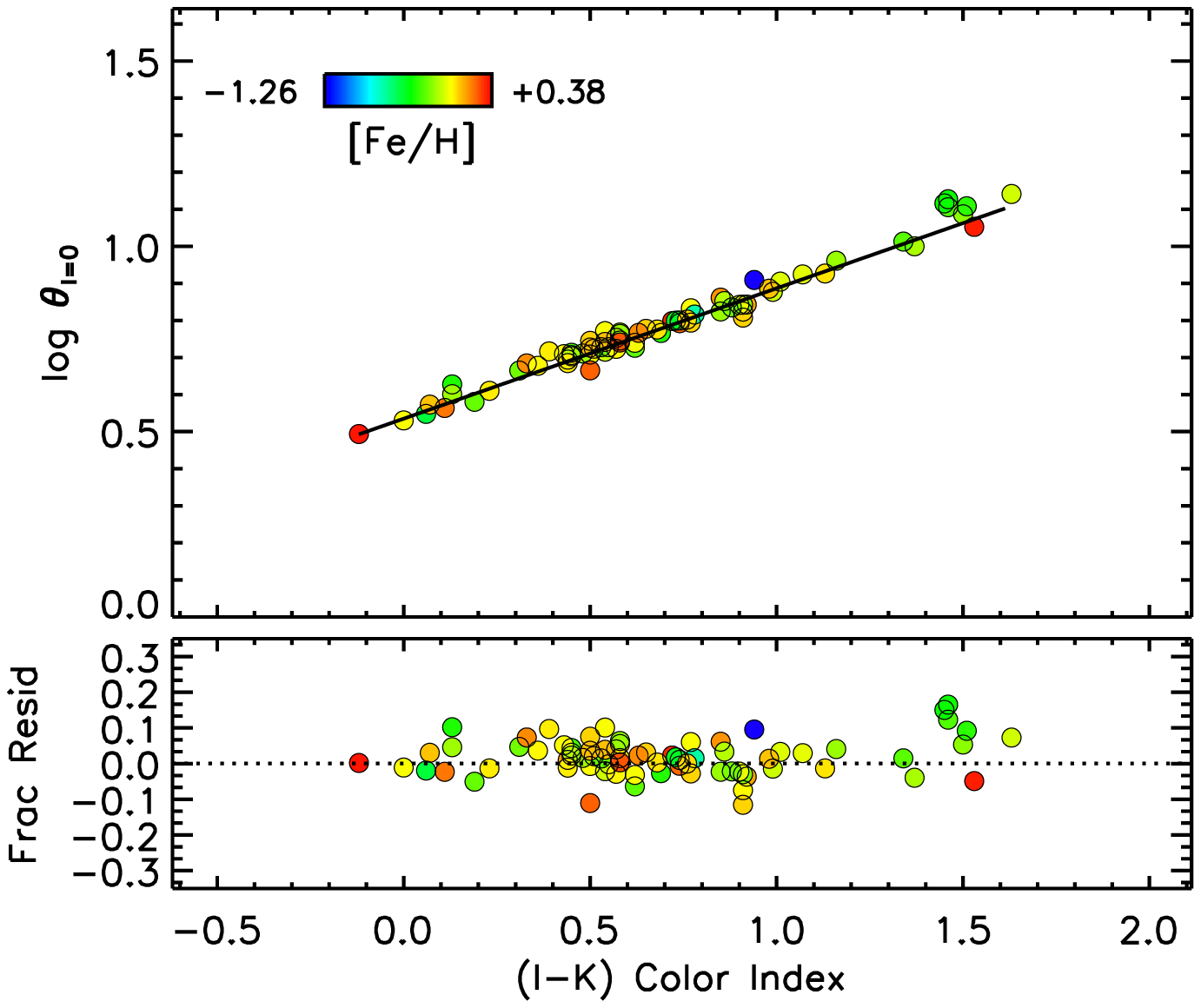, width=0.5\linewidth, clip=}	&
\epsfig{file=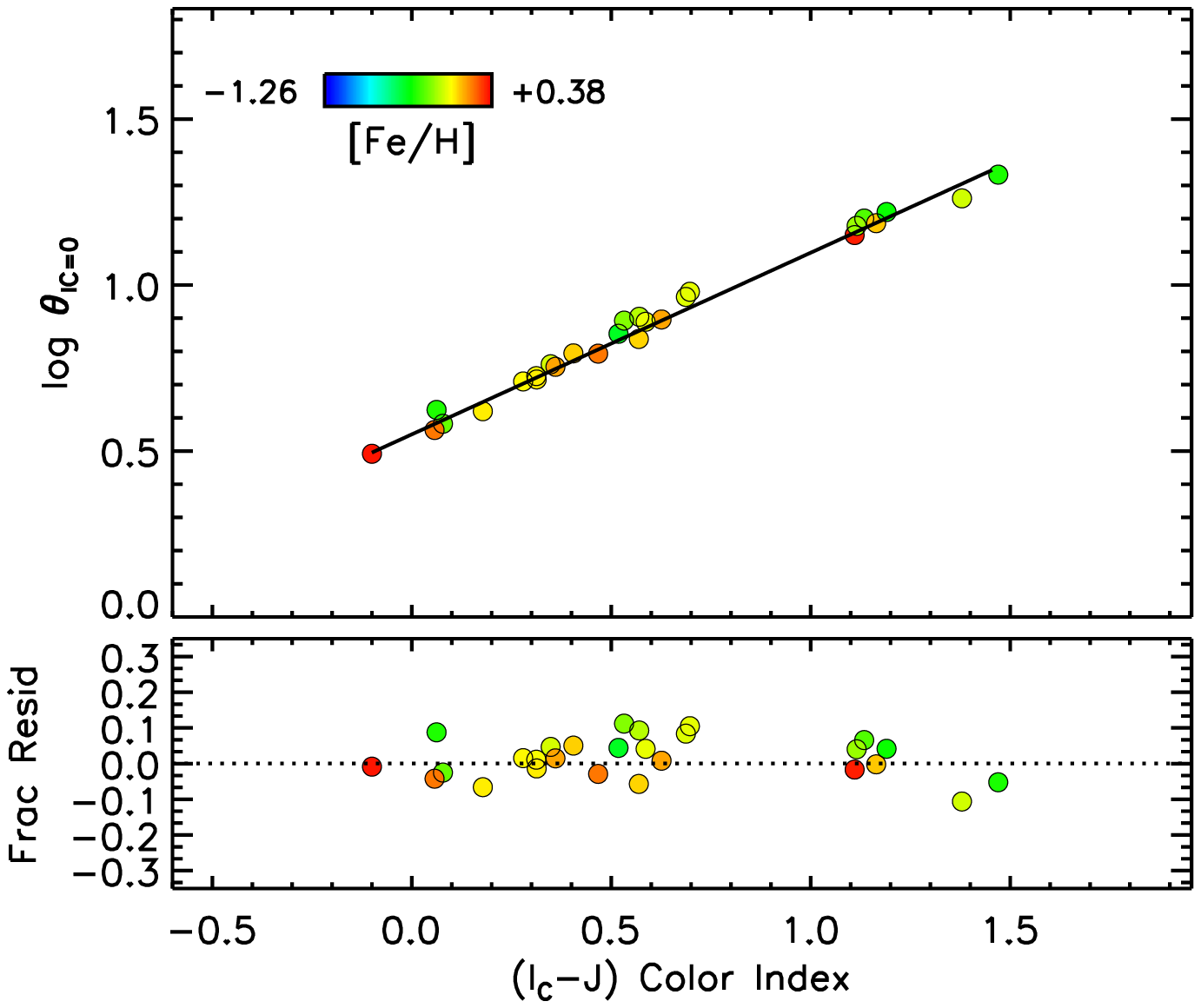, width=0.5\linewidth, clip=}	\\
\epsfig{file=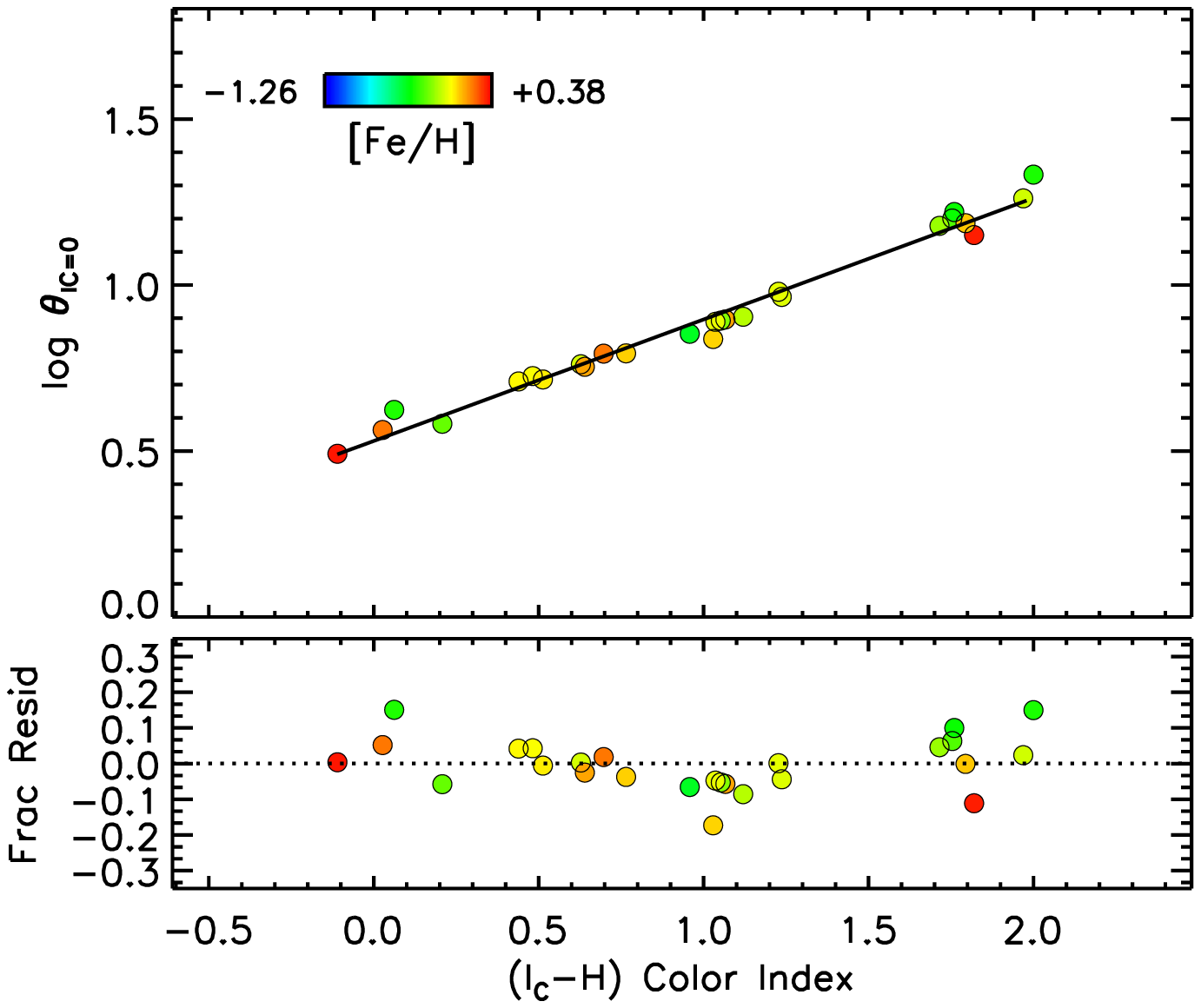, width=0.5\linewidth, clip=}	&
\epsfig{file=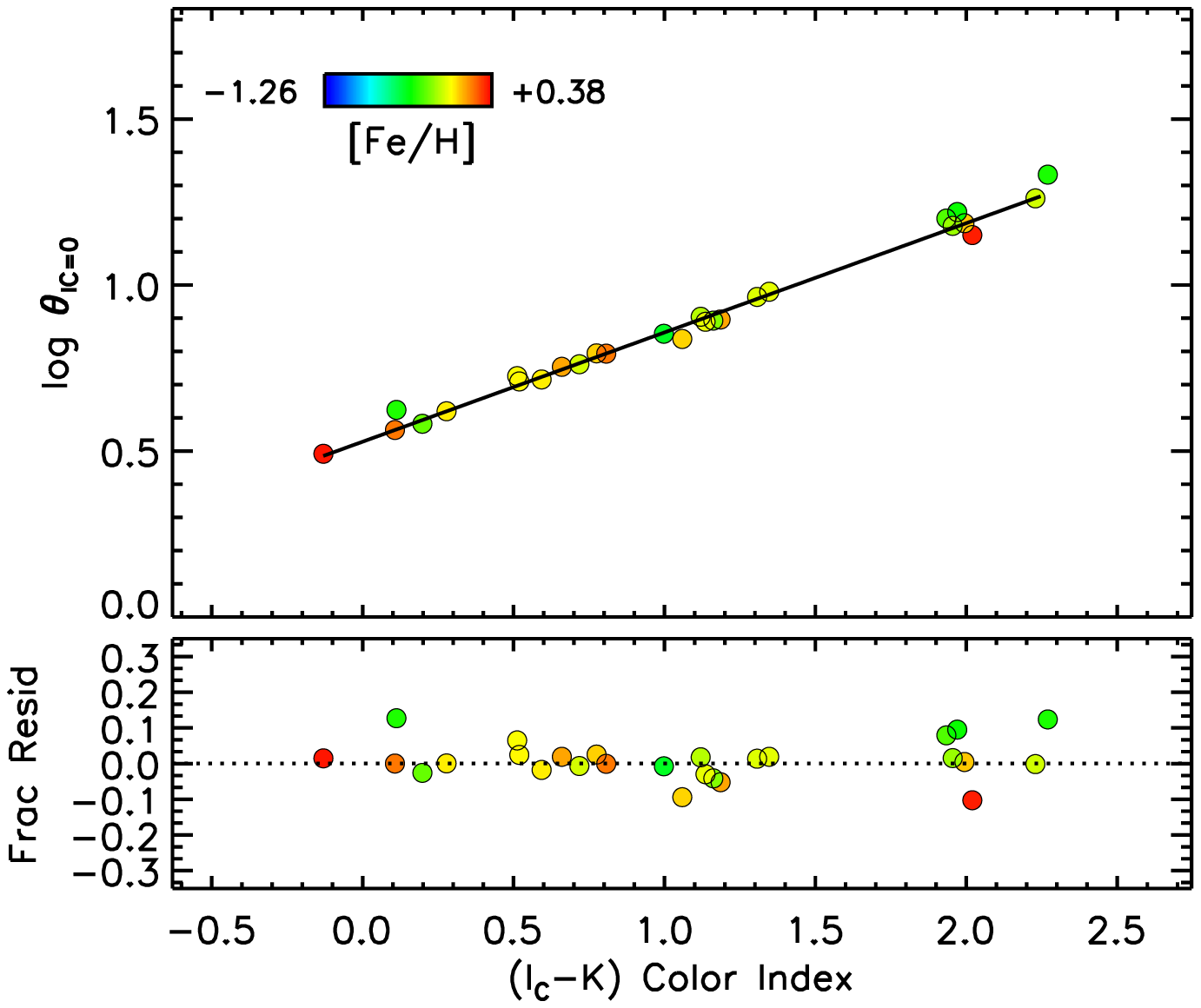, width=0.5\linewidth, clip=}	

 \end{tabular}
 \caption[Angular Diameters] {The top panel shows the zero-magnitude limb darkened angular diameter plotted against color index.  The solid black line plots the polynomial relation for the region that the relation holds true (Table\ref{tab:SB_poly3_coeffs}).  The color of the data point reflects the metallicity of the star as depicted in the legend.  The bottom panel shows the fractional residuals to our fit ($\theta_{\rm Obs.} - \theta_{\rm Fit})/\theta_{\rm Obs.}$, where the dotted line indicates zero deviation. See Section~\ref{sec:data} for details. }
 \label{fig:relations1}
 \end{figure}

\begin{figure}										
\centering
\begin{tabular}{cc}
\epsfig{file=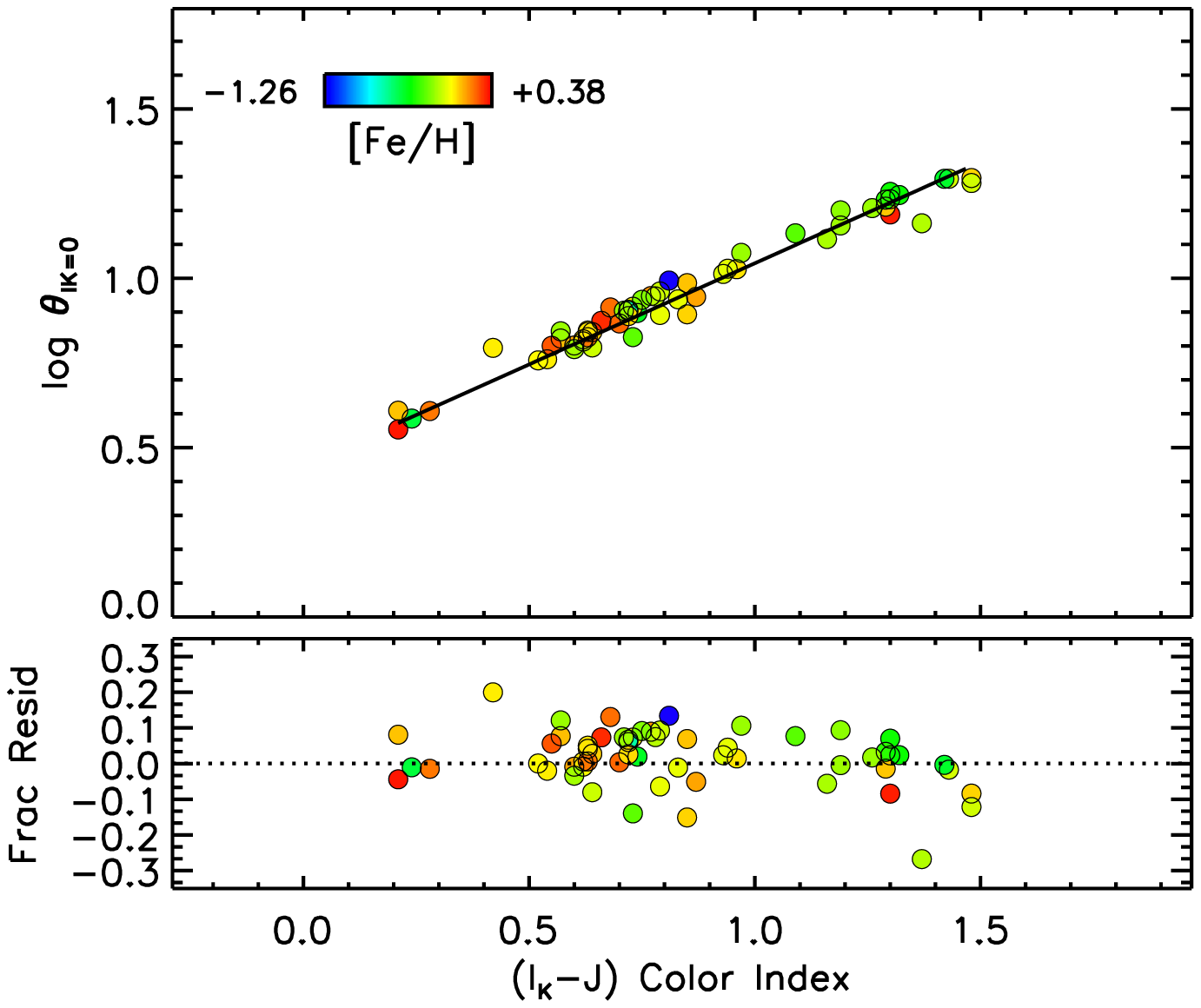, width=0.5\linewidth, clip=}	&
\epsfig{file=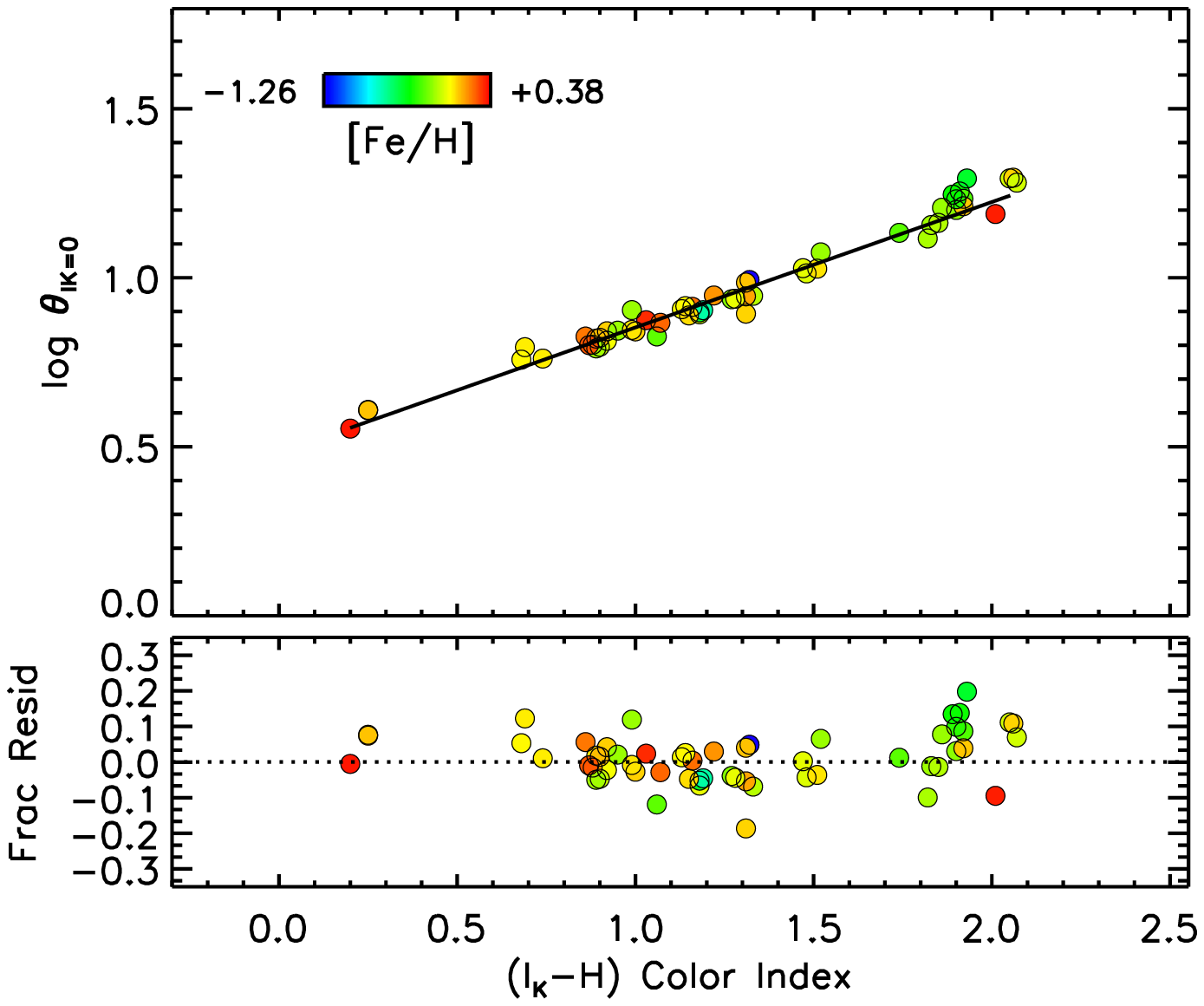, width=0.5\linewidth, clip=}	\\
\epsfig{file=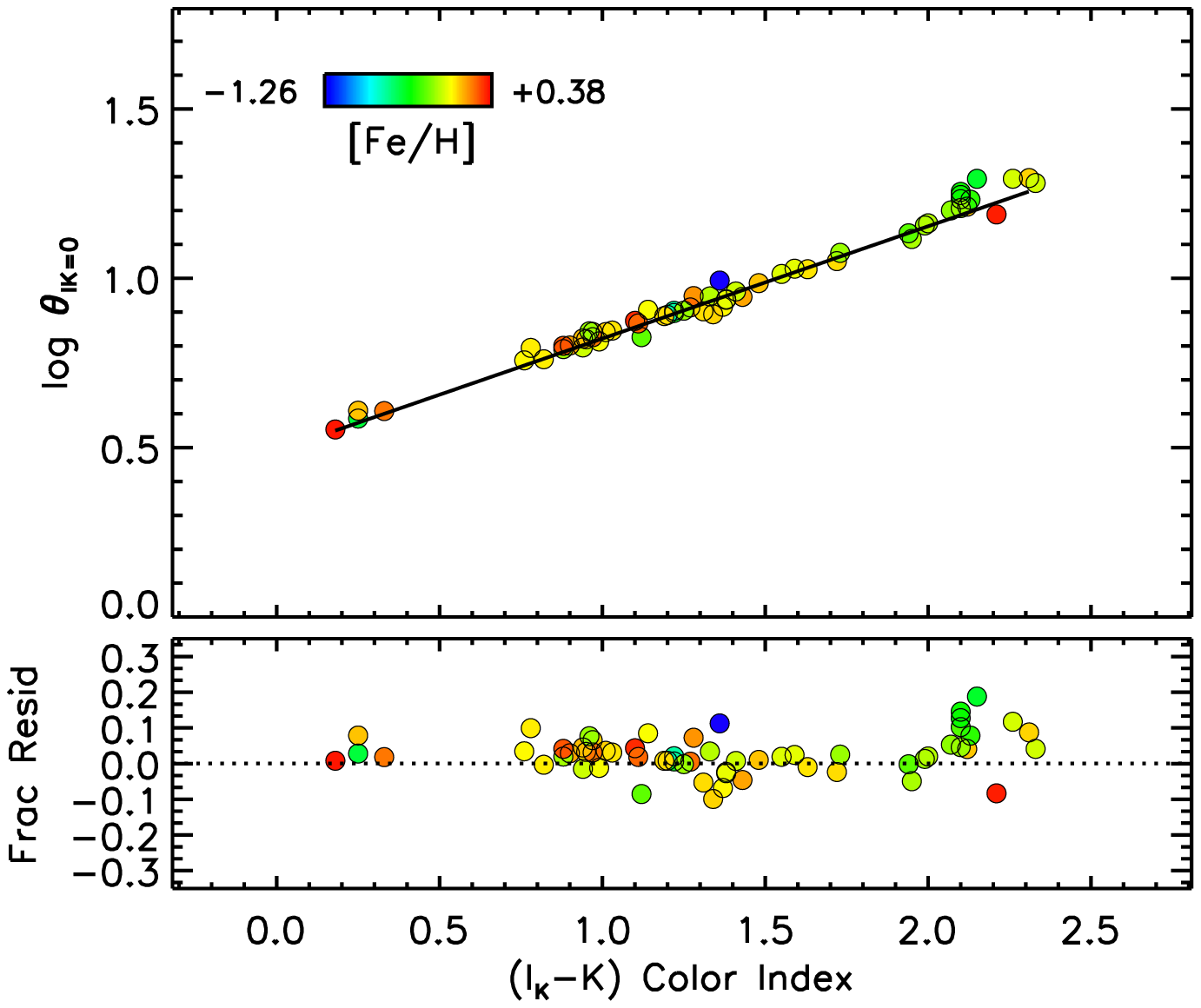, width=0.5\linewidth, clip=}	&
\epsfig{file=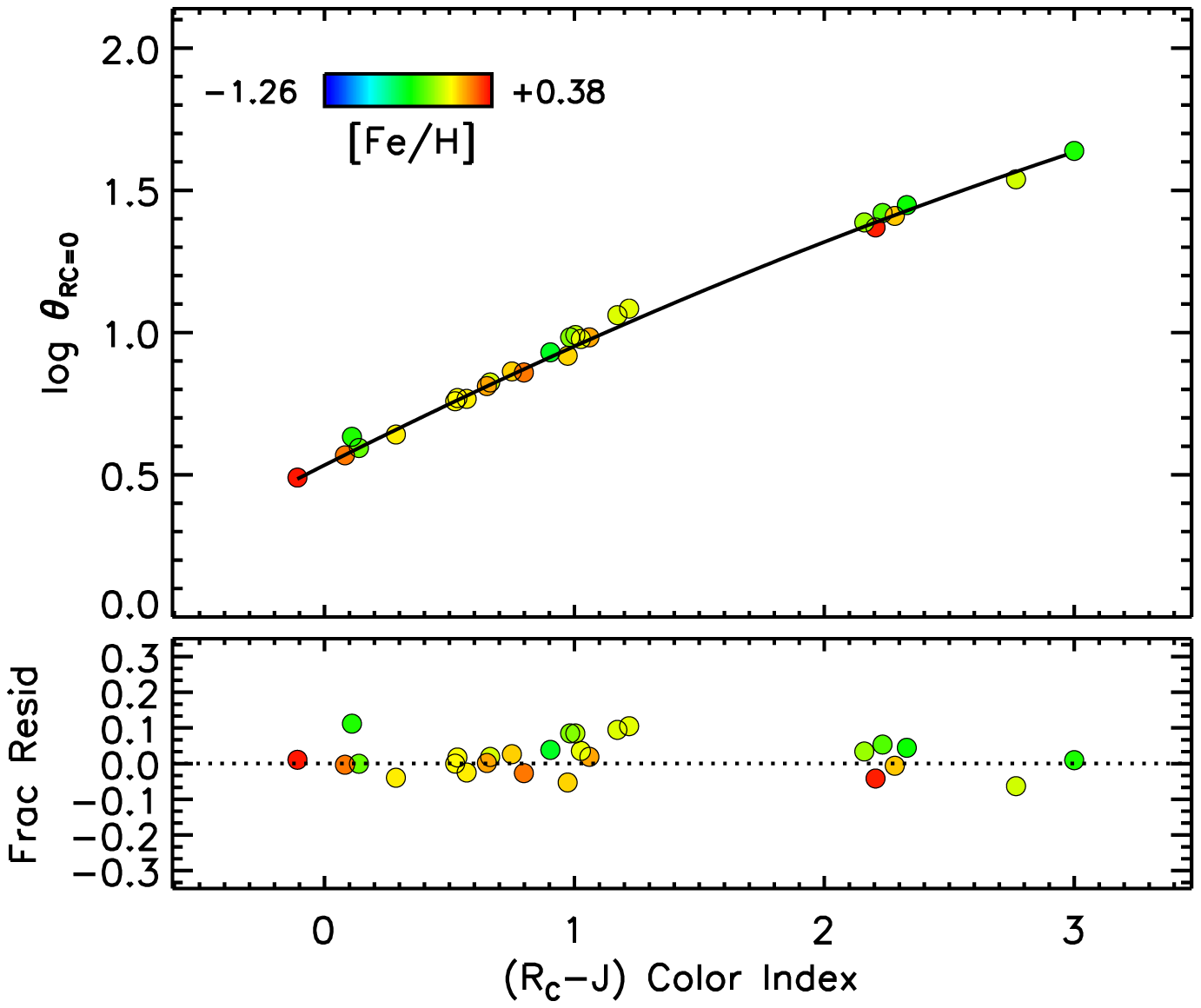, width=0.5\linewidth, clip=}	\\
\epsfig{file=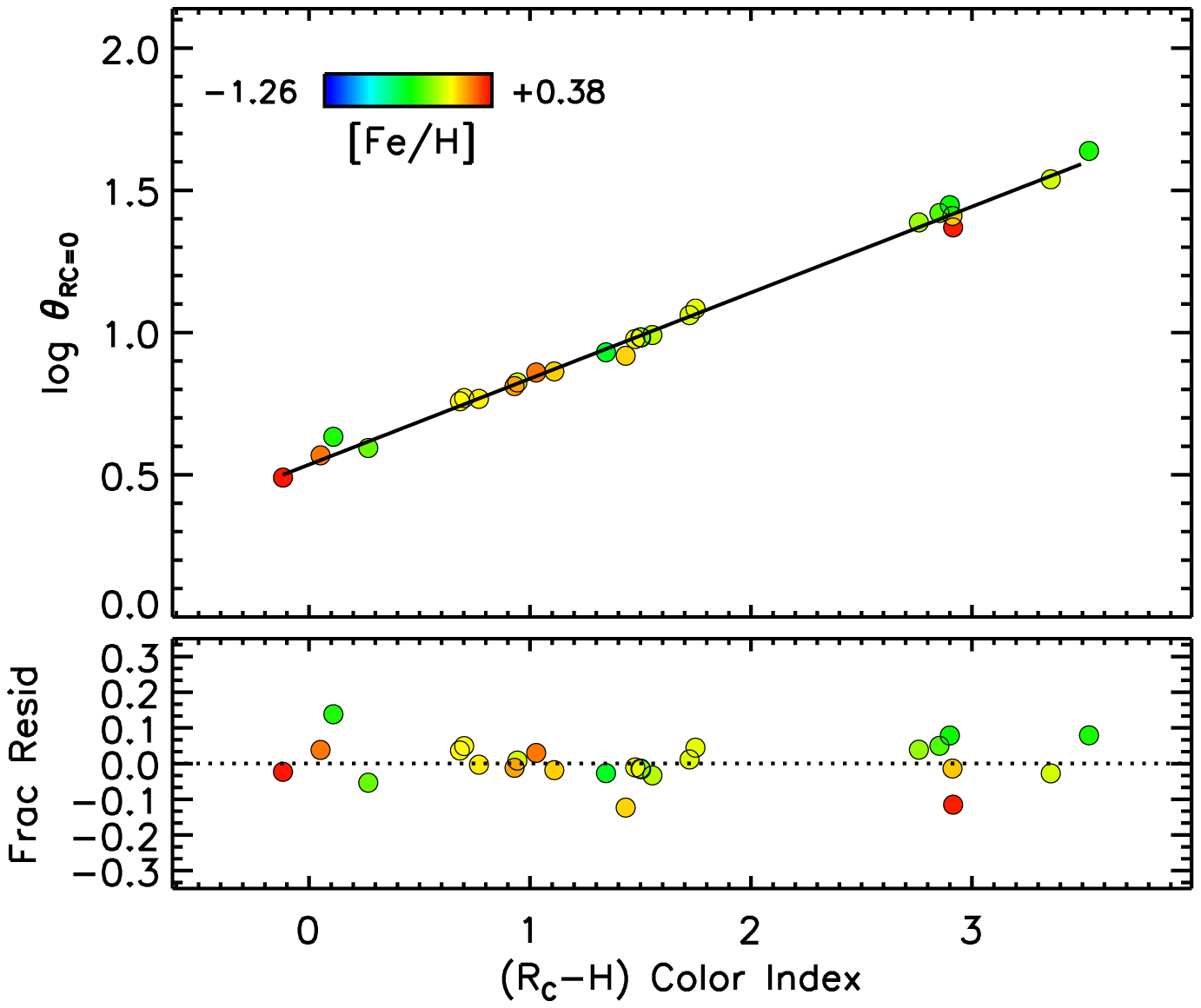, width=0.5\linewidth, clip=}	&
\epsfig{file=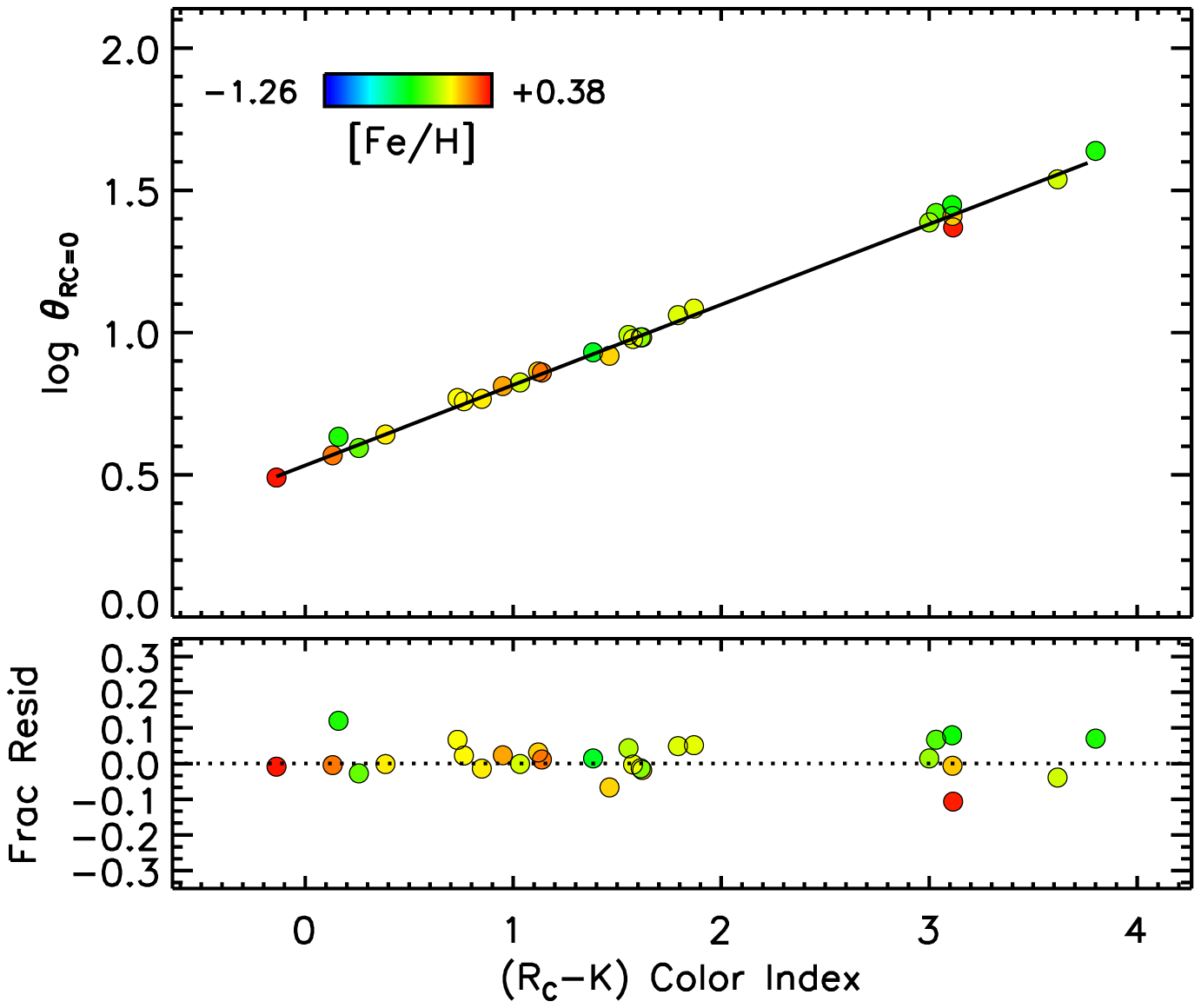, width=0.5\linewidth, clip=}	
 \end{tabular}
  \caption[Angular Diameters] {The top panel shows the zero-magnitude limb darkened angular diameter plotted against color index.  The solid black line plots the polynomial relation for the region that the relation holds true (Table\ref{tab:SB_poly3_coeffs}).  The color of the data point reflects the metallicity of the star as depicted in the legend. The bottom panel shows the fractional residuals to our fit ($\theta_{\rm Obs.} - \theta_{\rm Fit})/\theta_{\rm Obs.}$, where the dotted line indicates zero deviation.  See Section~\ref{sec:data} for details. }
  \label{fig:relations2}
  \end{figure}

\begin{figure}										
\centering
\begin{tabular}{cc}
\epsfig{file=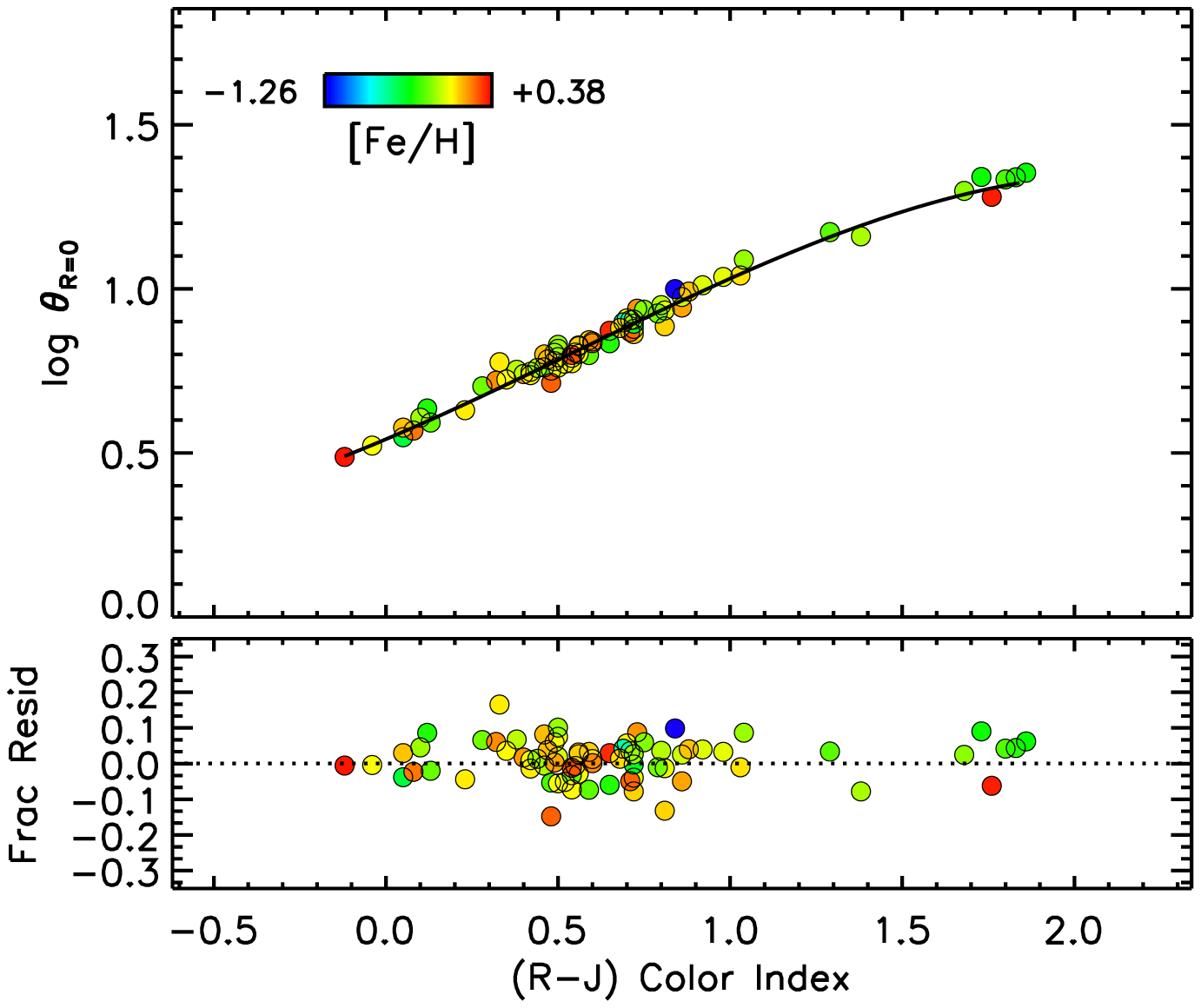, width=0.5\linewidth, clip=}	&
\epsfig{file=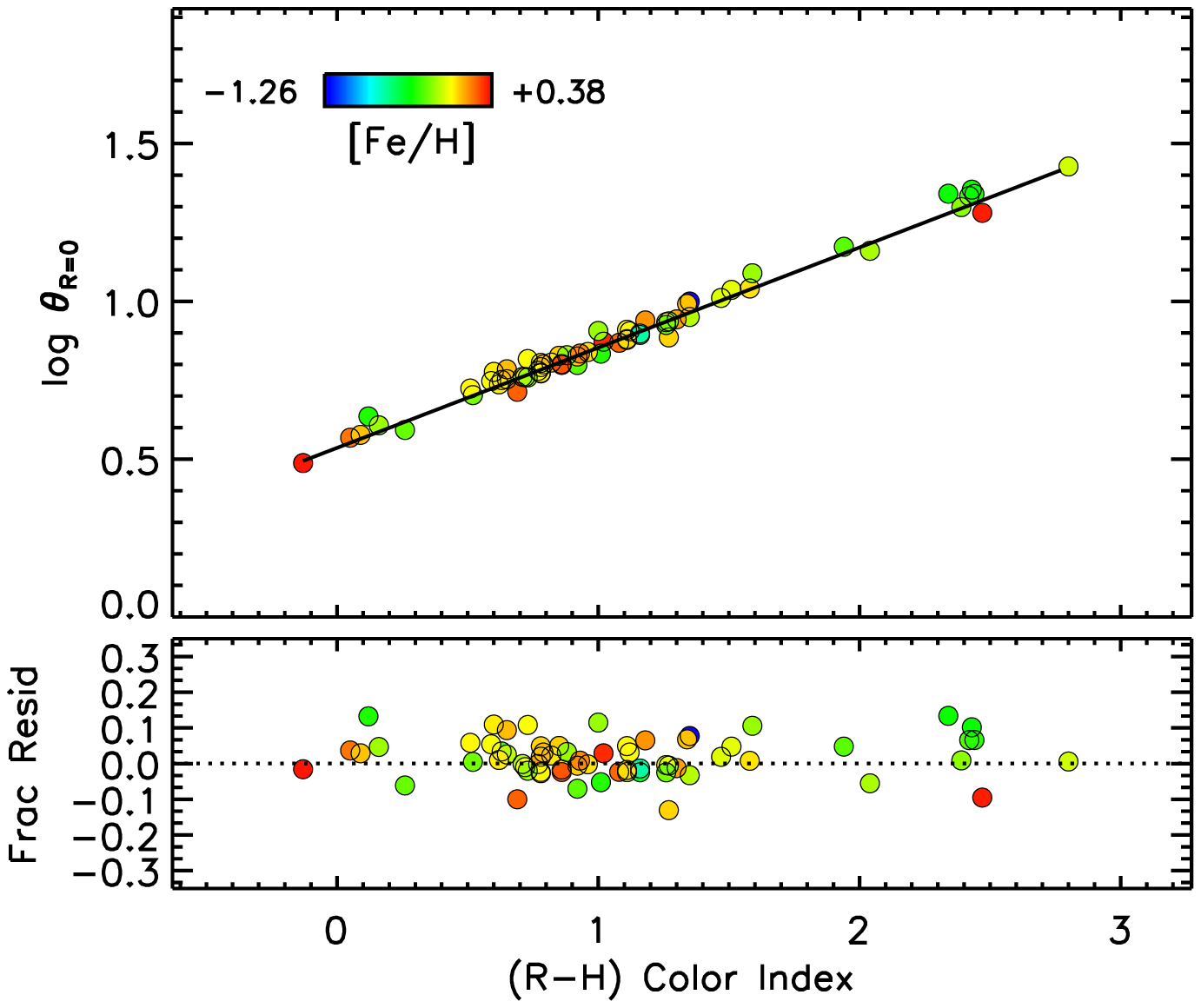, width=0.5\linewidth, clip=}	\\
\epsfig{file=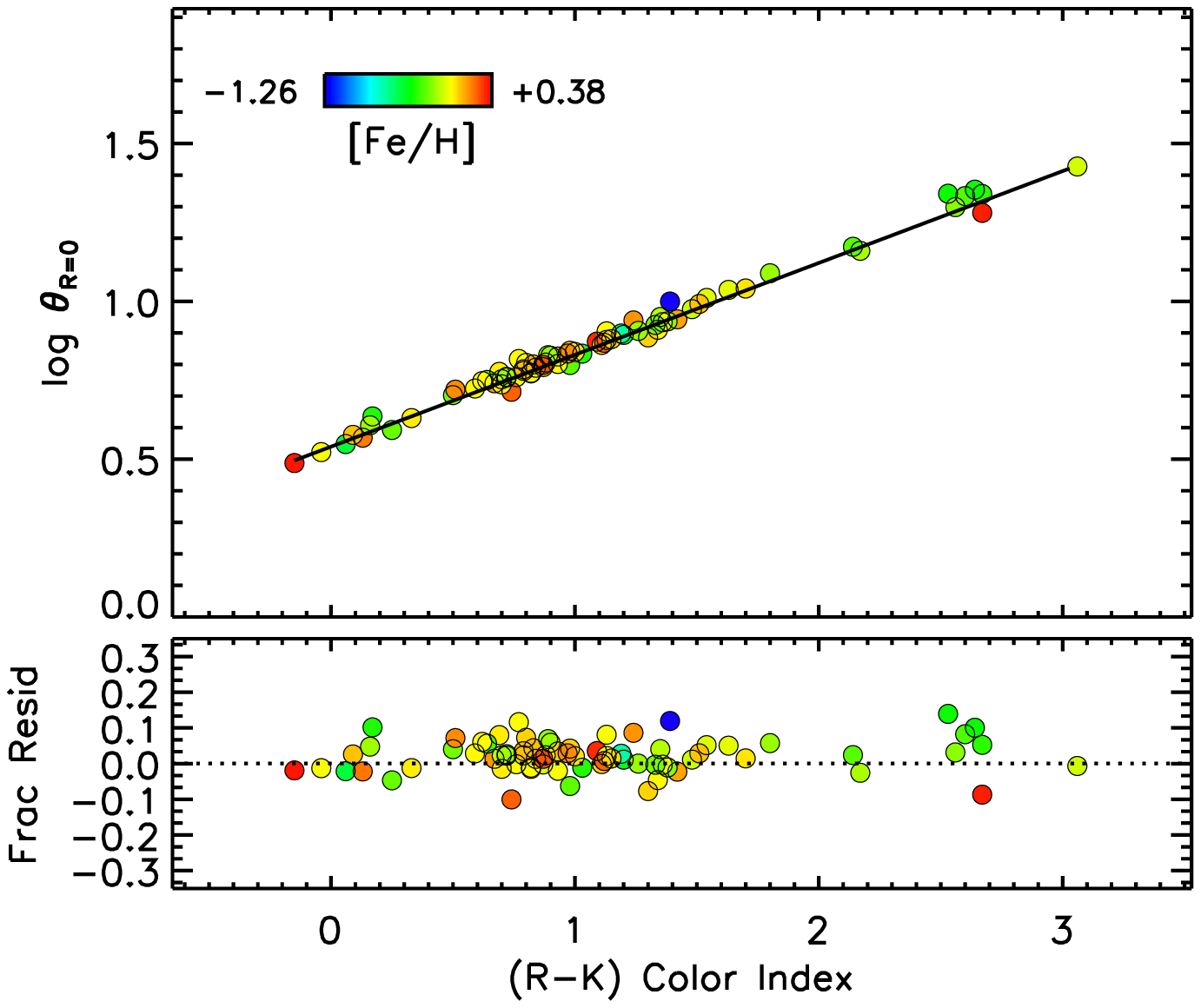, width=0.5\linewidth, clip=}	&
\epsfig{file=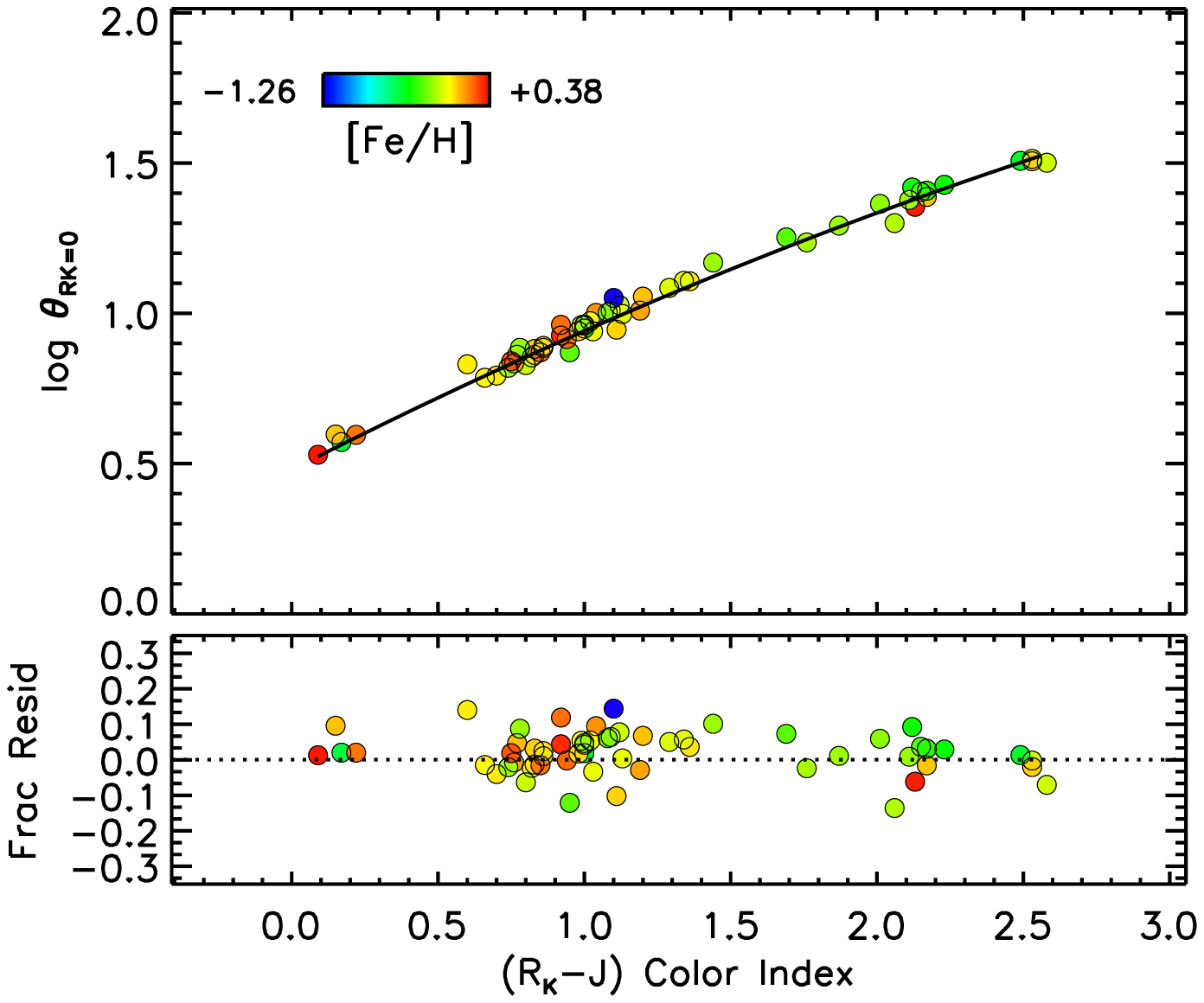, width=0.5\linewidth, clip=}	\\
\epsfig{file=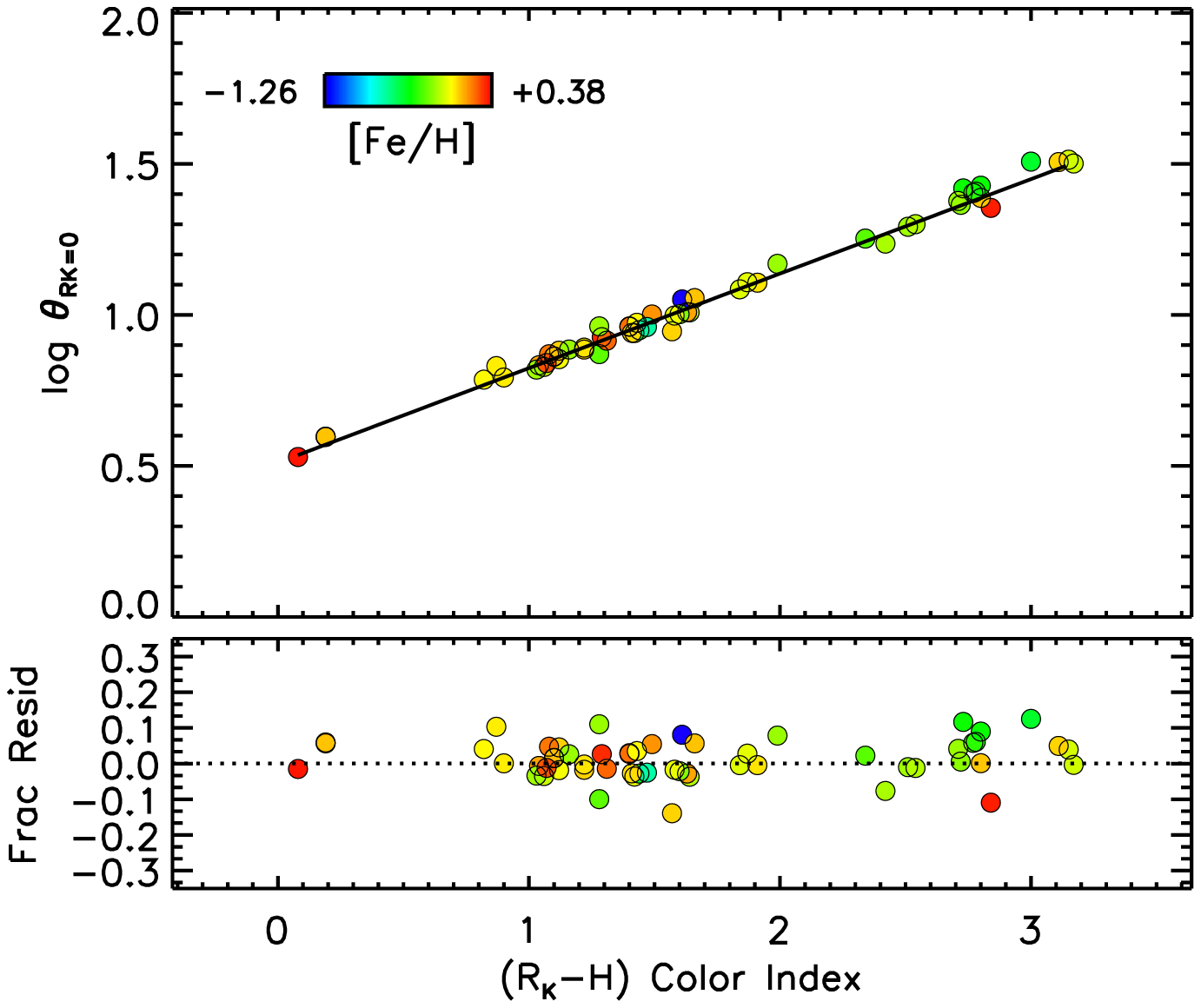, width=0.5\linewidth, clip=}	&
\epsfig{file=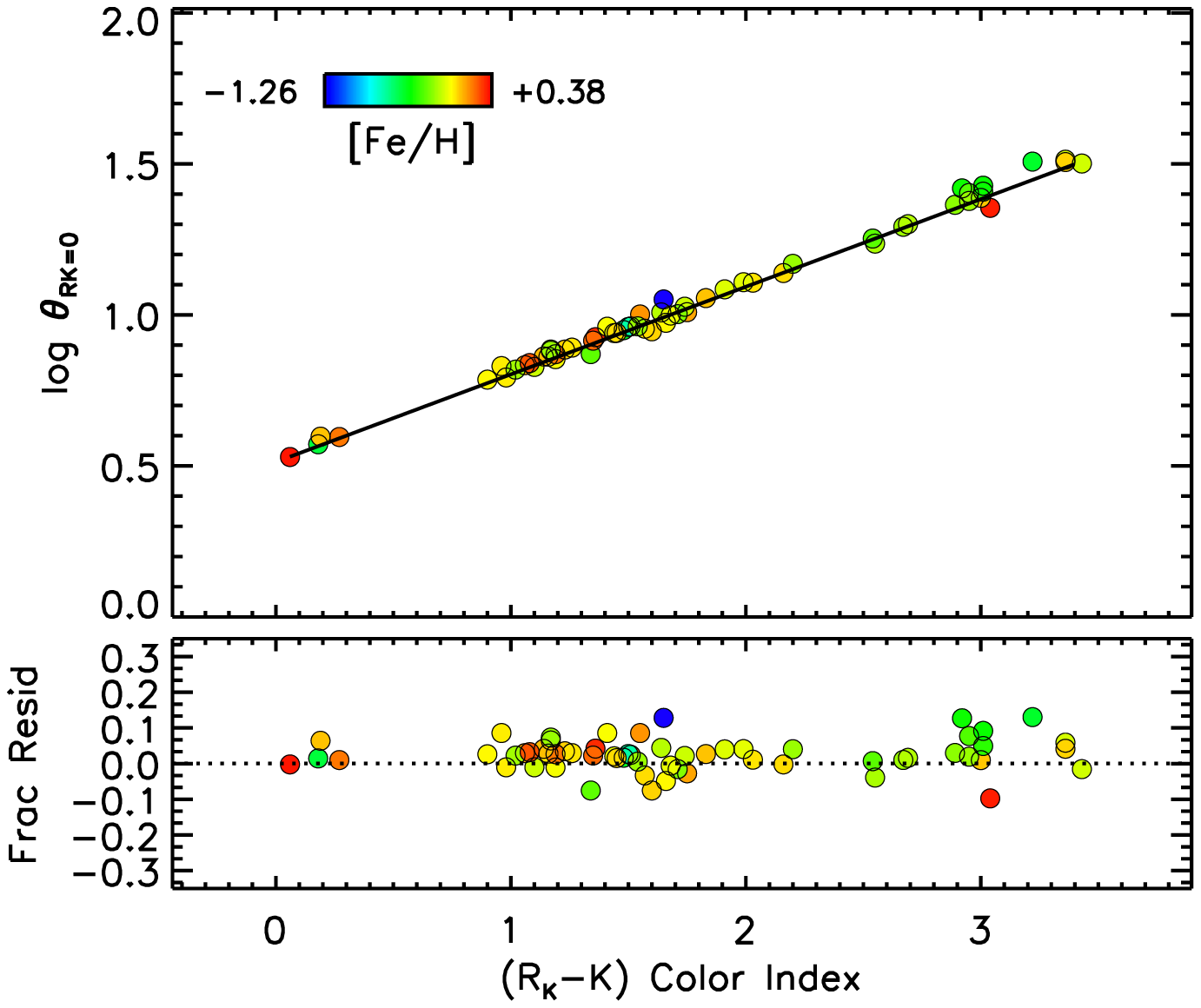, width=0.5\linewidth, clip=}	
 \end{tabular}
  \caption[Angular Diameters] {The top panel shows the zero-magnitude limb darkened angular diameter plotted against color index.  The solid black line plots the polynomial relation for the region that the relation holds true (Table\ref{tab:SB_poly3_coeffs}).  The color of the data point reflects the metallicity of the star as depicted in the legend. The bottom panel shows the fractional residuals to our fit ($\theta_{\rm Obs.} - \theta_{\rm Fit})/\theta_{\rm Obs.}$, where the dotted line indicates zero deviation.  See Section~\ref{sec:data} for details. }
  \label{fig:relations3}
  \end{figure}

\begin{figure}										
\centering
\begin{tabular}{cc}
\epsfig{file=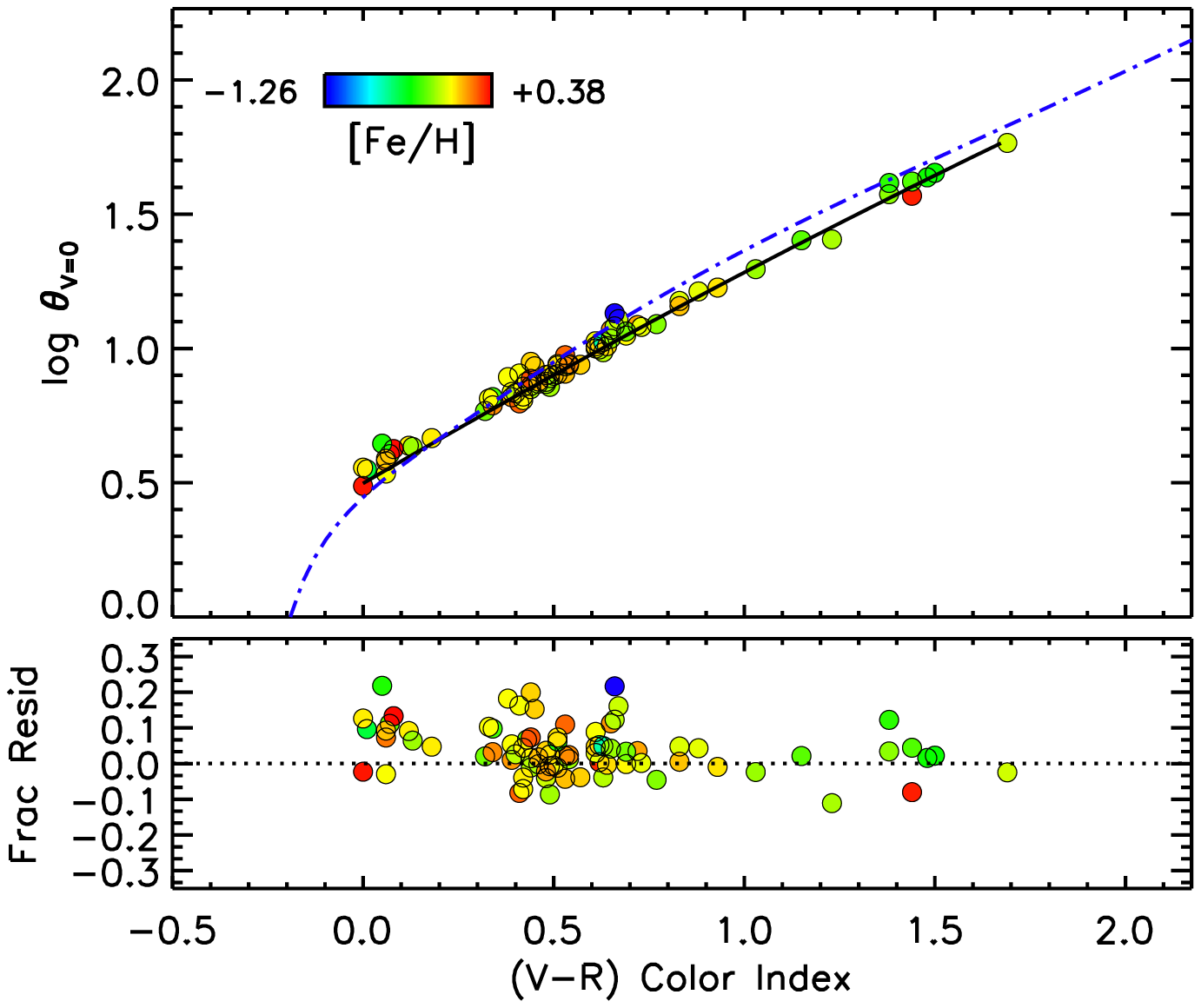, width=0.5\linewidth, clip=}	&
\epsfig{file=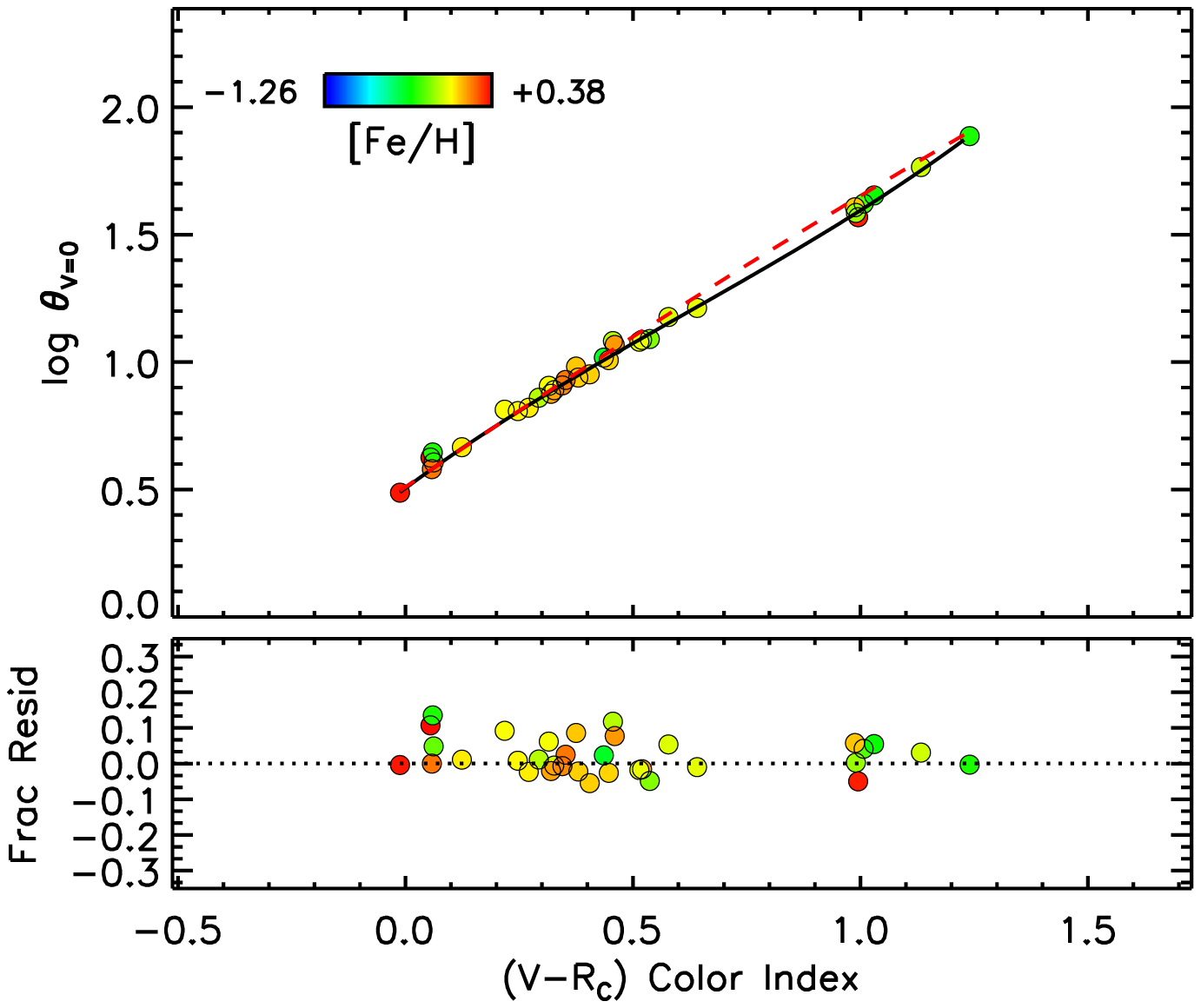, width=0.5\linewidth, clip=}	\\
\epsfig{file=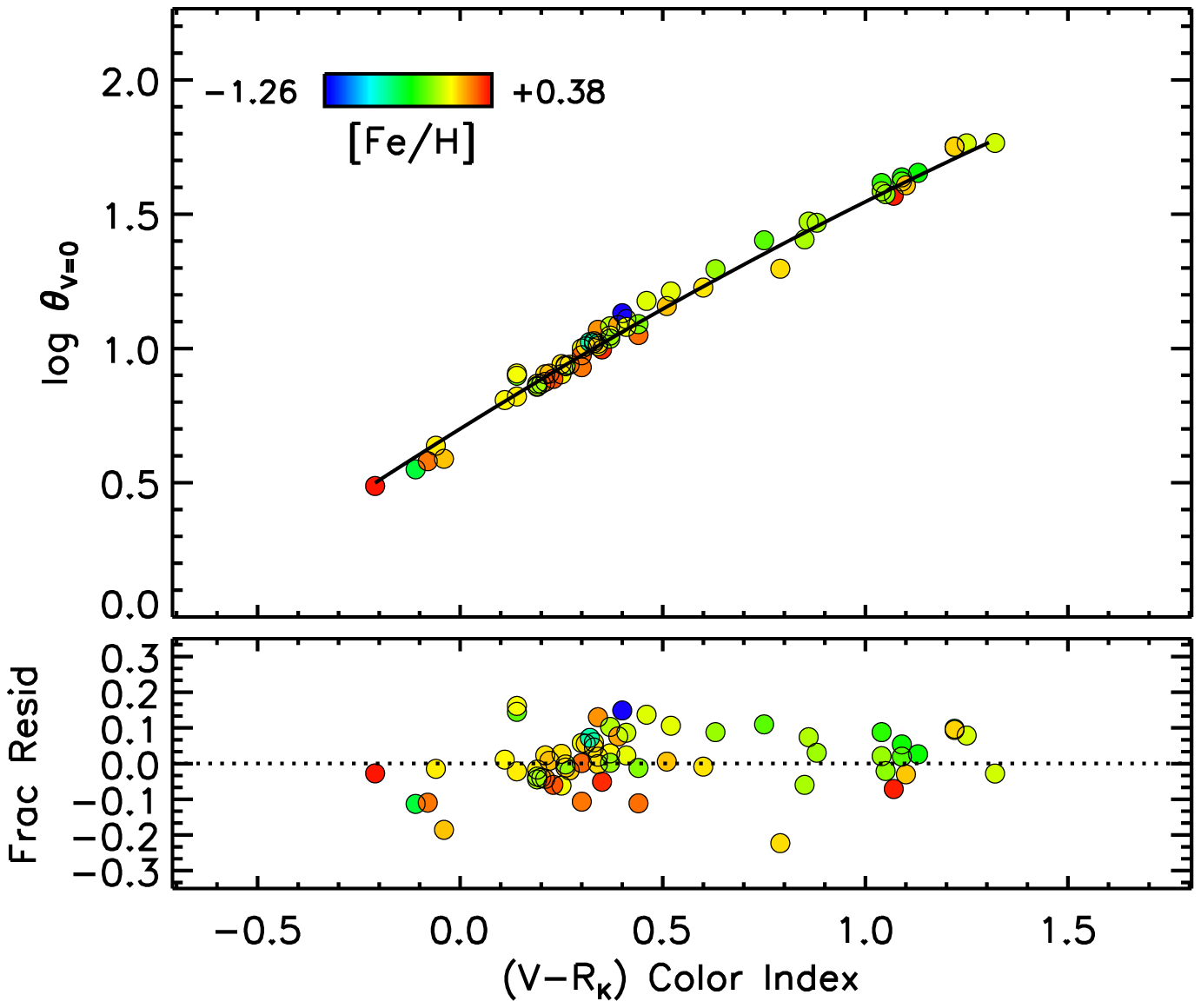, width=0.5\linewidth, clip=}	&
\epsfig{file=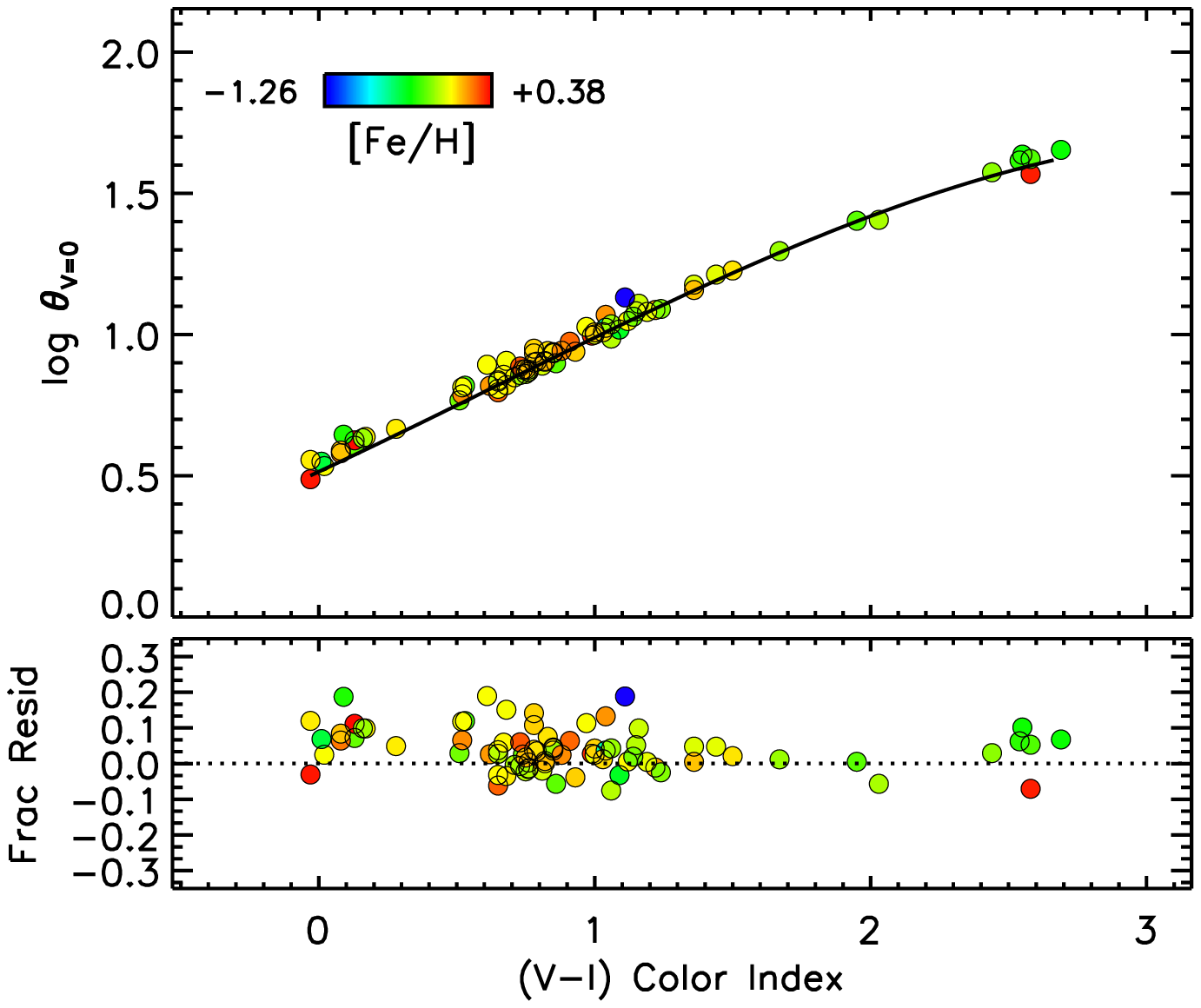, width=0.5\linewidth, clip=}	\\
\epsfig{file=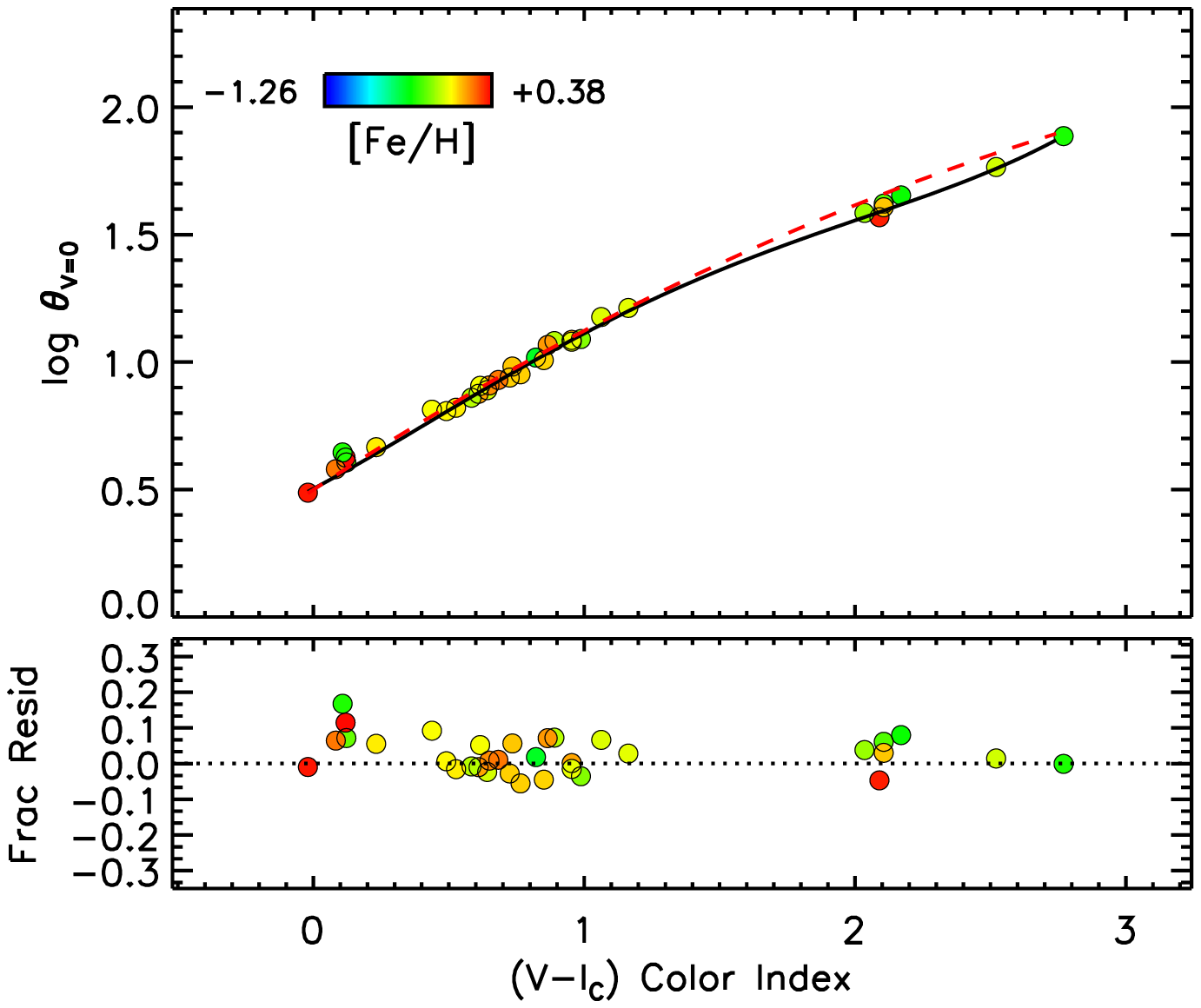, width=0.5\linewidth, clip=}	&
\epsfig{file=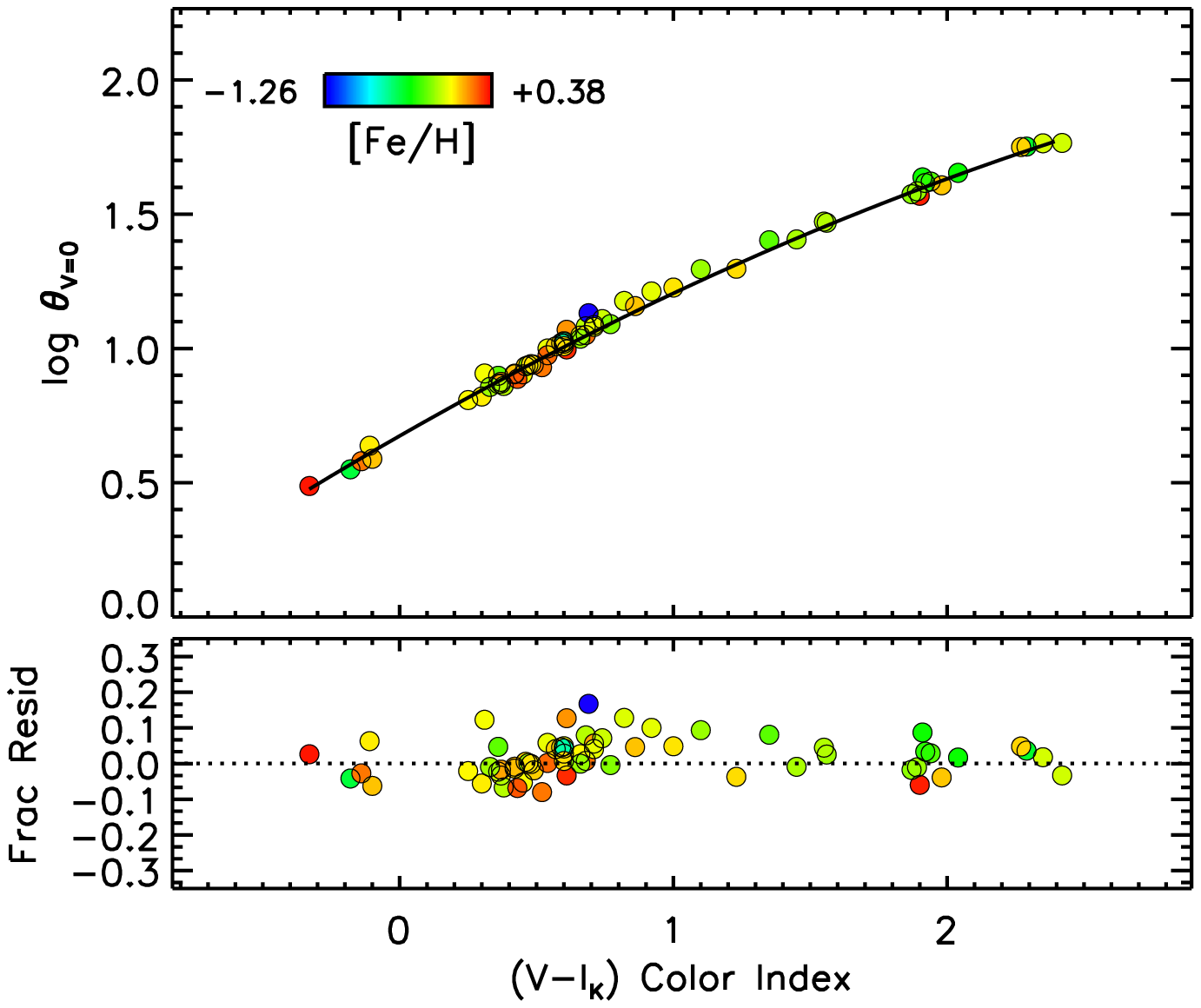, width=0.5\linewidth, clip=}	
 \end{tabular}
  \caption[Angular Diameters] {The top panel shows the zero-magnitude limb darkened angular diameter plotted against color index.  The solid black line plots the polynomial relation for the region that the relation holds true (Table\ref{tab:SB_poly3_coeffs}).  The color of the data point reflects the metallicity of the star as depicted in the legend.  A solution from \citet{ker08a} is plotted as a dashed red line, and a solution from \citet{bon06} is plotted as a blue dash-dotted line. The bottom panel shows the fractional residuals to our fit ($\theta_{\rm Obs.} - \theta_{\rm Fit})/\theta_{\rm Obs.}$, where the dotted line indicates zero deviation.  See Section~\ref{sec:data} for details. }
  \label{fig:relations4}
  \end{figure}

\begin{figure}										
\centering
\begin{tabular}{cc}
\epsfig{file=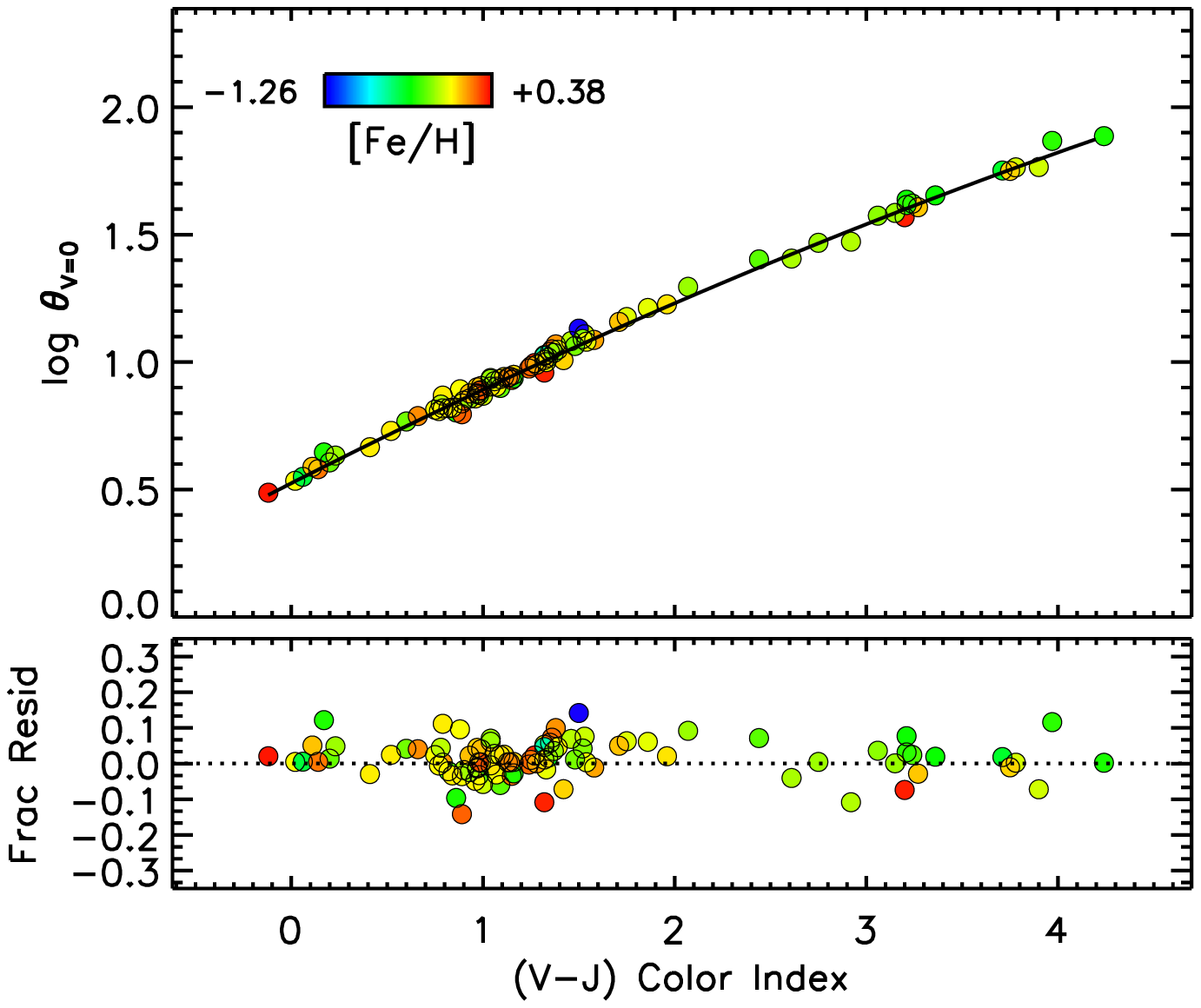, width=0.5\linewidth, clip=} &
\epsfig{file=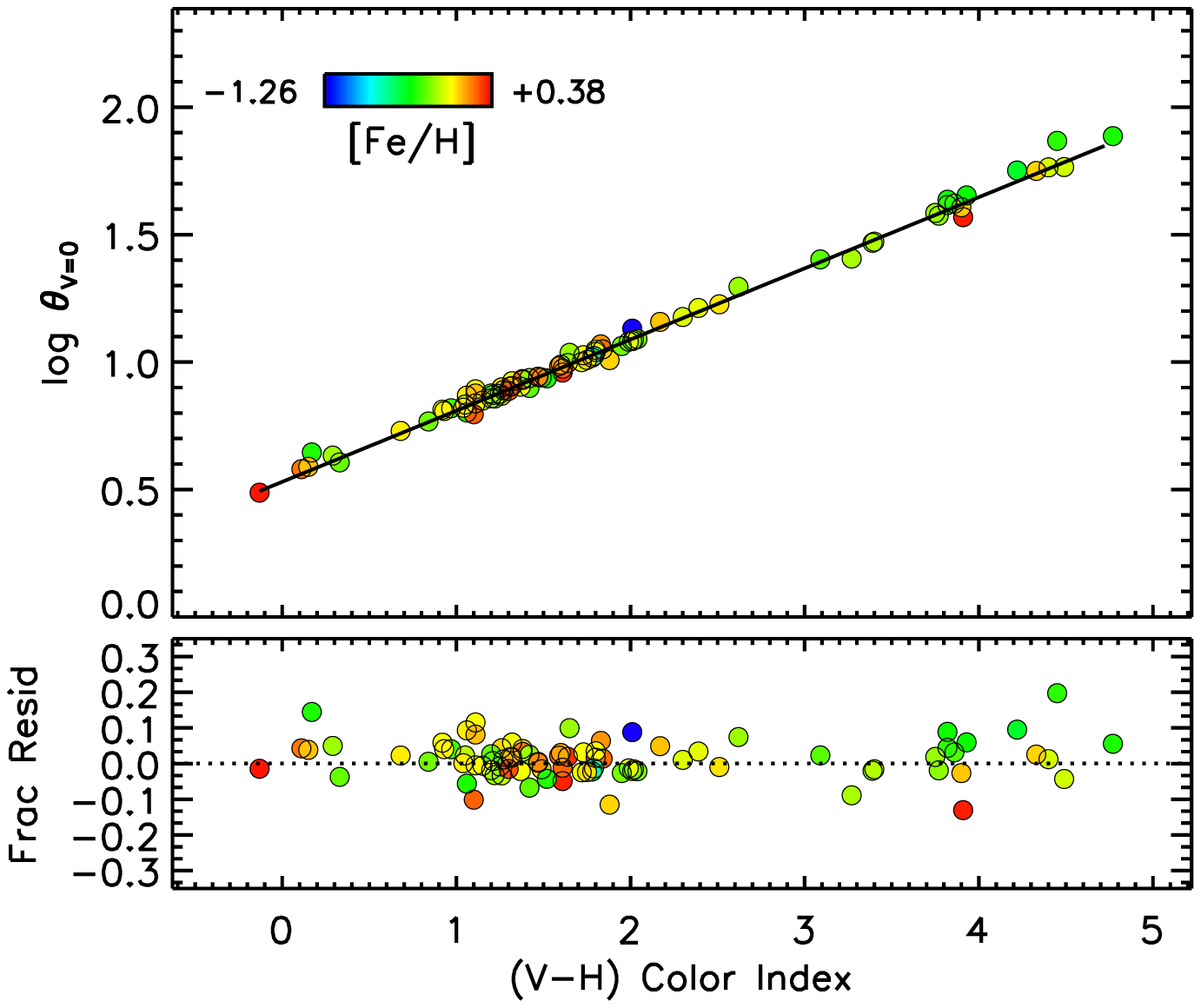, width=0.5\linewidth, clip=}	\\
\epsfig{file=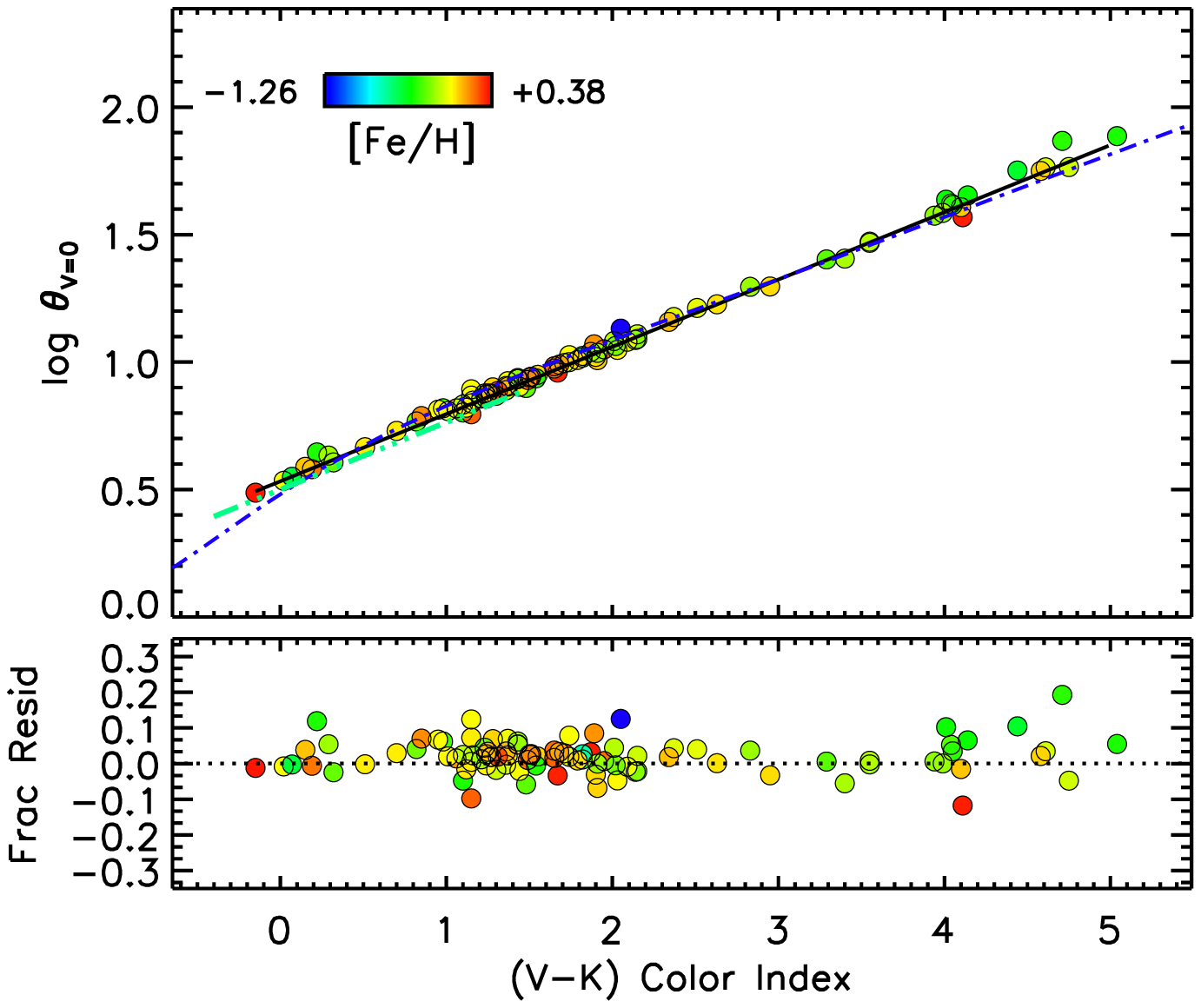, width=0.5\linewidth, clip=}	&
\epsfig{file=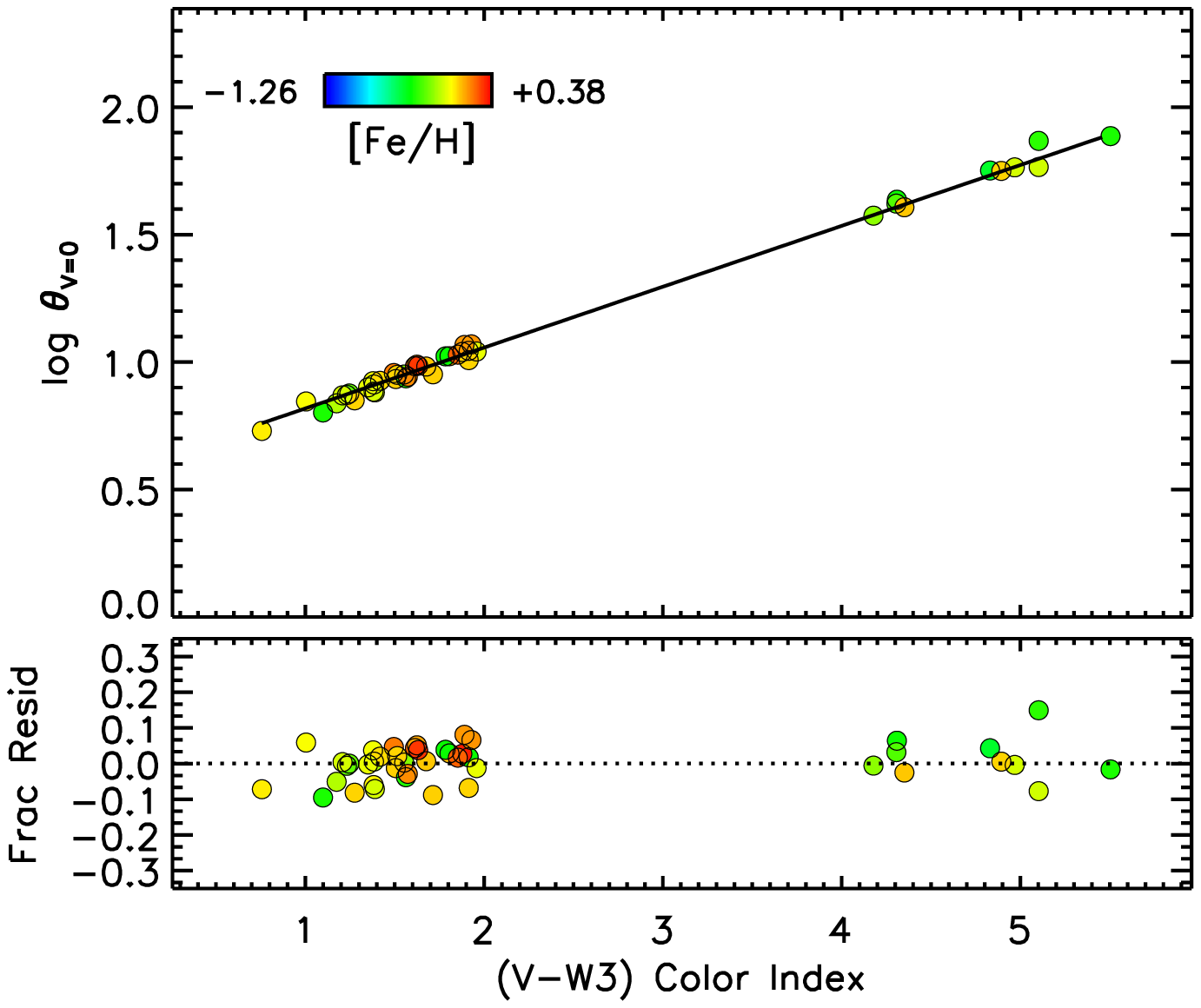, width=0.5\linewidth, clip=}	\\
\epsfig{file=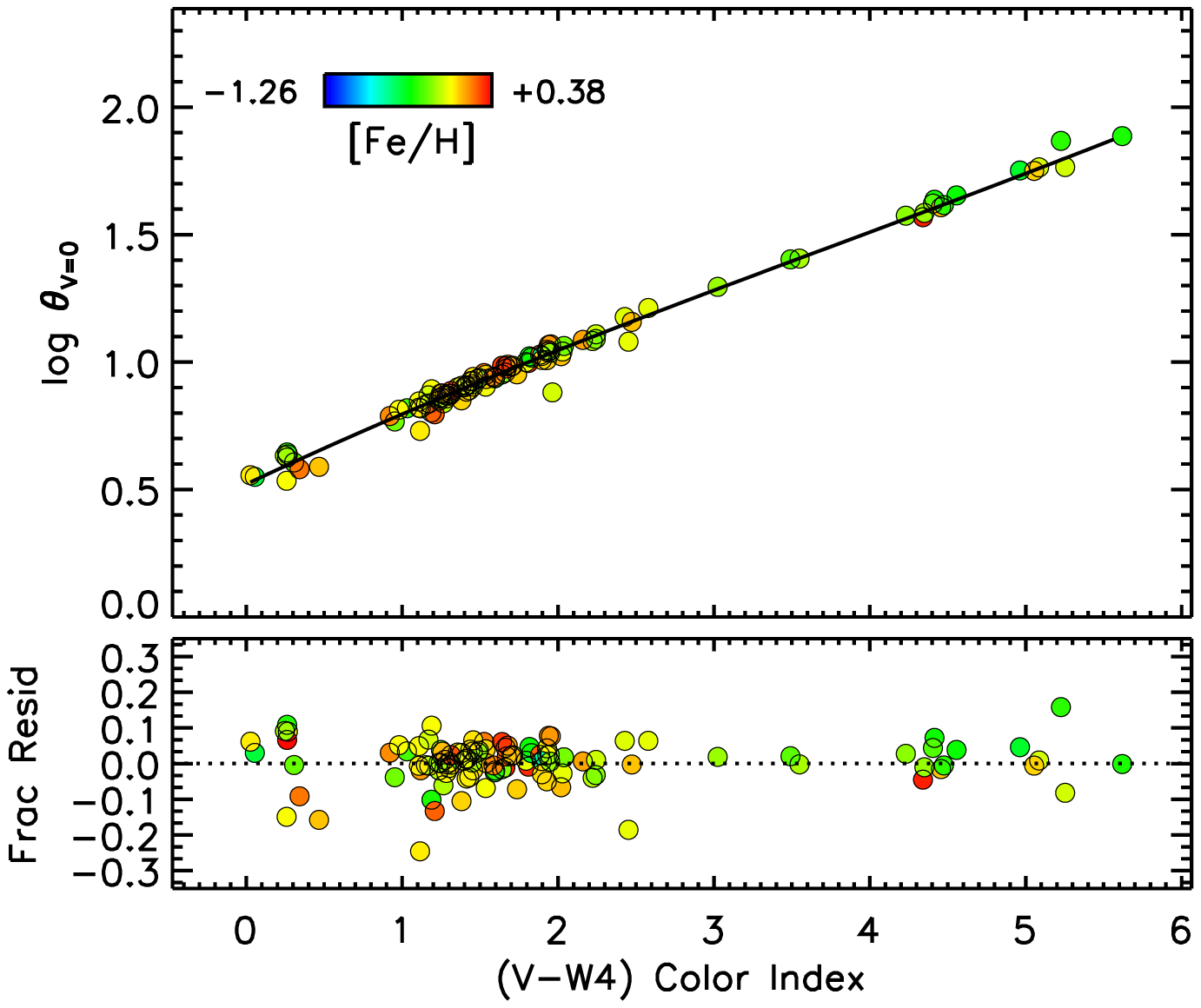, width=0.5\linewidth, clip=}	
 \end{tabular}
  \caption[Angular Diameters] {The top panel shows the zero-magnitude limb darkened angular diameter plotted against color index.  The solid black line plots the polynomial relation for the region that the relation holds true (Table\ref{tab:SB_poly3_coeffs}).  The color of the data point reflects the metallicity of the star as depicted in the legend. A solution from \citet{bon06} is plotted as a blue dash-dotted line. A solution from \citet{van99b} is plotted as a green triple dot-dashed line. The bottom panel shows the fractional residuals to our fit ($\theta_{\rm Obs.} - \theta_{\rm Fit})/\theta_{\rm Obs.}$, where the dotted line indicates zero deviation.  See Section~\ref{sec:data} for details. }
  \label{fig:relations5}
  \end{figure}

\begin{figure}										
\centering
\begin{tabular}{cc}
\epsfig{file=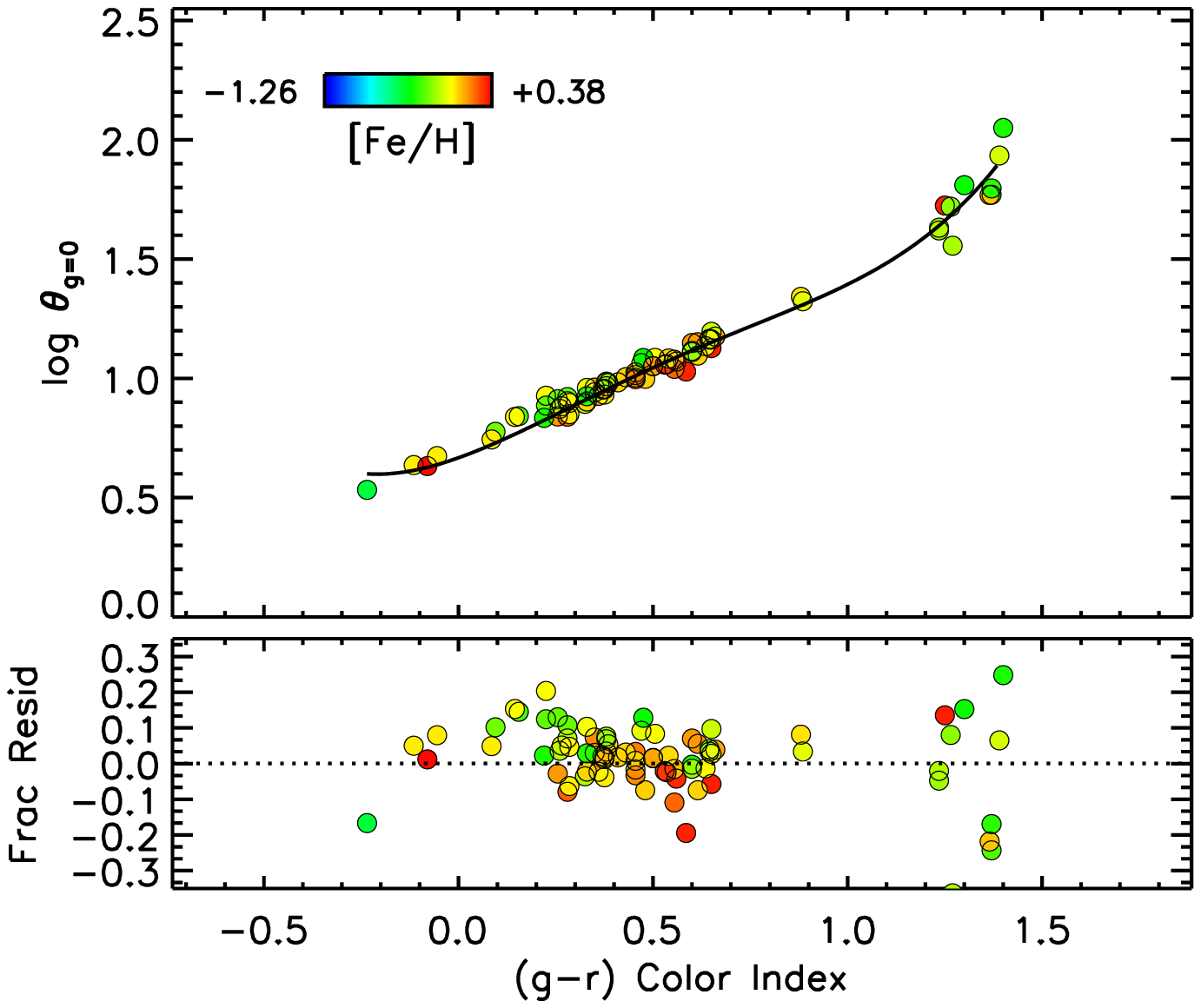, width=0.5\linewidth, clip=}	&
\epsfig{file=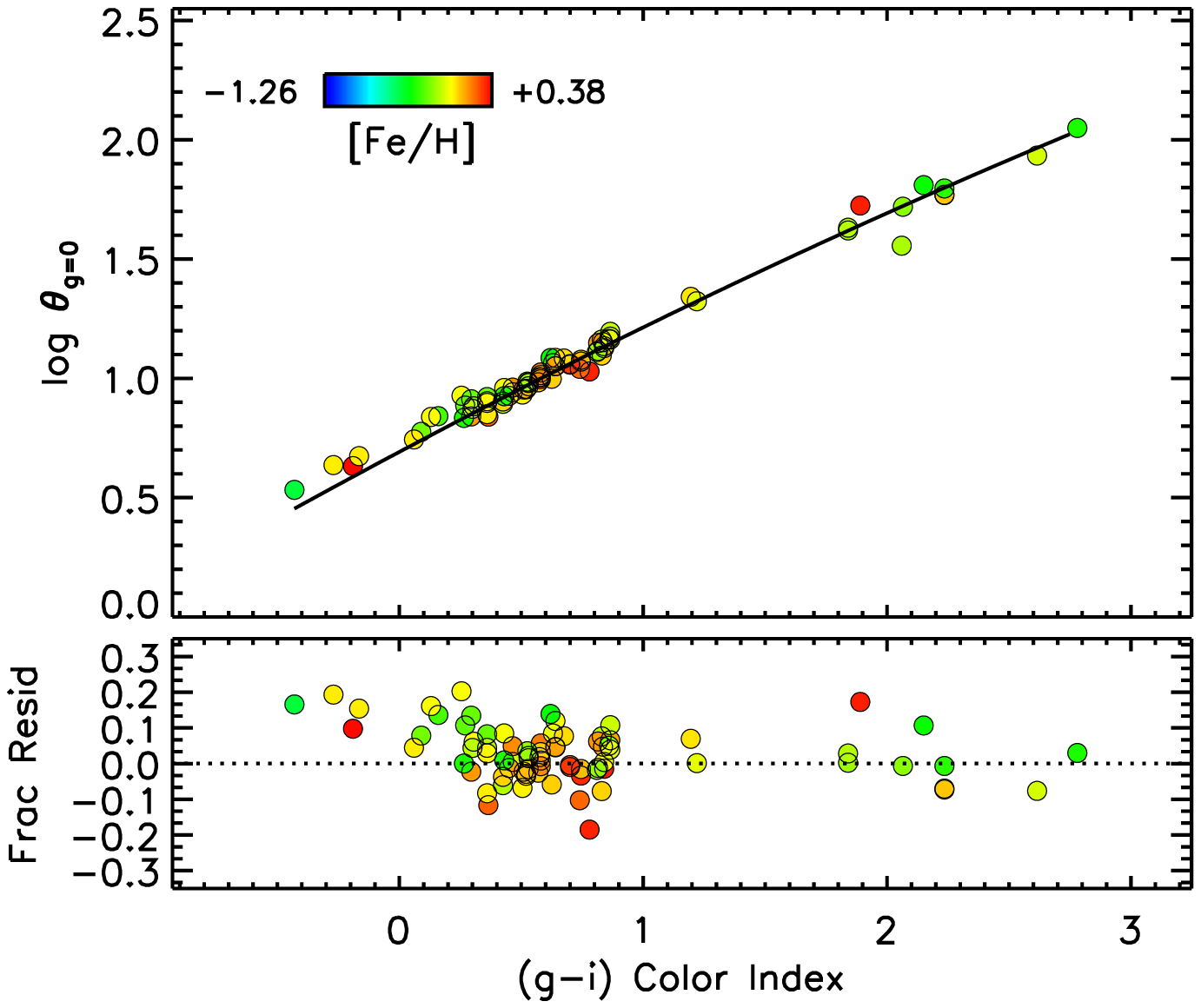, width=0.5\linewidth, clip=}	\\
\epsfig{file=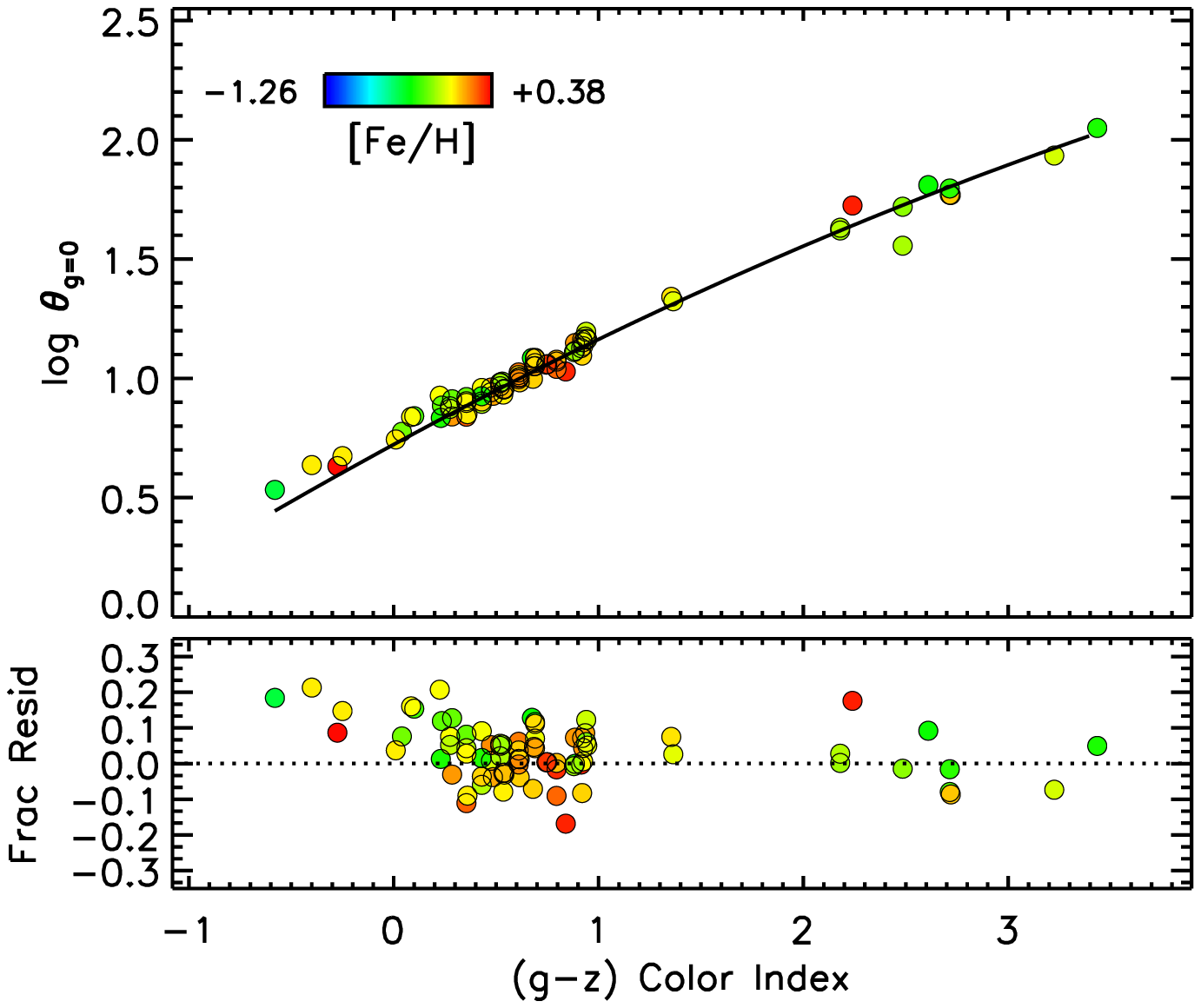, width=0.5\linewidth, clip=}	&
\epsfig{file=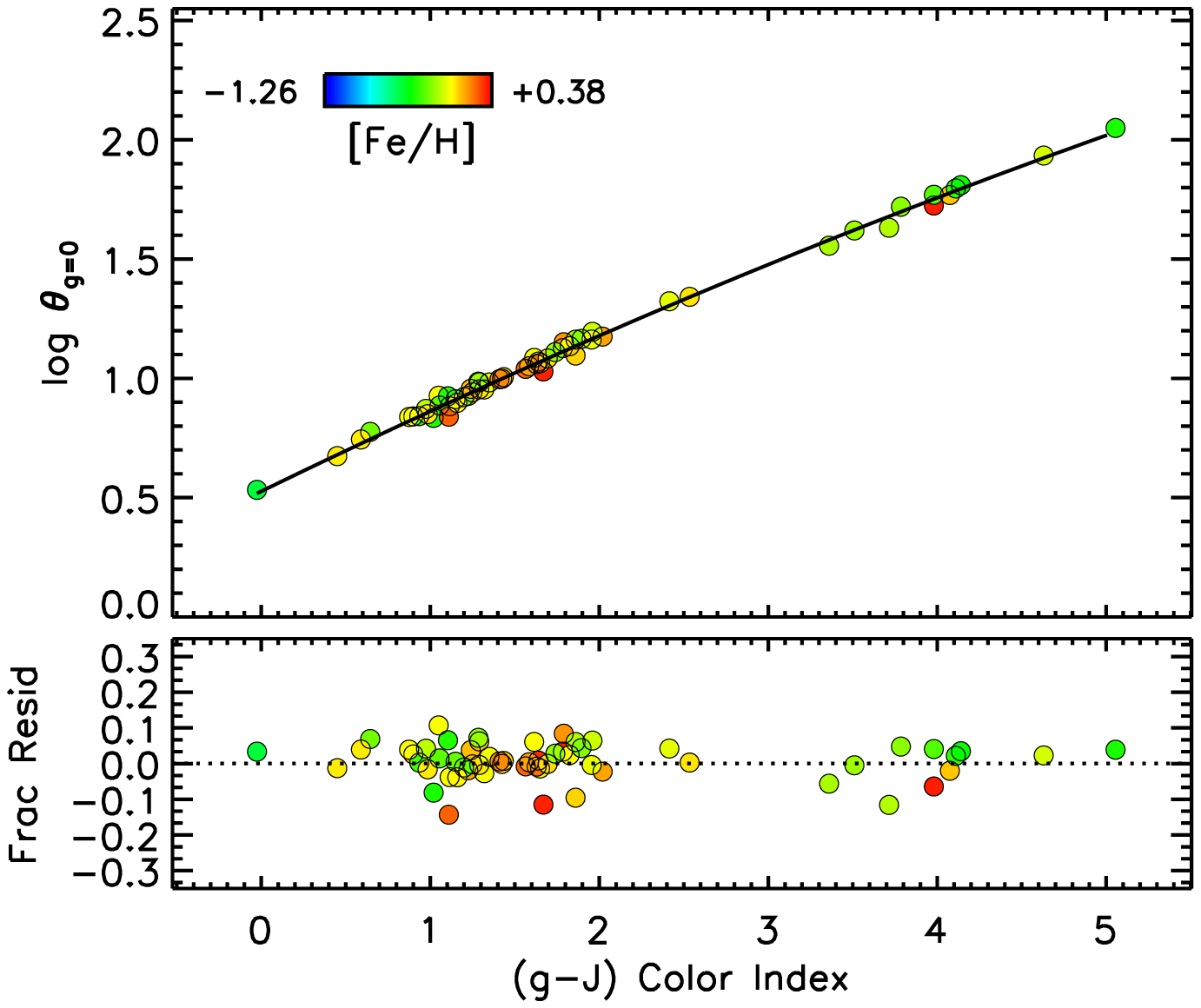, width=0.5\linewidth, clip=}	\\
\epsfig{file=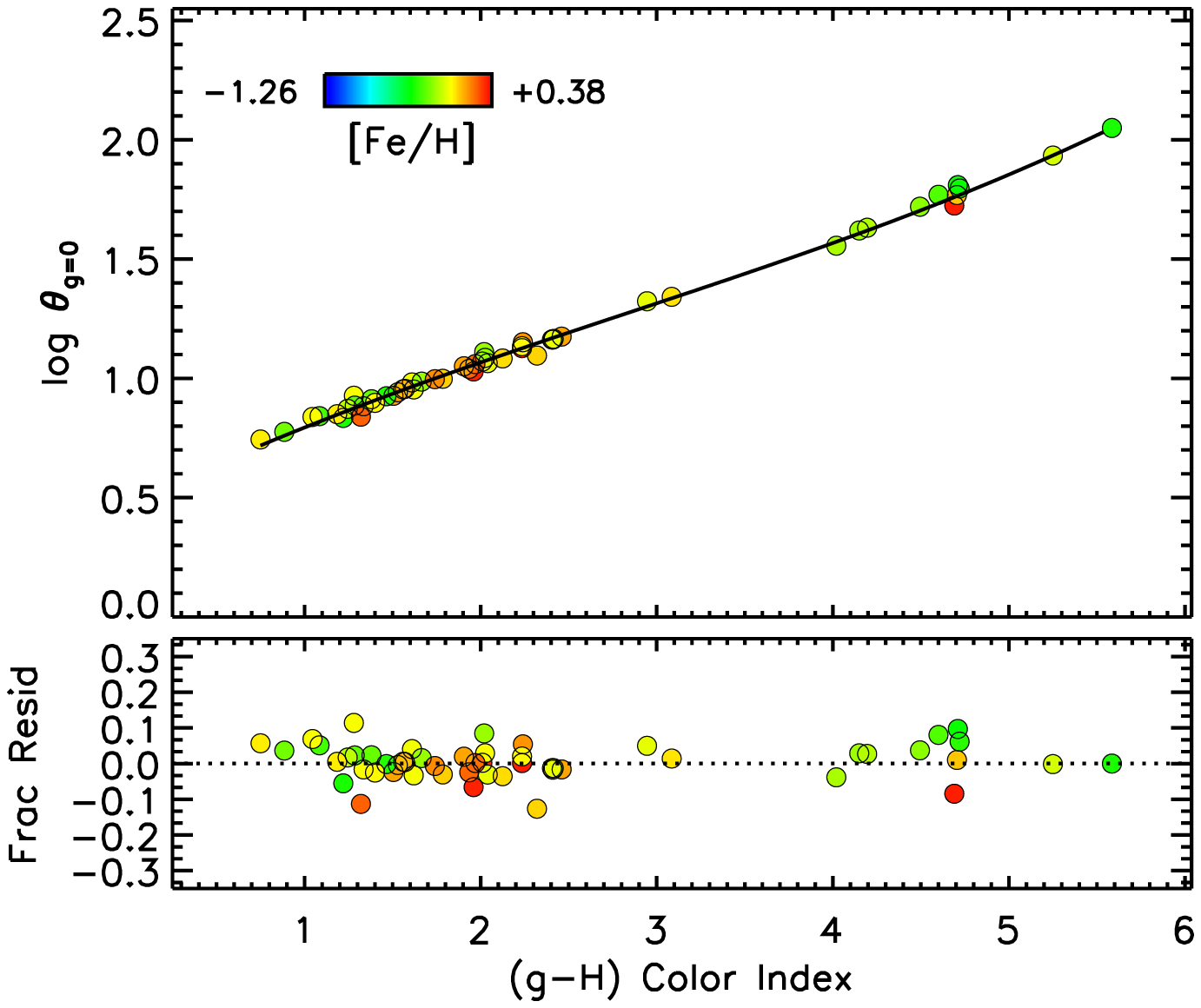, width=0.5\linewidth, clip=}	&
\epsfig{file=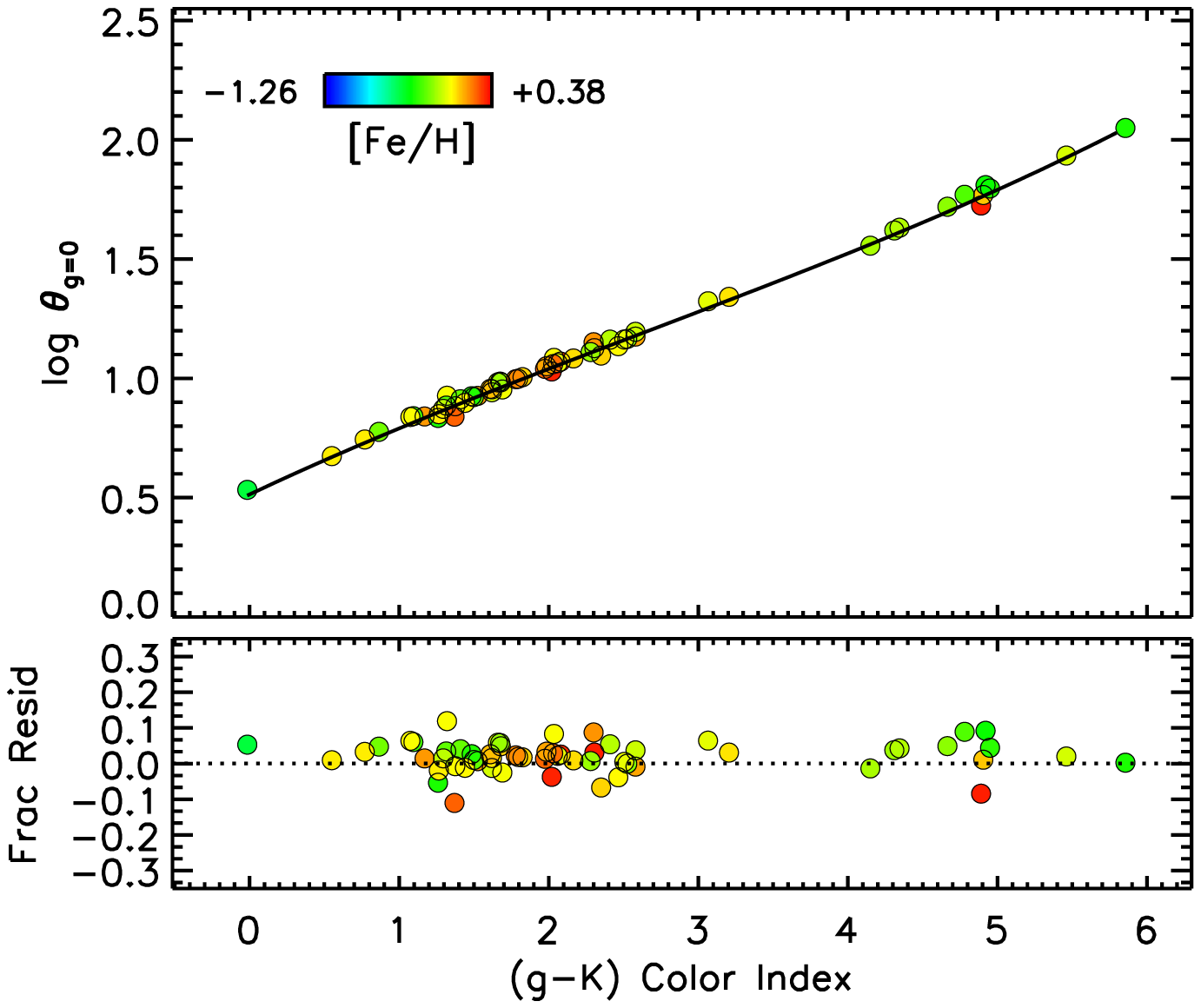, width=0.5\linewidth, clip=}
 \end{tabular}
  \caption[Angular Diameters] {The top panel shows the zero-magnitude limb darkened angular diameter plotted against color index.  The solid black line plots the polynomial relation for the region that the relation holds true (Table\ref{tab:SB_poly3_coeffs}).  The color of the data point reflects the metallicity of the star as depicted in the legend.  The bottom panel shows the fractional residuals to our fit ($\theta_{\rm Obs.} - \theta_{\rm Fit})/\theta_{\rm Obs.}$, where the dotted line indicates zero deviation.  See Section~\ref{sec:data} for details. }
  \label{fig:relations6}
  \end{figure}

\begin{figure}										
\centering
\begin{tabular}{cc}
\epsfig{file=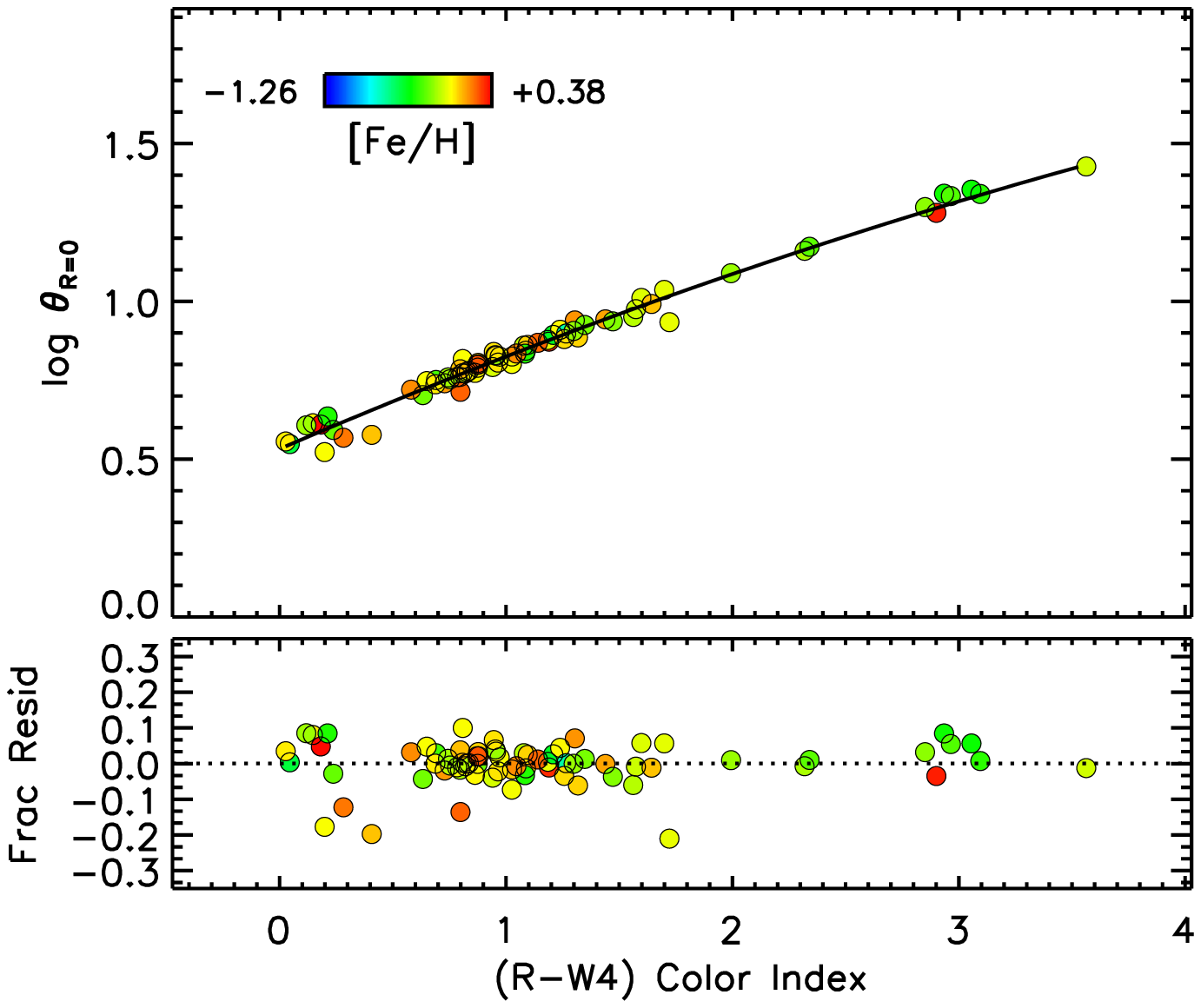, width=0.5\linewidth, clip=}	&
\epsfig{file=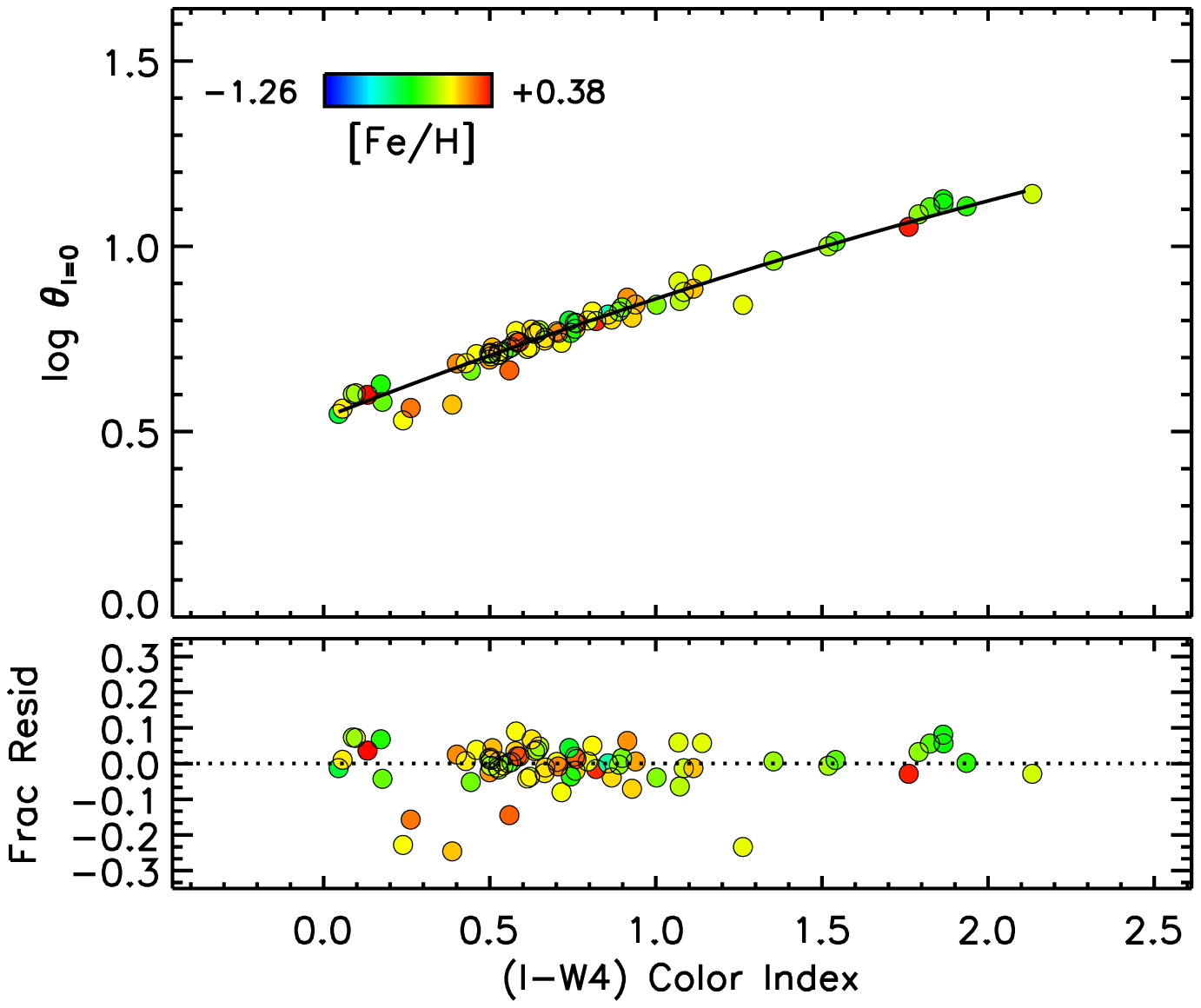, width=0.5\linewidth, clip=}	\\
\epsfig{file=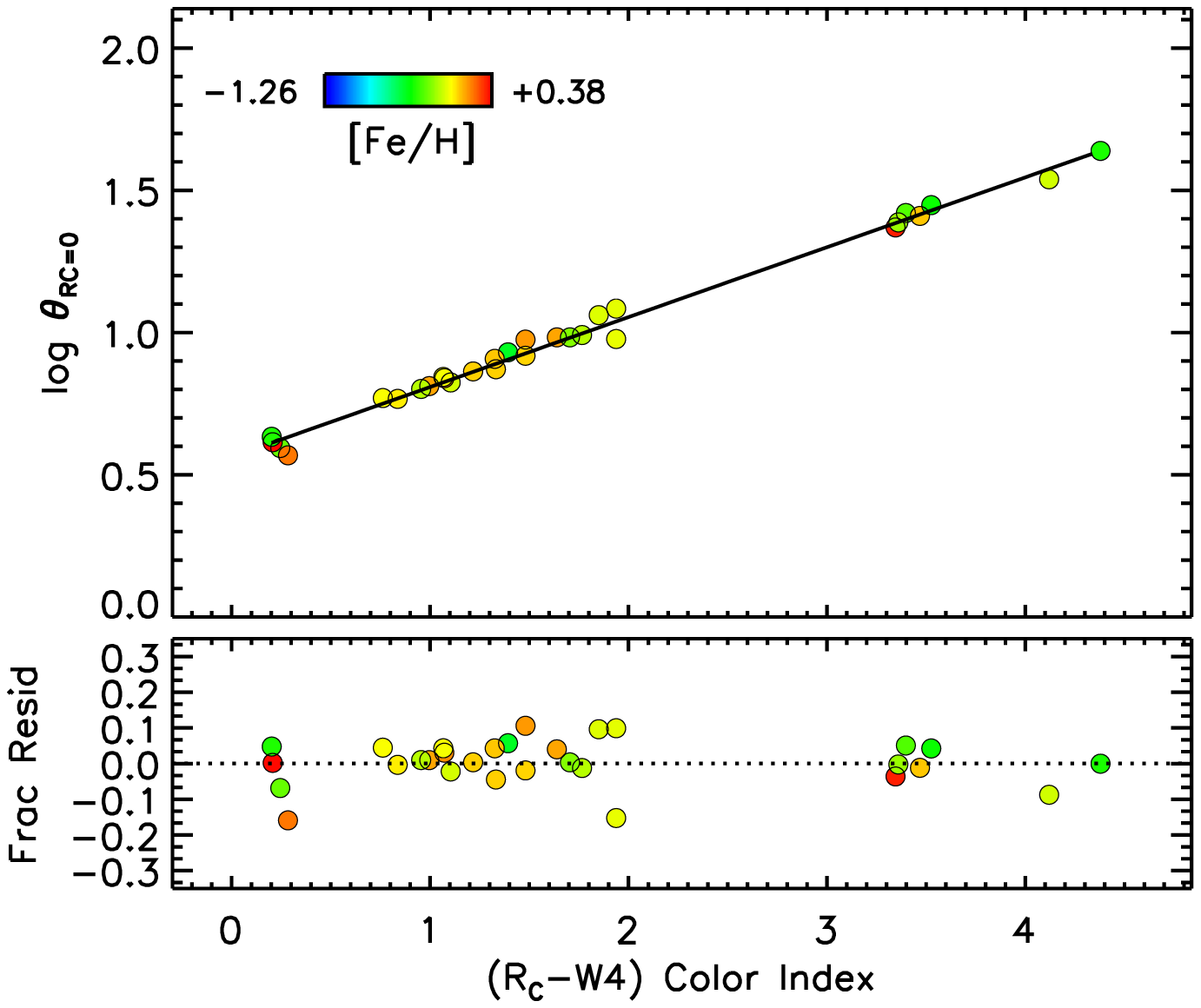, width=0.5\linewidth, clip=}	&
\epsfig{file=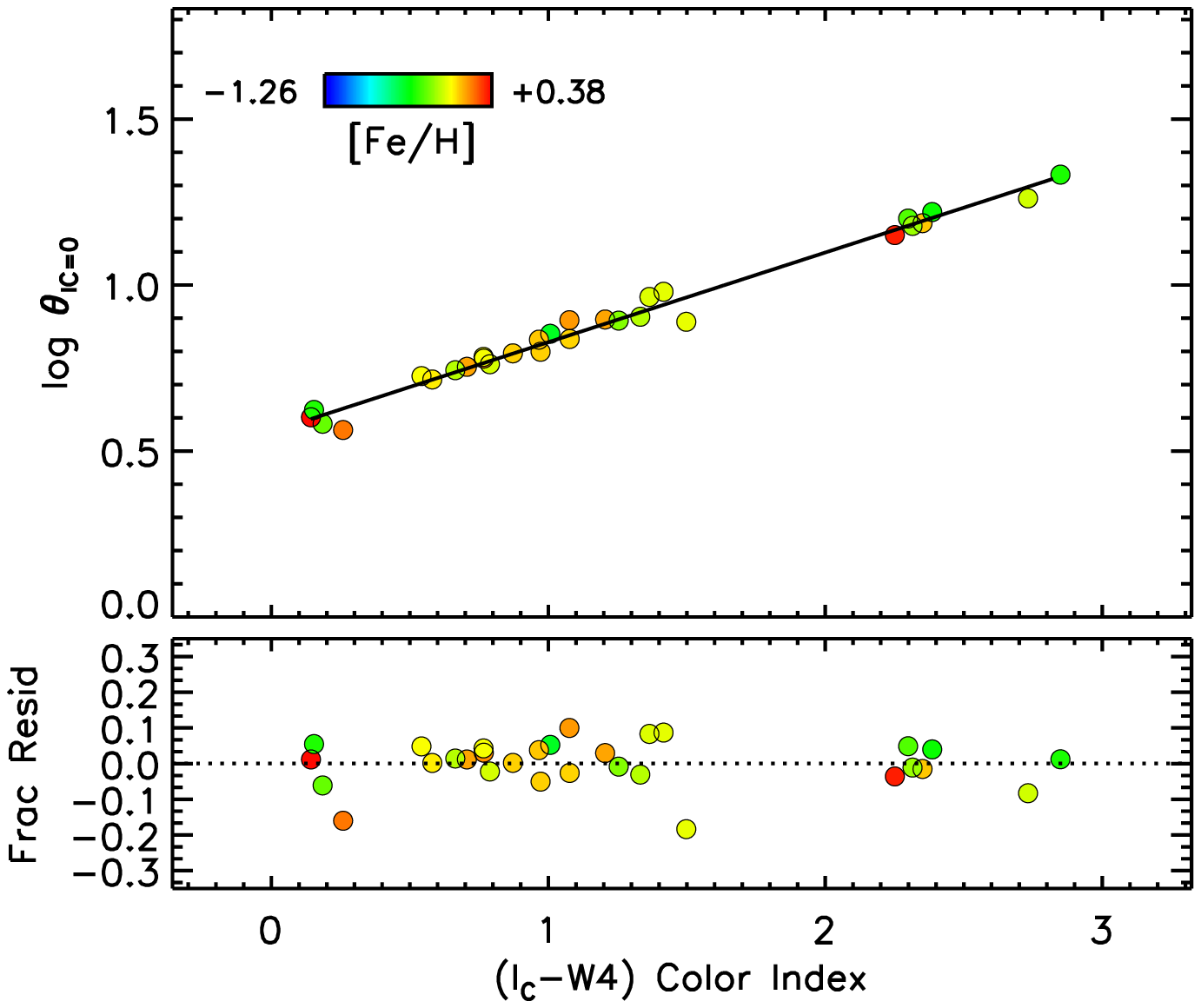, width=0.5\linewidth, clip=}	\\
\epsfig{file=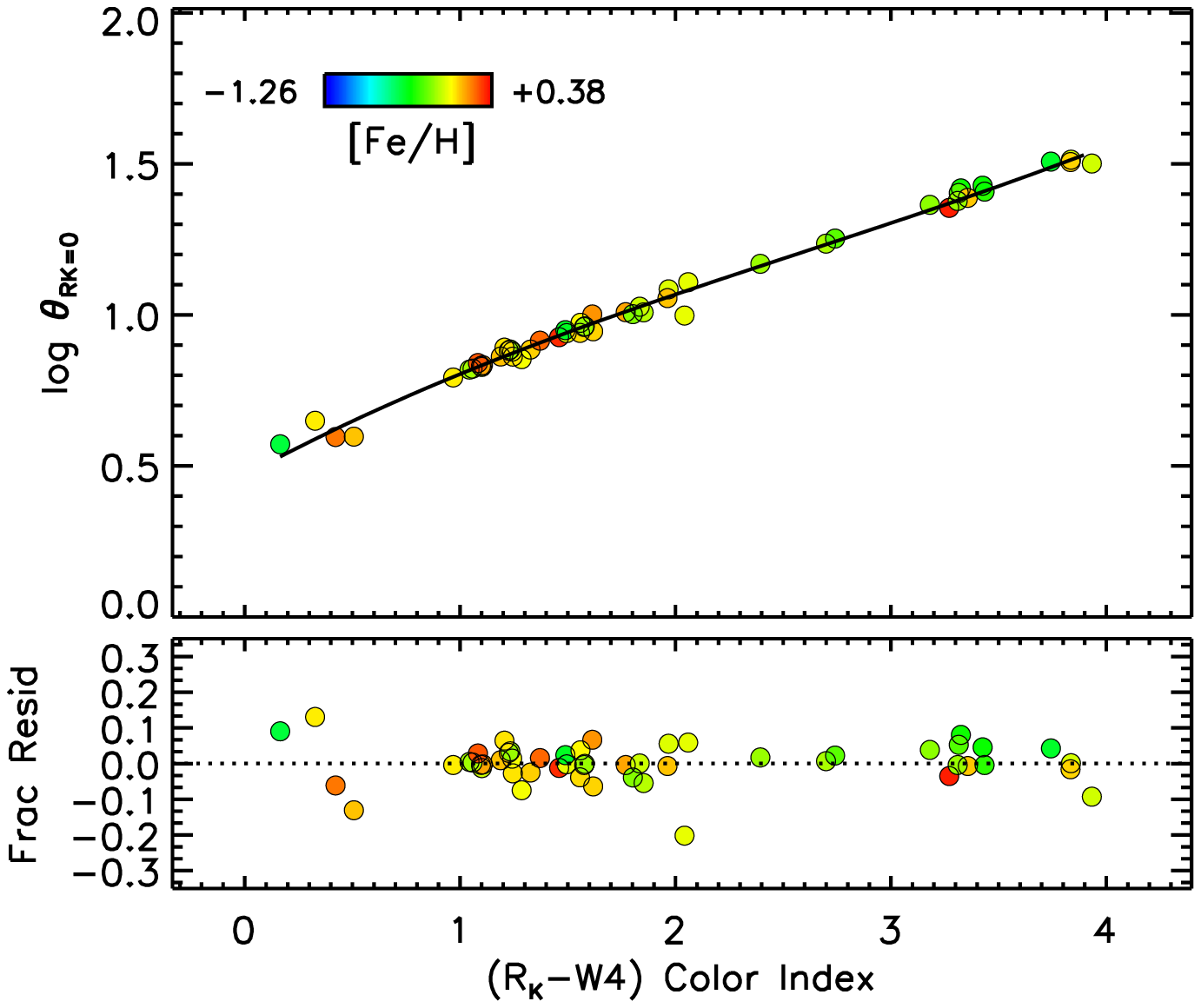, width=0.5\linewidth, clip=}	&
\epsfig{file=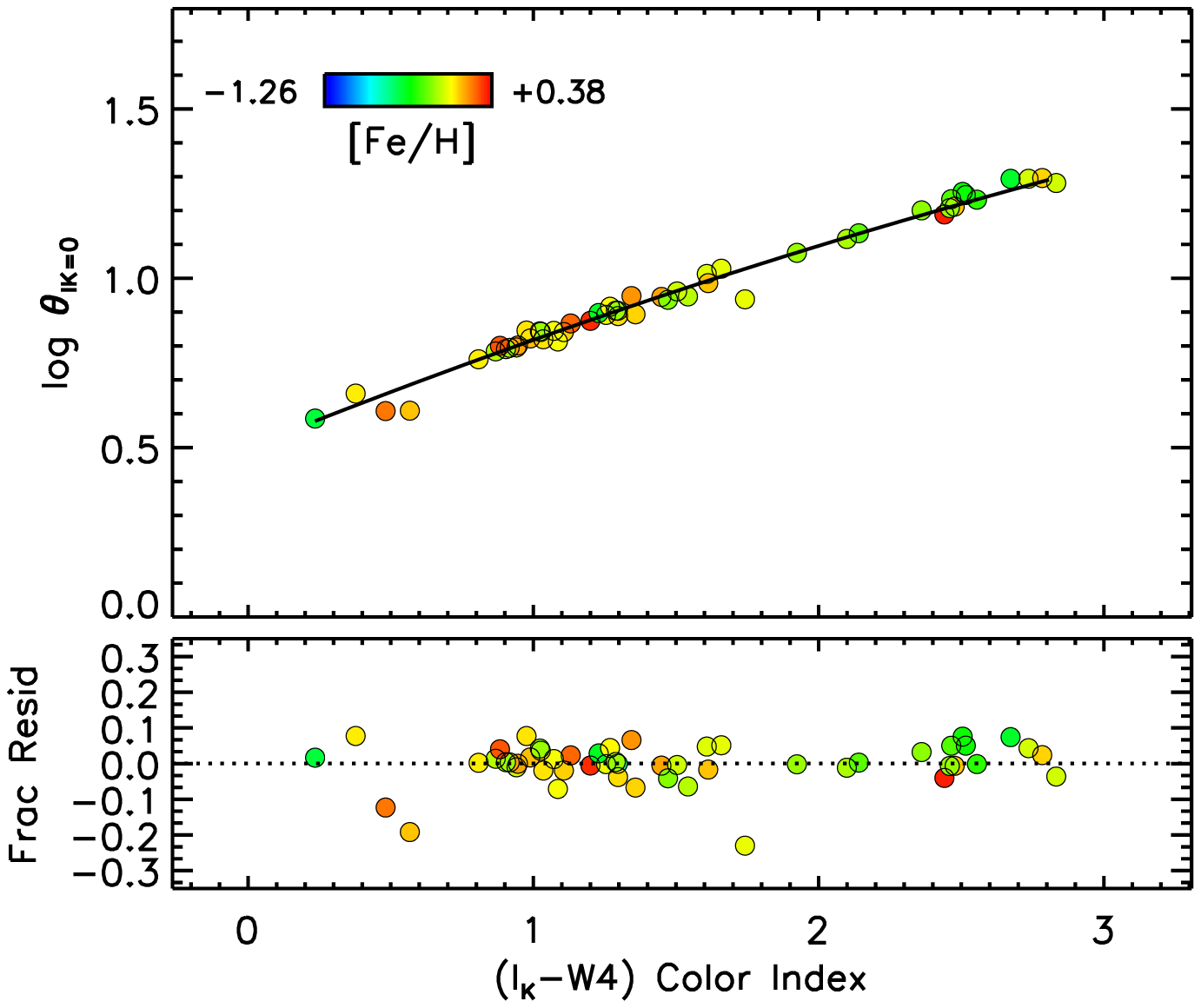, width=0.5\linewidth, clip=}	
 \end{tabular}
  \caption[Angular Diameters] {The top panel shows the zero-magnitude limb darkened angular diameter plotted against color index.  The solid black line plots the polynomial relation for the region that the relation holds true (Table\ref{tab:SB_poly3_coeffs}).  The color of the data point reflects the metallicity of the star as depicted in the legend.  A solution from \citet{ker08a} is plotted as a dashed red line. The bottom panel shows the fractional residuals to our fit ($\theta_{\rm Obs.} - \theta_{\rm Fit})/\theta_{\rm Obs.}$, where the dotted line indicates zero deviation.  See Section~\ref{sec:data} for details. }
  \label{fig:relations7}
  \end{figure}

\begin{figure}										
\centering
\begin{tabular}{cc}
\epsfig{file=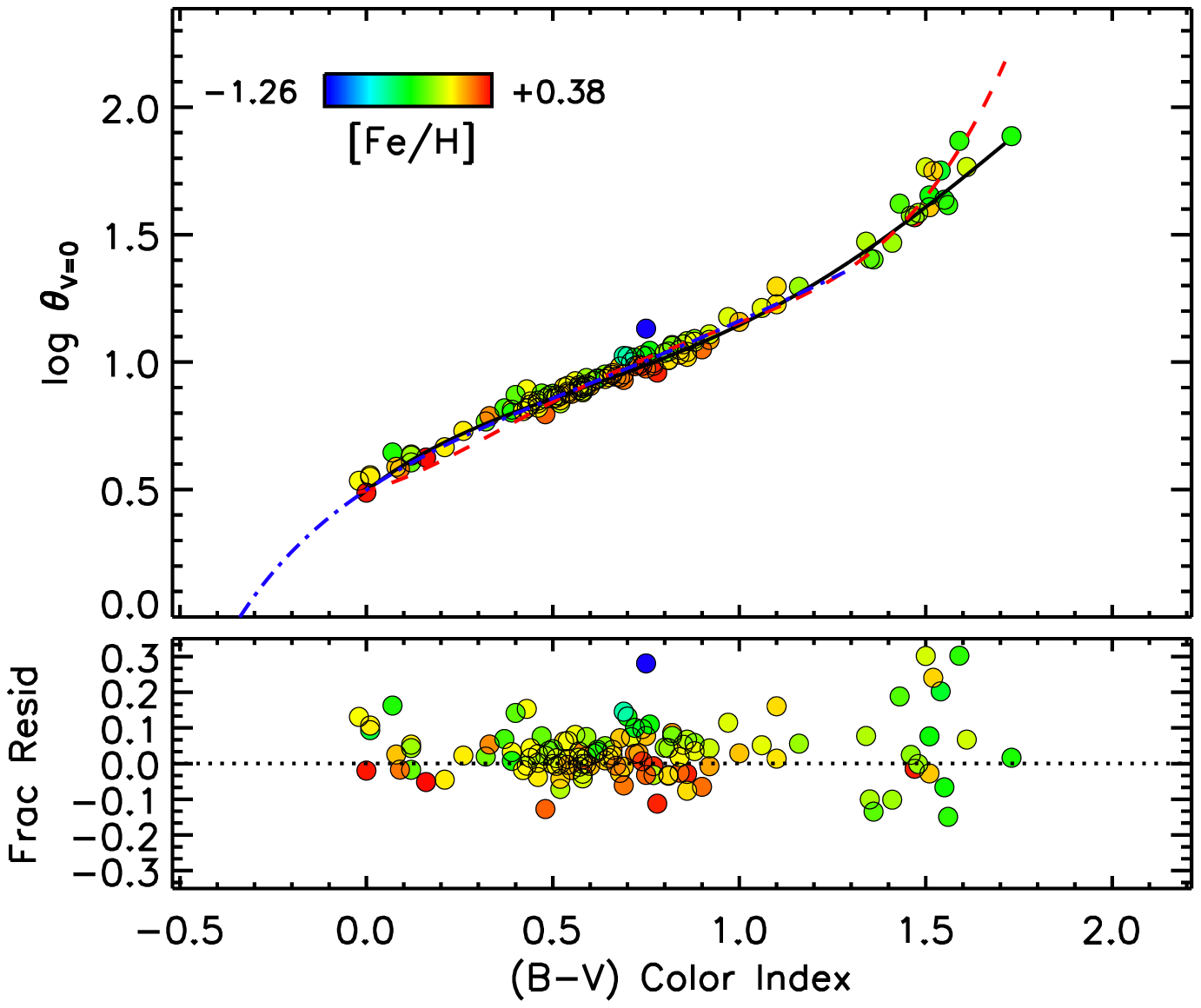, width=0.45\linewidth, clip=} &
\epsfig{file=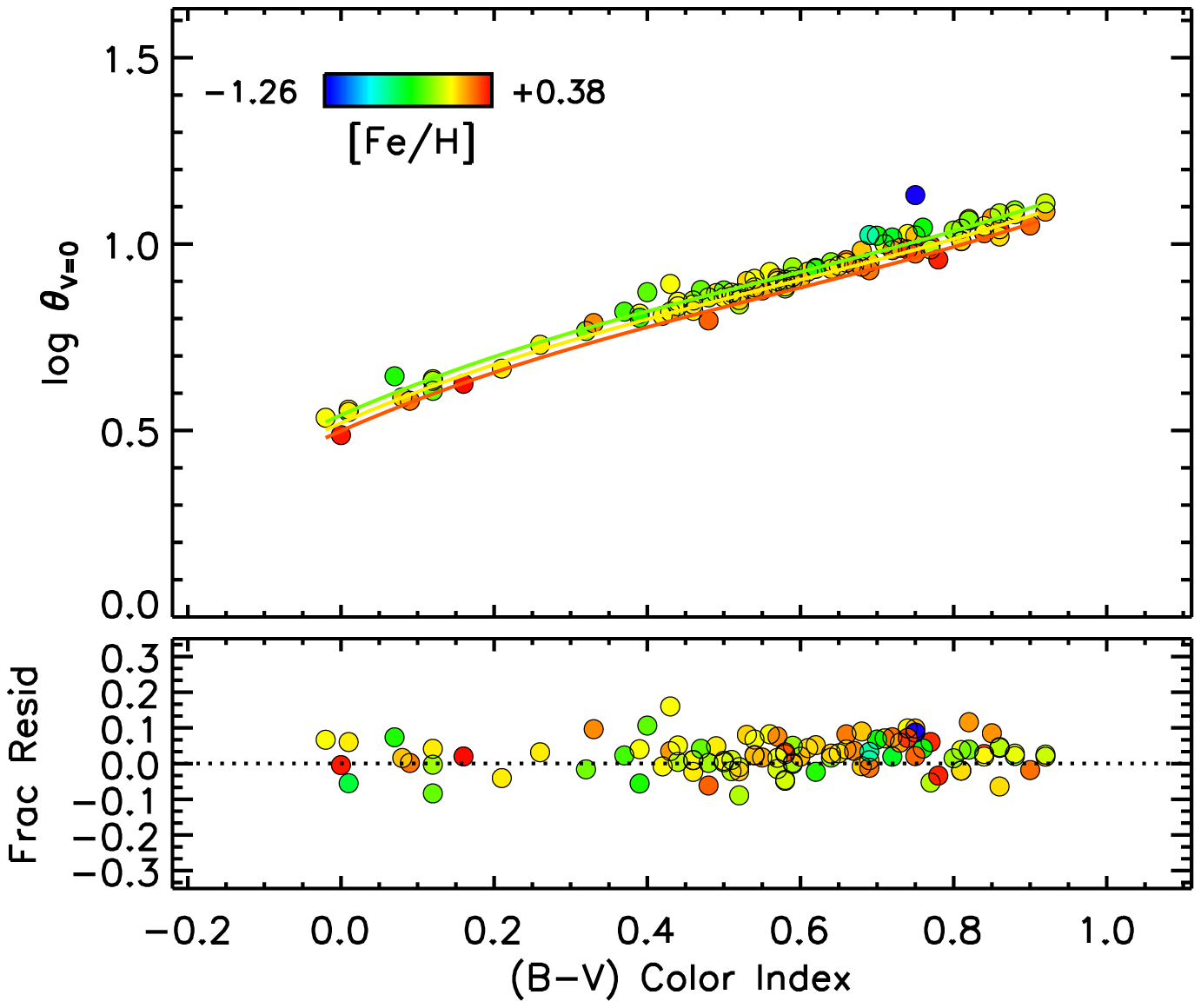, width=0.45\linewidth, clip=}	
 \end{tabular}
  \caption[ ] {The top panels show the zero-magnitude limb darkened angular diameter plotted against the $(B-V)$ color index.  The plot on the left is the metallicity independent solution where the solid black line plots the polynomial relation for the region that the relation holds true (Equation~\ref{eq:poly3}, Table~\ref{tab:SB_poly3_coeffs}).  A solution from \citet{ker08a} is plotted as a dashed red line, and the solution from \citet{bon06} is plotted as a blue dash-dotted line. The plot on the right shows the solution for the metallicity dependent solution (Equation~\ref{eq:poly3_cm_bmv}), where lines of constant metallicity are shown for [Fe/H]~$= +0.25, 0.0, -0.25$~dex (red, yellow, and green, respectively). In all plots, the color of the data point reflects the metallicity of the star as depicted in the legend.  The bottom panel shows the fractional residuals to our fits ($\theta_{\rm Obs.} - \theta_{\rm Fit})/\theta_{\rm Obs.}$, where the dotted line indicates zero deviation. See Figure~\ref{fig:BmV_V_FeH_residuals_log_SB} for close-up view on comparison of the residuals.  See Section~\ref{sec:data} and Section~\ref{sec:discussion} for details and discussion.}
  \label{fig:BmV_V_FeH}
  \end{figure}

\begin{figure}										
\centering
\begin{tabular}{cc}
\epsfig{file=gmr_g_log_SB.eps, width=0.45\linewidth, clip=} &
\epsfig{file=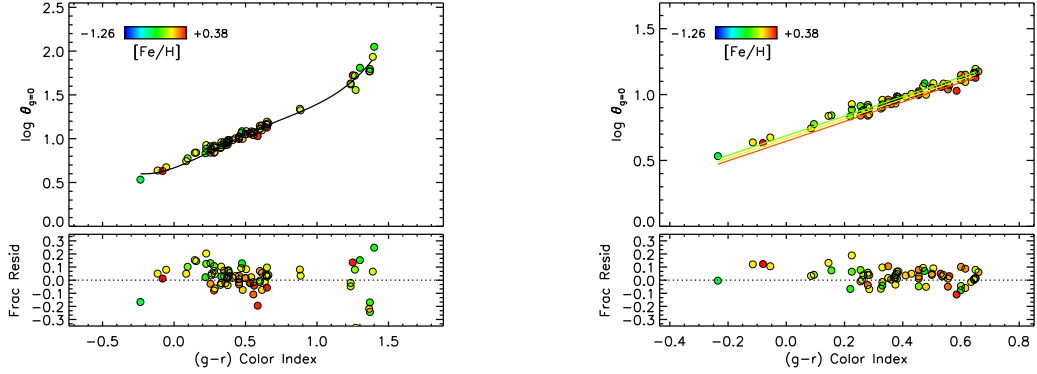, width=0.45\linewidth, clip=}	
 \end{tabular}
  \caption[ ] {The top panels show the zero-magnitude limb darkened angular diameter plotted against the $(g-r)$ color index.  The plot on the left is the metallicity independent solution where the solid black line plots the polynomial relation for the region that the relation holds true (Equation~\ref{eq:poly3}, Table~\ref{tab:SB_poly3_coeffs}).  The plot on the right shows the solution for the metallicity dependent colution (Equation~\ref{eq:poly3_cm_gmr}), where lines of constant metallicity are shown for [Fe/H]~$= +0.25, 0.0, -0.25$~dex (red, yellow, and green, respectively). In all plots, the color of the data point reflects the metallicity of the star as depicted in the legend.  The bottom panel shows the fractional residuals to our fits ($\theta_{\rm Obs.} - \theta_{\rm Fit})/\theta_{\rm Obs.}$, where the dotted line indicates zero deviation. See Figure~\ref{fig:BmV_V_FeH_residuals_log_SB} for close-up view on comparison of the residuals with respect to metallicity.  See Section~\ref{sec:data} and Section~\ref{sec:discussion} for details and discussion.}
  \label{fig:gmr_V_FeH}
  \end{figure}

\begin{figure}										
\centering
\begin{tabular}{cc}
\epsfig{file=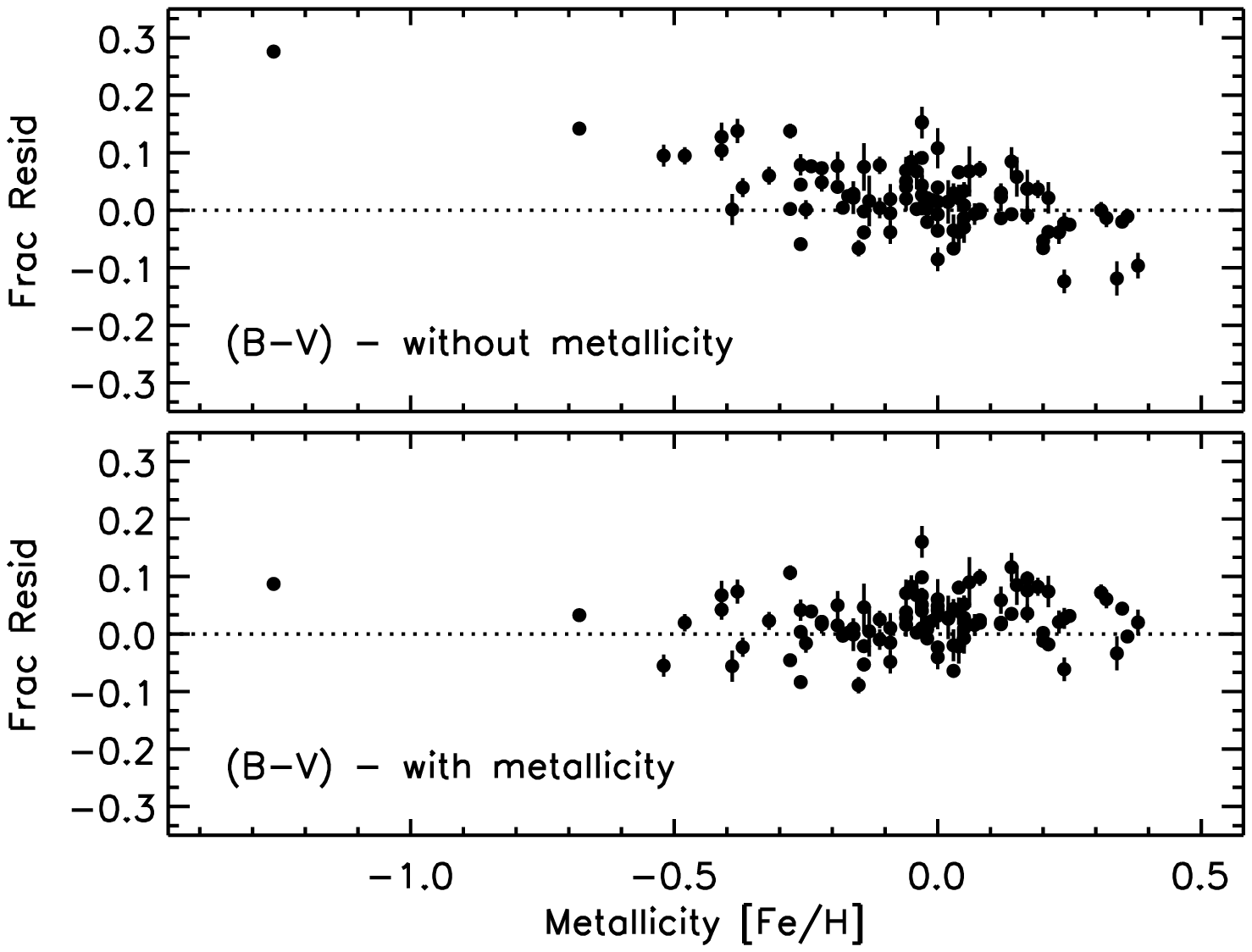, width=0.5\linewidth, clip=} &
\epsfig{file=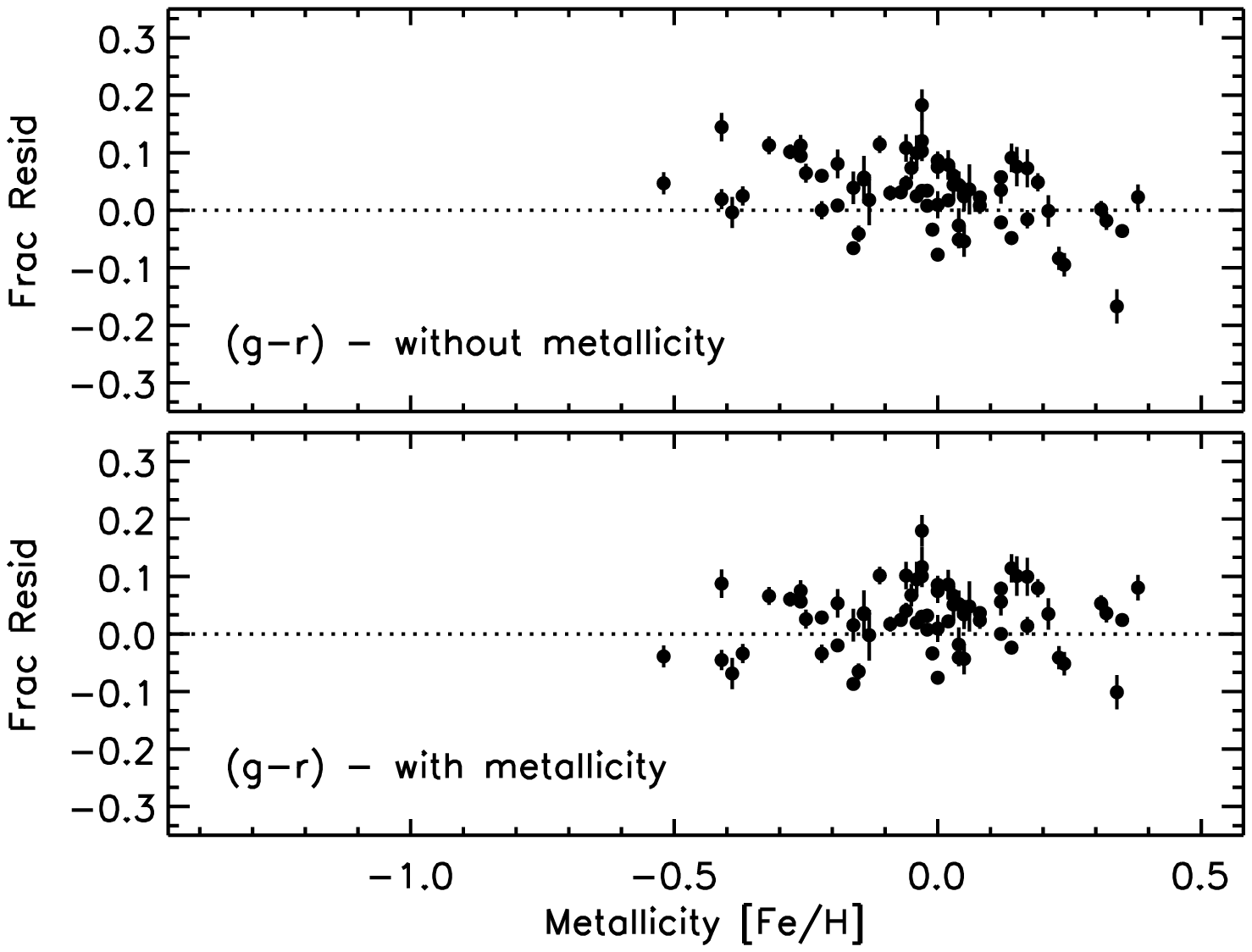, width=0.5\linewidth, clip=} 
 \end{tabular}
  \caption[ ] {Residuals to the model without ({\it top}) and with ({\it bottom}) metallicity dependence for the $(B-V), V$ and $(g-r), g$ relations versus stellar metallicity.  Data and fit for each solution are shown in Figure~\ref{fig:BmV_V_FeH} and \ref{fig:gmr_V_FeH}.  See \S~\ref{sec:photometry}, \S~\ref{sec:metallicity}, and \S~\ref{sec:discussion} for details.}
  \label{fig:BmV_V_FeH_residuals_log_SB}
  \end{figure}

\begin{figure}										
\centering
\epsfig{file=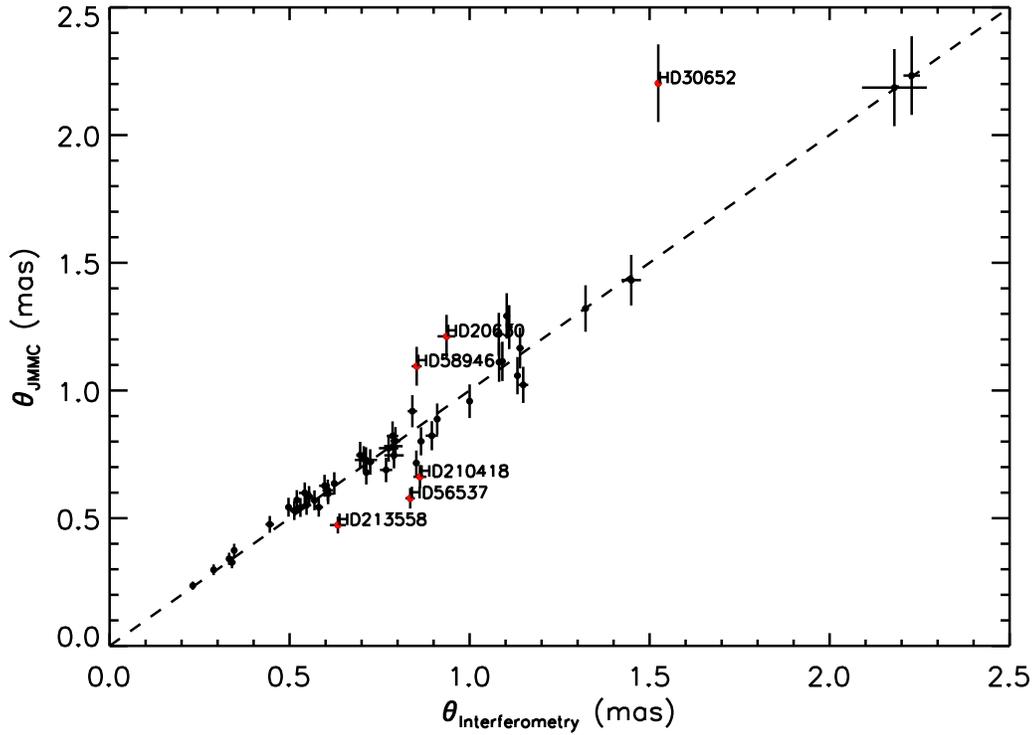, width=0.9\linewidth, clip=} 
  \caption[ ] {Interferometric angular diameters compared with those in the JMMC Catalog \citep{bon06}, showing agreement of $\theta_{\rm Interferometry}/\theta_{\rm JMMC} = 0.999 \pm 0.095$. The dashed line shows a 1:1 agreement.  We show stars with differences greater than 3 sigma in {\it red} and label their HD numbers in the plot window.  See Section~\ref{sec:discussion} for details.}
  \label{fig:Compare_JMMC_diameters_new}
  \end{figure}

In this paper, we take advantage of a large dataset of directly measured, interferometric stellar radii to calibrate surface brightness relations that are able to predict angular sizes of stars through photometry in two bands. We provide the polynomial coefficients for each model in Table~\ref{tab:SB_poly3_coeffs} for a large number of commonly used broad-band colors. All solutions and data are shown in Figures \ref{fig:relations1} through \ref{fig:relations7}.  Compared to previously published work on this topic, we find generally good agreement in the calculated relations with some exceptions, as we discuss in \S \ref{sec:discussion}.  We show a that the $(B-V)$ and $(g-r)$ colors show a clear improvement when including metallicity as an extra term (Figure \ref{fig:BmV_V_FeH} and \ref{fig:gmr_V_FeH}; Equation~\ref{eq:poly3_cm_bmv} and \ref{eq:poly3}).  This is the first time surface brightness relations have shown a dependence on stellar metallicity from empirical data (see discussion in \S \ref{sec:discussion}).

The well-understood procedure to obtain multi-band photometry on stars that are much too faint and small for their radii to be measured interferometrically enables a relatively straightforward and empirical prediction of their angular diameters.  The formulations  presented here expand the parameter space of the color - radius relations in \citet{boy12b} beyond low-mass K- and M-dwarfs because they are insensitive to stellar evolution. Thus, with the knowledge of a star's trigonometric parallax values, this work provides access to physical radii for the full span of A-M type stars.


\acknowledgments

TSB acknowledges support provided through NASA grant ADAP12-0172.  GvB is supported in part through NASA grant NNX13AF01G, and NSF grant AST-1212203.  The authors would like to thank Nicolas Nardetto and Jean-Baptiste Le Bouquin, and Pierre Kervella for useful discussions on the topic. This research has made use of the Jean-Marie Mariotti Center \texttt{SearchCal} service \footnote{Available at http://www.jmmc.fr/searchcal} co-developped by FIZEAU and LAOG/IPAG, and of CDS Astronomical Databases SIMBAD and VIZIER \footnote{Available at http://cdsweb.u-strasbg.fr/}. This research has made use of the SIMBAD literature database, operated at CDS, Strasbourg, France, and of NASA's Astrophysics Data System. This research has made use of the VizieR catalogue access tool, CDS, Strasbourg, France.  This publication makes use of data products from the Wide-field Infrared Survey Explorer, which is a joint project of the University of California, Los Angeles, and the Jet Propulsion Laboratory/California Institute of Technology, funded by the National Aeronautics and Space Administration.

\clearpage
\bibliographystyle{apj}            
\bibliography{apj-jour,paper}      


\newpage

\begin{landscape}
\begin{deluxetable}{lcccccccccc}
\tabletypesize{\tiny}
\tablewidth{0pt}
\tablecaption{Solutions to Angular Diameter Relations\label{tab:SB_poly3_coeffs}}
\tablehead{
\colhead{\textbf{ }} &
\colhead{\textbf{Color}} &
\colhead{\textbf{\# of}} &
\colhead{\textbf{Range}} &
\colhead{\textbf{ }} &
\colhead{\textbf{ }} &
\colhead{\textbf{ }} &
\colhead{\textbf{ }} &
\colhead{\textbf{ }} &
\colhead{\textbf{Reduced}} &
\colhead{\textbf{$\sigma$}} 	 \\
\colhead{\textbf{$\lambda$}} &
\colhead{\textbf{Index}} &
\colhead{\textbf{Points}}	&
\colhead{\textbf{(mag)}}	&
\colhead{\textbf{$a_0 \pm \sigma$}} &
\colhead{\textbf{$a_1 \pm \sigma$}}	&
\colhead{\textbf{$a_2 \pm \sigma$}}	&
\colhead{\textbf{$a_3 \pm \sigma$}}	&
\colhead{\textbf{$a_4 \pm \sigma$}}	&
\colhead{\textbf{$\chi^{2}$}} &
\colhead{\textbf{(\%)}}
}

\startdata

              $V$ & $(B-V)$ &    124 &  [$-0.02-1.73$]  & $ 0.49612 \pm  0.00111$ & $ 1.11136 \pm  0.00939$ & $-1.18694 \pm  0.02541$ & $ 0.91974 \pm  0.02412$ & $-0.19526 \pm  0.00738$ &     73.8 &      7.8 \\ 
                          $V$ & $(V-R_J)$ &     81 &  [$ 0.00-1.69$]  & $ 0.49743 \pm  0.00100$ & $ 0.82790 \pm  0.00304$ & $-0.04227 \pm  0.00196$ & \nodata & \nodata &     60.2 &      7.0  \\ 
                          $V$ & $(V-I_J)$ &     80 &  [$-0.03-2.69$]  & $ 0.51435 \pm  0.00101$ & $ 0.45435 \pm  0.00349$ & $ 0.04152 \pm  0.00390$ & $-0.02124 \pm  0.00108$ & \nodata &     45.4 &      5.9 \\ 
                          $V$ & $(V-R_C)$ &     34 &  [$-0.01-1.24$]  & $ 0.50524 \pm  0.00109$ & $ 1.31557 \pm  0.00788$ & $-0.49134 \pm  0.01850$ & $ 0.26584 \pm  0.01131$ & \nodata &     28.1 &      4.9 \\ 
                          $V$ & $(V-I_C)$ &     34 &  [$-0.02-2.77$]  & $ 0.50659 \pm  0.00103$ & $ 0.56448 \pm  0.00793$ & $ 0.17460 \pm  0.01647$ & $-0.16268 \pm  0.01002$ & $ 0.03292 \pm  0.00184$ &     25.7 &      5.1 \\ 
                          $V$ & $(V-R_K)$ &     64 &  [$-0.21-1.32$]  & $ 0.70071 \pm  0.00056$ & $ 0.93899 \pm  0.00252$ & $-0.09351 \pm  0.00229$ & \nodata & \nodata &    170.0 &      7.7 \\ 
                          $V$ & $(V-I_K)$ &     64 &  [$-0.33-2.42$]  & $ 0.67424 \pm  0.00060$ & $ 0.58321 \pm  0.00150$ & $-0.05227 \pm  0.00073$ & \nodata & \nodata &     67.5 &      5.3\\ 
                            $V$ & $(V-J)$ &     95 &  [$-0.12-4.24$]  & $ 0.52464 \pm  0.00086$ & $ 0.38167 \pm  0.00108$ & $-0.01431 \pm  0.00026$ & \nodata & \nodata &     41.3 &      5.1 \\ 
                            $V$ & $(V-H)$ &     86 &  [$-0.13-4.77$]  & $ 0.53019 \pm  0.00059$ & $ 0.27917 \pm  0.00030$ & \nodata & \nodata & \nodata &     45.4 &      5.3 \\ 
                            $V$ & $(V-K)$ &     97 &  [$-0.15-5.04$]  & $ 0.53246 \pm  0.00057$ & $ 0.26382 \pm  0.00028$ & \nodata & \nodata & \nodata &     34.2 &      4.6 \\ 
             					$V$ & $(V-W3)$ &     44 &  [$ 0.76-5.50$]  & $ 0.57935 \pm  0.00188$ & $ 0.23879 \pm  0.00054$ & \nodata & \nodata & \nodata &      8.6 &      5.1 \\ 
              $V$ & $(V-W4)$ &    111 &  [$ 0.03-5.62$]  & $ 0.52073 \pm  0.00214$ & $ 0.28979 \pm  0.00283$ & $-0.01641 \pm  0.00112$ & $ 0.00144 \pm  0.00013$ & \nodata &     20.9 &      7.2 \\

              $g$ & $(g-r)$ &     79 &  [$-0.23-1.40$]  & $ 0.66728 \pm  0.00203$ & $ 0.58135 \pm  0.01180$ & $ 0.88293 \pm  0.03470$ & $-1.41005 \pm  0.04331$ & $ 0.67248 \pm  0.01736$ &    155.3 &      9.7 \\ 
                            $g$ & $(g-i)$ &     79 &  [$-0.43-2.78$]  & $ 0.69174 \pm  0.00125$ & $ 0.54346 \pm  0.00266$ & $-0.02149 \pm  0.00097$ & \nodata & \nodata  &  111.1 &      9.2 \\ 
                            $g$ & $(g-z)$ &     79 &  [$-0.58-3.44$]  & $ 0.72292 \pm  0.00108$ & $ 0.46563 \pm  0.00203$ & $-0.02499 \pm  0.00061$ & \nodata & \nodata &    123.1 &      9.5 \\ 
                            $g$ & $(g-J)$ &     60 &  [$-0.02-5.06$]  & $ 0.52662 \pm  0.00226$ & $ 0.34439 \pm  0.00216$ & $-0.00920 \pm  0.00039$ & \nodata & \nodata &     19.4 &      4.9 \\ 
                            $g$ & $(g-H)$ &     53 &  [$ 0.75-5.59$]  & $ 0.46684 \pm  0.00571$ & $ 0.36437 \pm  0.00667$ & $-0.04206 \pm  0.00229$ & $ 0.00493 \pm  0.00024$ &      \nodata  & 23.1  &      4.7 \\ 
                            $g$ & $(g-K)$ &     60 &  [$-0.01-5.86$]  & $ 0.51356 \pm  0.00443$ & $ 0.29555 \pm  0.00506$ & $-0.02164 \pm  0.00171$ & $ 0.00272 \pm  0.00018$ & \nodata &     20.7 &      4.2 \\ 

          $R_J$ & $(R_J-J)$ &     74 &  [$-0.12-1.86$]  & $ 0.54161 \pm  0.00081$ & $ 0.44407 \pm  0.00370$ & $ 0.11255 \pm  0.00668$ & $-0.06697 \pm  0.00278$ & \nodata &     50.9 &      5.6 \\ 
                    $R_J$ & $(R_J-H)$ &     66 &  [$-0.13-2.80$]  & $ 0.53572 \pm  0.00066$ & $ 0.31753 \pm  0.00058$ & \nodata & \nodata & \nodata &     53.2 &      5.5 \\ 
                    $R_J$ & $(R_J-K)$ &     75 &  [$-0.15-3.06$]  & $ 0.53954 \pm  0.00063$ & $ 0.29108 \pm  0.00052$ & \nodata & \nodata & \nodata &     37.2 &      4.6 \\ 
                      $R_J$ & $(R_J-W4)$ &     74 &  [$ 0.03-3.56$]  & $ 0.53243 \pm  0.00161$ & $ 0.30816 \pm  0.00236$ & $-0.01557 \pm  0.00067$ & \nodata & \nodata &     23.0 &      5.9 \\ 
                    $R_C$ & $(R_C-J)$ &     27 &  [$-0.11-3.00$]  & $ 0.53356 \pm  0.00093$ & $ 0.44321 \pm  0.00178$ & $-0.02541 \pm  0.00063$ & \nodata & \nodata &     49.0 &      4.7 \\ 
                    $R_C$ & $(R_C-H)$ &     26 &  [$-0.12-3.53$]  & $ 0.53563 \pm  0.00075$ & $ 0.30231 \pm  0.00048$ & \nodata & \nodata & \nodata &     62.3 &      5.6 \\ 
                    $R_C$ & $(R_C-K)$ &     27 &  [$-0.14-3.80$]  & $ 0.53282 \pm  0.00075$ & $ 0.28271 \pm  0.00045$ & \nodata & \nodata & \nodata &     45.7 &      4.7 \\ 
                      $R_C$ & $(R_C-W4)$ &     30 &  [$ 0.20-4.38$]  & $ 0.56261 \pm  0.00110$ & $ 0.24575 \pm  0.00050$ & \nodata & \nodata & \nodata &     32.2 &      6.3 \\ 
                    $R_K$ & $(R_K-J)$ &     59 &  [$ 0.09-2.58$]  & $ 0.47966 \pm  0.00129$ & $ 0.49606 \pm  0.00227$ & $-0.03434 \pm  0.00087$ & \nodata & \nodata &     105.6 &      5.8 \\ 
                    $R_K$ & $(R_K-H)$ &     57 &  [$ 0.08-3.17$]  & $ 0.51110 \pm  0.00077$ & $ 0.31285 \pm  0.00050$  & \nodata & \nodata & \nodata &      78.8 &      5.3 \\ 
                    $R_K$ & $(R_K-K)$ &     61 &  [$ 0.06-3.43$]  & $ 0.51325 \pm  0.00076$ & $ 0.29002 \pm  0.00046$ & \nodata & \nodata & \nodata &     53.7 &      4.5 \\ 
                      $R_K$ & $(R_K-W4)$ &     52 &  [$ 0.17-3.93$]  & $ 0.46844 \pm  0.00520$ & $ 0.38664 \pm  0.00902$ & $-0.05846 \pm  0.00479$ & $ 0.00748 \pm  0.00077$ & \nodata &     19.8 &      5.4 \\ 
                    $I_J$ & $(I_J-J)$ &     75 &  [$-0.09-0.78$]  & $ 0.56179 \pm  0.00061$ & $ 0.80862 \pm  0.00204$ & \nodata & \nodata & \nodata &     99.1 &      9.1 \\
                    $I_J$ & $(I_J-H)$ &     66 &  [$-0.10-1.37$]  & $ 0.53904 \pm  0.00086$ & $ 0.36206 \pm  0.00270$ & $ 0.03569 \pm  0.00201$ & \nodata & \nodata &     69.9 &      6.6 \\ 
                    $I_J$ & $(I_J-K)$ &     75 &  [$-0.12-1.63$]  & $ 0.53510 \pm  0.00066$ & $ 0.35175 \pm  0.00088$  & \nodata & \nodata & \nodata &     42.9 &      5.2 \\ 
                      $I_J$ & $(I_J-W4)$ &     74 &  [$ 0.04-2.13$]  & $ 0.53766 \pm  0.00171$ & $ 0.34918 \pm  0.00392$ & $-0.02832 \pm  0.00180$ & \nodata & \nodata &     27.5 &      6.6 \\ 
                    $I_C$ & $(I_C-J)$ &     27 &  [$-0.10-1.47$]  & $ 0.55013 \pm  0.00074$ & $ 0.54738 \pm  0.00117$ & \nodata & \nodata & \nodata &     58.1 &      5.6 \\ 
                    $I_C$ & $(I_C-H)$ &     26 &  [$-0.11-2.00$]  & $ 0.53026 \pm  0.00077$ & $ 0.36595 \pm  0.00079$ & \nodata & \nodata & \nodata &     97.6 &      7.4 \\ 
                    $I_C$ & $(I_C-K)$ &     27 &  [$-0.13-2.27$]  & $ 0.52800 \pm  0.00077$ & $ 0.32919 \pm  0.00071$ & \nodata & \nodata & \nodata &     64.9 &      5.4 \\ 
                      $I_C$ & $(I_C-W4)$ &     30 &  [$ 0.14-2.85$]  & $ 0.55786 \pm  0.00114$ & $ 0.27009 \pm  0.00076$ & \nodata & \nodata & \nodata &     30.5 &      6.4 \\ 
                    $I_K$ & $(I_K-J)$ &     59 &  [$ 0.21-1.48$]  & $ 0.44692 \pm  0.00097$ & $ 0.59745 \pm  0.00125$ & \nodata & \nodata & \nodata &    146.4 &      7.8 \\ 
                    $I_K$ & $(I_K-H)$ &     57 &  [$ 0.20-2.07$]  & $ 0.48163 \pm  0.00090$ & $ 0.37123 \pm  0.00078$ & \nodata & \nodata & \nodata &    113.7 &      7.1 \\ 
                    $I_K$ & $(I_K-K)$ &     61 &  [$ 0.18-2.33$]  & $ 0.49053 \pm  0.00087$ & $ 0.33156 \pm  0.00069$ & \nodata & \nodata & \nodata &     73.8 &      5.5 \\ 
                      $I_K$ & $(I_K-W4)$ &     52 &  [$ 0.23-2.83$]  & $ 0.50028 \pm  0.00345$ & $ 0.33757 \pm  0.00484$ & $-0.01992 \pm  0.00145$ & \nodata & \nodata &     23.6 &      5.9 \\

\enddata

\tablecomments{Solutions to Equation~\ref{eq:poly3}. The $V,(B-V)$ solution in this table (marked with a $^{\dagger}$) does not include metallicity, and for this color index we advise the use of the alternate metallicity-dependent solution presented in the text (Equation~\ref{eq:poly3_cm_bmv}).  Refer to Section ~\ref{sec:data} for details.}

\end{deluxetable}

 \end{landscape}
 


\newpage

\newpage


\newpage


\newpage


\newpage


\newpage


\newpage


\newpage


\newpage
  
\newpage

\newpage

\newpage


\end{document}